\documentclass[prd,aps,groupedaddress,showpacs,preprintnumbers,
preprint,
nofootinbib]{revtex4}%
\usepackage{graphicx}
\usepackage{amsmath}
\usepackage{bm}
\usepackage{epsfig}
\usepackage{amssymb}

\newcommand{\slsh}{\!\!\!\slash}

\begin{document}
\vspace*{1cm}
\title{Factorization in exclusive quarkonium production}
\author{Geoffrey~T.~Bodwin}
\affiliation{High Energy Physics Division, Argonne National Laboratory,\\
9700 South Cass Avenue, Argonne, Illinois 60439, USA}
\author{Xavier \surname{Garcia i Tormo}}
\affiliation{High Energy Physics Division, Argonne National Laboratory,\\
9700 South Cass Avenue, Argonne, Illinois 60439, USA}
\affiliation{Department of Physics, University of Alberta,
Edmonton, Alberta, Canada T6G 2G7 \footnote{Current address}}
\author{Jungil~Lee}
\affiliation{Department of Physics, Korea University, Seoul 136-701, Korea}

\date{\today}

\preprint{\begin{tabular}{l}ANL-HEP-PR-09-97  \\Alberta Thy 15-09\end{tabular}}
\pacs{12.38.-t, 12.38.Bx, 14.40.Pq}

\begin{abstract}
We present factorization theorems for two exclusive heavy-quarkonium
production processes: production of two quarkonia in $e^+e^-$
annihilation and production of a quarkonium and a light meson in
$B$-meson decays. We describe the general proofs of factorization and
supplement them with explicit one-loop analyses, which illustrate some
of the features of the soft-gluon cancellations. We find that violations
of factorization are generally suppressed relative to the factorized
contributions by a factor $v^2m_c/Q$ for each $S$-wave charmonium and 
a factor $m_c/Q$ for each $L$-wave charmonium with $L>0$. Here, $v$ is
the velocity of the heavy quark or antiquark in the quarkonium rest
frame, $Q=\sqrt{s}$ for $e^+e^-$ annihilation, $Q=m_B$ for $B$-meson
decays, $\sqrt{s}$ is the $e^+e^-$ center-of-momentum energy, $m_c$
is the charm-quark mass, and $m_B$ is the $B$-meson mass.
There are modifications to the suppression factors if quantum-number
restrictions apply for the specific process.
\end{abstract}

\maketitle

\tableofcontents

\section{Introduction \label{sec:int}}
A crucial step in the calculation of the amplitudes for hard-scattering
hadronic processes is the separation of the effects of the strong
interactions into short-distance and long-distance contributions. The
short-distance contributions are, by virtue of asymptotic freedom in
quantum chromodynamics (QCD), perturbatively calculable, while the
long-distance contributions are parametrized in terms of inherently
nonperturbative quantities. These separations are usually embodied in
factorization theorems for the processes. In the
case of hard-scattering processes that involve
heavy-quarkonium states, it has been proposed that the
effective theory nonrelativistic QCD (NRQCD) could be used to describe
the separation of perturbative effects that produce a heavy-quark pair
from the nonperturbative effects that bring about the evolution of the
heavy-quark pair into the quarkonium bound state \cite{Bodwin:1994jh}.
Recently, progress has been made in understanding factorization issues
in inclusive heavy-quarkonium production processes
\cite{Nayak:2005rt,Nayak:2005rw,Nayak:2007mb,Nayak:2007zb}. However, a 
proof of factorization to all orders in QCD perturbation theory is still 
lacking for inclusive quarkonium production. In the present paper we 
discuss factorization for exclusive quarkonium production.

The exclusive production of double-charmonium states in $e^+e^-$
annihilation has provided an important testing ground in which to
compare predictions of theoretical models of charmonium production with
experimental measurements. Measurements of the cross sections for
double-charmonium production by the Belle \cite{Abe:2004ww} and BABAR
\cite{Aubert:2005tj} collaborations have, in several instances,
disagreed with theoretical predictions 
\cite{Braaten:2002fi,Liu:2002wq,Hagiwara:2003cw,Bodwin:2007ga}
and have led to a re-examination of  the bases for those predictions.

The exclusive decays of $B$ mesons into a light meson plus a charmonium
state are also of interest, partly because they could provide new
constraints on the Cabibbo-Kobayashi-Maskawa (CKM) matrix and enhance
our understanding of the origins of CP violation. However, the
nonperturbative effects of the strong interactions are significant in
such processes and must be taken into account in order to make reliable
QCD-based calculations of the process rates. Factorization theorems for
these processes would provide a first-principles framework within which
to take into account the strong-interaction effects. In the case of
exclusive decays of $B$ mesons into a light meson plus a charmonium state,
several factorization theorems have been proposed
\cite{Beneke:2000ry,Chay:2000xn,Bobeth:2007sh}.

In this paper, we present proofs, valid to all orders in QCD
perturbation theory, of factorization theorems for the exclusive
quarkonium-production processes mentioned above, giving details of the
proofs that were summarized in Ref.~\cite{Bodwin:2008nf}. These are
the first proofs of factorization theorems for quarkonium production.
We also present explicit calculations at one-loop order that
illustrate key features of the general arguments. Although our
analyses are for the specific cases of $B$ decays and $e^+e^-$
annihilation, the techniques that we describe should apply to other
exclusive quarkonium-production processes, and may also shed light on
factorization in inclusive quarkonium production. However, we note that,
because we consider exclusive two-body quarkonium-production processes,
rather than inclusive quarkonium production, we avoid the issues raised
in Ref.~\cite{Nayak:2005rw,Nayak:2005rt} concerning light particles that
are comoving with a quarkonium and the issues raised in
Ref.~\cite{Nayak:2007mb,Nayak:2007zb} concerning the
color-transfer-enhancement mechanism that appears when an additional
heavy quark is comoving with a quarkonium.

In the analysis of Ref.~\cite{Bodwin:2008nf}, it was assumed that gluons
cannot have transverse momentum components that are smaller than the QCD
scale, $\Lambda_{\rm QCD}$. The possibility that external on-shell lines
can emit gluons of arbitrarily low energy was discussed in detail in
Ref.~\cite{Bodwin:2009cb}, and it will be considered here as well. The
factorization theorems stated in Ref.~\cite{Bodwin:2008nf} remain
unchanged.

In the case of the exclusive  production of double-charmonium states in
$e^+e^-$ annihilation, we will argue that the production amplitude can be
written in the following factorized form: 
\begin{equation}
\mathcal{A}\left(e^+e^-\to\gamma^*\to H_1+H_2\right)
=\sum_{ij}A_{ij}\left<H_1\vert\mathcal{O}_i\vert 0\right>
\left<H_2\vert\mathcal{O}_j\vert 0\right>. 
\label{ep-em-fact}
\end{equation} 
The factors $\left<H_{n}\vert\mathcal{O}_i\vert 0\right>$ are NRQCD matrix
elements, which describe the nonperturbative evolution of the charm-quark 
and the charm-antiquark ($c\bar c$) pair into a charmonium state
$H_{n}$. The sum over the matrix elements is organized as an expansion in
powers of $v$, the relative velocity between the $c$ and the $\bar
c$ in the charmonium rest frame. (For charmonium, $v^2\approx 0.3$.) The
quantity $A_{ij}$ is a short-distance coefficient, which contains the
amplitude for an $e^+e^-$ pair to annihilate through a virtual photon into
two $c\bar c$ pairs in the color and angular-momentum states of
the NRQCD operators
$\mathcal{O}_i$ and $\mathcal{O}_j$. 

In the case of $e^+e^-$ annihilation, we define the hard-scattering
scale $Q\equiv\sqrt{s}$ to be the center-of-momentum (CM) energy of the
$e^+e^-$ pair. We will argue that the factorized form in
Eq.~(\ref{ep-em-fact}) holds up to corrections of relative order
$f_1f_2$, where $f_l=v^2m_c/Q$ for an $S$-wave charmonium $H_l$,
$f_l=m_c/Q$ for an $L$-wave charmonium $H_l$ with $L>0$, and $m_c$ is the
charm-quark mass.  As we will discuss in detail, these suppression
factors are modified if quantum-number restrictions apply for the
specific process.

In the case of exclusive decays of $B$ mesons into a light meson plus a
charmonium state, we will argue that the decay amplitude can be written in
the following factorized form\footnote{The two terms in the
factorization formula (\ref{B-meson-fact}) are analogous to the two
terms in the factorization formula in Eq.~(4) of
Ref.~\cite{Beneke:2000ry} for the case of decays into two light mesons.}:
\begin{equation}
\mathcal{A}\left(B\to H_{1}+K\right)=\sum_{ife}
F^{B\to K}_{f(i,e)}(M_1^2) A_{ie}
\langle H_{1}\vert\mathcal{O}_i\vert
0\rangle +\sum_{ije}A'_{ije}\otimes \Phi_{Kj}\otimes \Phi_{B1}
\langle H_{1}\vert\mathcal{O}_i\vert 0\rangle.
\label{B-meson-fact}
\end{equation}
Again, the factors $\left<H_{1}\vert\mathcal{O}_i\vert 0\right>$ are
the NRQCD matrix elements, which describe the nonperturbative evolution
of the $c\bar c$ pair into a charmonium state $H_{1}$. The quantities
$F_f^{B\to K}$, $\Phi_{Kj}$, and $\Phi_{B1}$ are also nonperturbative
objects, which we describe below. The quantities $A_{ie}$ and $A'_{ije}$
are short-distance coefficients. They contain the amplitude for the
electroweak vertex to produce a $c\bar c$ pair in the color and
angular-momentum state of $\mathcal{O}_i$. We approximate the
electroweak vertex as a local four-fermion vertex. The sum over $e$ is
over the various operators in the electroweak effective action. The sum
over $f$ is over the allowed form factors that result from shrinking the
hard subdiagram (to be described later) to a local vertex with respect
to the $B$-meson-to-light-meson transition process. The symbol $\otimes$
represents the convolution of a short-distance coefficient with the
light-cone distributions of the light meson and the $B$ meson.

The first term of Eq.~(\ref{B-meson-fact}) contains a
$B$-meson-to-light-meson form factor
\begin{equation}
F^{B\to K}_{f(i,e)}(M_1^2)=
\langle K\vert
\bar{\Psi}_l\Gamma_{f(i,e)}\Psi_b\vert B\rangle.
\label{form-factor}
\end{equation}
Here, $B$ and $K$ denote the $B$ meson and the light meson,
respectively, and $M_1$ is the charmonium mass. The quantity
$\Gamma_{f(i,e)}$ is the product of a Dirac matrix and a color matrix
that arises when one shrinks the hard-scattering subdiagram  to
a point with respect to the $B$-meson-to-light-meson transition
amplitude.  It is understood that the fields $\bar{\Psi}_l$
and $\Psi_b$ are in a color-singlet state.
Following Ref.~\cite{Beneke:2000ry}, we define $F_{f}^{B\to K}$ in the
first term in Eq.~(\ref{B-meson-fact}) to be the ``physical'' meson form
factor, which contains both hard and soft contributions. Then, in the
second term in Eq.~(\ref{B-meson-fact}), one must omit from the
short-distance coefficients the hard contributions that are already
contained in the first term in Eq.~(\ref{B-meson-fact}).

The second term of Eq.~(\ref{B-meson-fact}) involves the light-cone
distribution amplitude(s) of the light meson $\Phi_{Kj}$, which are 
defined by the expression
\begin{eqnarray}
\label{eq:PHI-K}
&&
\frac{p_{K}^-}{\pi}
\int_{-\infty}^{+\infty}
dx^+ {\rm exp}[ - i(2y-1) p_{K}^-x^+]
 \langle  K(p_{K})\vert \bar{\Psi}_{\alpha}(x^+)
P[x^+,-x^+]
\Psi_{\beta}(-x^+)\vert 
 0\rangle
\nonumber\\
&&\quad\quad
\equiv\sum_j \Phi_{Kj}(y)
\left[\Gamma_{Kj}
\right]_{\alpha\beta},
\end{eqnarray}
and the light-cone distribution of the $B$ meson $\Phi_{B1}$, which is
defined by the expression
\begin{eqnarray}
\label{eq:PHI-B}
&&
\frac{p_B^+}{2\pi}
\int_{-\infty}^{+\infty}
dx^- {\rm exp}[ i\xi p_B^+x^-]
 \langle 0\vert \bar{\Psi}_{l\beta}(x^-)
P[x^-,0]
\Psi_{b\alpha}(0)\vert 
 B(p_B)\rangle
\nonumber\\
&&\quad\quad\equiv\sum_m \Phi_{Bm}(\xi) 
\left[\Gamma_{Bm}\right]_{\alpha\beta}\nonumber\\
&&\quad\quad \approx
-\frac{if_B}{4}
\left\{
(/\!\!\!p_b+m_b)\gamma_5
\left[\Phi_{B1}(\xi)+/\!\!\!n_-\Phi_{B2}(\xi)
\right]
\right\}_{\alpha\beta}.
\end{eqnarray}
Here, $\Psi$ is the quark field, $\alpha$ and $\beta$ are Dirac indices, 
 $\bar{\Psi}$ and $\Psi$ in each matrix element are understood to be
in a color-singlet state,
and the $\Gamma_{Kj}$ and the $\Gamma_{Bm}$ are Dirac-matrix structures for
the light meson and the $B$ meson, respectively.\footnote{
For example, for the leading-twist distributions of the pseudoscalar 
meson $P$, the longitudinally polarized vector meson $V$, and the 
transversely polarized vector meson $V_\perp$, $\Gamma_{Kj}$
is $i(f_P/4)q\slsh\gamma_5$,  $-i(f_V/4)q\slsh$, and
$-i(f_{V_\perp}/8)[\epsilon\slsh,q\slsh]$, respectively. Here,
$f_P$, $f_V$, and $f_{V_\perp}$ are the meson decay constants.}
We define light-cone variables $k=(k^+,k^-,\bm{k}_\perp)$ in terms of
Cartesian components as $k^+=(1/\sqrt{2})(k^0+k^z)$ 
and $k^-=(1/\sqrt{2})(k^0-k^z)$.
In Eq.~(\ref{eq:PHI-B}),
$n_-$ is the vector $n_-=\sqrt{2}(0,1,\bm{0}_\perp)$, 
and we have retained only the leading-twist $B$ meson
light-cone distributions. 
We take the spatial components of $p_K$ to lie along the minus
$z$ direction, and we take the $B$ meson to be at rest.
The expression
$[y,x]$ in Eqs.~(\ref{eq:PHI-K}) and (\ref{eq:PHI-B}) 
is the exponentiated line integral of the gauge field:
\begin{equation}
[y,x]=\exp\left[\int_{x}^y igT_a A_\mu^a dx^\mu\right].
\label{eikonal-line}
\end{equation}
$P$ indicates
path ordering, $T_a$ is a generator of color SU(3), and
$A_\mu^a$ is the gluon field. 

In the case of $B$-meson decays, we define the hard-scattering scale
$Q$ to be the $B$-meson mass $m_B$. We will argue that
the factorized form in Eq.~(\ref{B-meson-fact}) holds up to corrections
of relative order $f_1$, where $f_1=v^2m_c/Q$ for an $S$-wave quarkonium
$H_1$ and $f_1=m_c/Q$ for an $L$-wave quarkonium $H_1$ with
$L>0$.\footnote{Ref.~\cite{Beneke:2008pi} presents an analysis of the
process $B \to \chi_{cJ} K$ in the limit $m_b\to \infty$ with $m_c/m_b$
fixed, where $m_b$ is the bottom-quark mass.  The use of the term
``factorization'' in that paper has, therefore, a different meaning than
in the present paper, in which we take $m_c/m_b$ to be a small parameter.}
As in the $e^+e^-$-annihilation case, these suppression factors are
modified if quantum-number restrictions apply for the specific process.
This result was suggested previously in
Ref.~\cite{Beneke:2000ry}. However, there it was conjectured only that
the violations of factorization vanish in the limit $m_c\to 0$.

The remainder of this paper is organized as follows. In
Sec.~\ref{sec:model}, we specify models for the production amplitudes. In
Sec.~\ref{sec:fact}, we outline the proofs of factorization for the
processes under consideration. There we describe the momentum regions
that are leading in the hard-scattering scale, the momentum regions in
which loop integrands become singular, the diagrammatic topologies of
the leading and singular regions, the approximations that are
appropriate to contributions involving momenta that are soft or
collinear, the factorization of the soft and collinear singular
regions, the subsequent construction of the factorized form, and the
corrections to the factorized form. We illustrate general features of
the factorization proof with explicit one-loop examples in
Sec.~\ref{sec:1lex}. In Sec.~\ref{sec:softapp} we describe the
implementation of the soft approximation at the one-loop level.
Sections~\ref{sec:DoubleH} and \ref{sec:Bdecays} contain one-loop
examples for $e^+e^-$ annihilation and $B$ decays, respectively. We
summarize and discuss our results in Sec.~\ref{sec:concl}. The
Appendix contains the expressions for the quark-antiquark
spin-projection operators that we use in
Sec.~\ref{sec:1lex}.

\section{Model for the amplitude \label{sec:model}} 

We carry out our analyses in the rest frame of the $B$ meson and in the
CM frame of the $e^+e^-$ pair, choosing the three-momentum of the
quarkonium $H_1$ to be in the positive $z$ direction and choosing the
three-momentum of the light meson $K$ or the quarkonium $H_2$  to be in
the negative $z$ direction. We take the constituents of each meson and
quarkonium to be on the mass shell. We also assume that, for each meson
and quarkonium, there is an integration over the relative momentum of
the constituents, weighted by a meson wave function, and subject to the
mass-shell constraints.

We model the $B$ meson as an on-shell ``active'' bottom quark, which
participates in the electroweak interaction, and an on-shell
``spectator'' light antiquark, which does not participate in the
electroweak interaction. We take the quark and antiquark to be in a
color-singlet state. We take the bottom quark to have momentum $p_b$
and mass
$m_b$, with (in Cartesian coordinates)
\begin{equation}
p_b=\left(\sqrt{m_b^2+\bm{q}_B^{2}},
\bm{q}_B\right)
\sim\left(m_b,\Lambda_{\rm QCD}\right).
\end{equation}
We take the spectator antiquark to have momentum $p_l$, with
\begin{equation}
p_l=\left(\vert \bm{q}_B\vert,-\bm{q}_B\right)
\sim\left(\Lambda_{\rm QCD},\Lambda_{\rm QCD}\right).
\end{equation}
 The momentum of the $B$ meson, $p_B$, is given by the sum of $p_b$ and 
$p_l$:
\begin{equation}
p_B=p_b+p_l=\left(m_B,\bm{0}\right)=\left(\sqrt{m_b^2+\bm{q}_B^{\; 2}}
+\vert \bm{q}_B\vert,
\bm{0}\right)\sim\left(m_b,\bm{0}\right).
\end{equation}

Similarly, we model the light meson $K$ as an on-shell active light quark
and an on-shell spectator light antiquark, with the quark and antiquark
in a color-singlet state. We can write the quark momentum, $p_{k_{q}}$,
and the antiquark momentum, $p_{k_{\bar{q}}}$, as 
\begin{subequations}
\begin{eqnarray}
p_{k_{q}} & = & \frac{1}{2}p_K+r_k,\label{eq:pkmpa}\\
p_{k_{\bar{q}}} & = & \frac{1}{2}p_K-r_k,\label{eq:pkmpb}
\end{eqnarray}
\end{subequations}
with $p_K\cdot r_k=0$. In the rest frame of the light meson, we denote
the vectors that are associated with the light meson with a hat.
Then, we have
\begin{subequations}
\begin{eqnarray}
\hat{p}_K&=&\left(m_K,\bm{0}\right)=\left(2\vert 
\hat{\bm{r}}_{k}\vert,\bm{0}\right),\\
\hat{r}_k&=&(0,\hat{\bm{r}}_k),\\
\hat{p}_{k_{q}} & = & \left(\vert\hat{\bm{r}}_k\vert,
\hat{\bm{r}}_k\right),\\
\hat{p}_{k_{\bar{q}}} & = & 
\left(\vert\hat{\bm{r}}_k\vert,-\hat{\bm{r}}_k\right).
\end{eqnarray}
\end{subequations}
The quantity $\hat{\bm{r}}_k$ is of order $\Lambda_{\rm QCD}$.

The boosts from the light-meson rest frame to the $B$-meson rest frame are
given, for an arbitrary momentum $k$, by
\begin{subequations}
\begin{eqnarray}
\hat{k}^+ & \to & \frac{E_K-P_{\rm CM}}{m_K}\,\hat{k}^+,\\
\hat{k}^- & \to & \frac{E_K+P_{\rm CM}}{m_K}\,\hat{k}^-,\\
\hat{k}_{\perp} & \to & \hat{k}_{\perp}.
\end{eqnarray}
\end{subequations}
Here, $P_{\rm CM}$ is the magnitude of the three-momentum of
either $H_1$ or $K$ in the $B$-rest frame,
\begin{subequations}
\begin{eqnarray}
P_{\rm CM}&=&\frac{\lambda^{1/2}(s,M_1^2,m_K^2)}{2\sqrt{s}}
\sim m_b\;,\label{P_cm}
\\
\lambda(x,y,z)&=&x^2+y^2+z^2-2(xy+yz+zx)\;,
\end{eqnarray}
\end{subequations}
$E_K$ is the energy of the meson $K$,
\begin{equation}
E_{K}=\sqrt{P_{\rm CM}^2+m_K^2}\sim m_b\;.
\end{equation}
$M_1\sim m_c$ in Eq.~(\ref{P_cm}) is the heavy-quarkonium mass,
which we will define in our model below. Therefore, in the $B$-meson
rest frame we have
\begin{subequations}
\begin{eqnarray}
p_K  &=&\left(\frac{1}{\sqrt{2}}\left[E_{K}-P_{\rm CM}\right],
\frac{1}{\sqrt{2}}\left[E_{K}+P_{\rm CM}\right],\bm{0}_{\perp}\right) 
 \sim  \left(\frac{\Lambda_{\rm QCD}^2}{m_b},m_b,\bm{0}_{\perp}\right),
\label{pk}
\\
r_k  &=&\left(
\frac{\hat{r}_k^{z}}{\sqrt{2}}\,\frac{E_{K}-P_{\rm CM}}{m_K},
-\frac{\hat{r}_k^{z}}{\sqrt{2}}\,\frac{E_{K}+P_{\rm CM}}{m_K},
\hat{\bm{r}}_{k\perp}\right) 
 \sim  \left(\frac{\Lambda_{\rm QCD}^2}{m_b},m_b,
\Lambda_{\rm QCD}\right).
\end{eqnarray}
\end{subequations}

It is now convenient to define a momentum fraction $y$ and a
vector $q_k$ that has zero minus component. In terms of these 
quantities, the momenta of the quark and the antiquark are
\begin{subequations}
\begin{eqnarray}
p_{k_{q}} & = & yp_K+q_k,\\
p_{k_{\bar{q}}} & = & (1-y)p_K-q_k\equiv \bar{y}p_K-q_k.
\end{eqnarray}
\end{subequations}
$y$ is the fraction of minus component of the momentum of the meson 
that is carried by the quark:
\begin{equation}
y=\frac{1}{2}+\frac{r_k^-}{p_K^-}.
\end{equation}
Hence,
\begin{equation}
q_k=\left(\frac{1}{2}-y\right)p_K+r_k
=\left(2\frac{\hat{r}_k^{z}}{\sqrt{2}}
\,\frac{E_{K}-P_{\rm CM}}{m_K},0,
\hat{\bm{r}}_{k\perp}\right)
\sim\left(\frac{\Lambda_{\rm QCD}^2}{m_b},0,\Lambda_{\rm QCD}\right).
\label{qk}
\end{equation}

Finally, we model the charmonium states as an on-shell charm quark and
an on-shell charm antiquark in a color-singlet state, with the
momentum of the charm quark equal to $p_{iq}$ and the momentum of
the charm antiquark equal to $p_{i\bar{q}}$. We take
\begin{subequations}
\begin{eqnarray}
p_{iq}&=&\frac{1}{2}P_i+q_i,\\
p_{i\bar q}&=&\frac{1}{2}P_i-q_i,
\end{eqnarray}
\end{subequations}
where $P_i$ is the quarkonium momentum and $P_i\cdot q_i=0$.  In the
quarkonium rest frame, we denote vectors that are associated with
the quarkonium with a hat. The quantity $\hat{q}_i$ has only spatial
components, whose magnitudes are of order $m_cv$. Hence,
\begin{subequations}
\label{eq:def-Pandq}
\begin{eqnarray}
\hat{P}_i&=&(M_i,\bm{0})
=\left( 2\sqrt{m_c^2+\hat{\bm{q}}_i^{2}},\bm{0} \right),\\
\hat{q}_i&=&(0,\hat{\bm{q}}_i).
\end{eqnarray}
\end{subequations}
In the case in which the quarkonium~$i$ is in a spin-triplet state, we 
also define a spin-polarization vector $\epsilon_i$. In the 
quarkonium~$i$ rest frame, $\epsilon_i$ has  spatial 
components of order unity and temporal component zero:
\begin{equation}
\hat{\epsilon}_{i}=(0,\hat{\bm{\epsilon}}_i),
\end{equation}
which implies that $P_i\cdot\epsilon_i=0$.

The boost from the rest frame of the quarkonium with momentum $P_1$ to
the $B$-meson rest frame or the $e^+e^-$ CM frame is
\begin{subequations}\label{eq:boost1}%
\begin{eqnarray}\label{eq:boost1a}
\hat{k}^+ & \to & \frac{E_1+P_{\rm CM}}{M_1}\,\hat{k}^+,\\
\hat{k}^- & \to & \frac{E_1-P_{\rm CM}}{M_1}\,\hat{k}^-, \\
\hat{k}_{\perp} & \to & \hat{k}_{\perp}.\label{eq:boost1c}
\end{eqnarray}
\end{subequations}
The boost from the rest frame of the 
quarkonium with momentum $P_2$ to the $e^+e^-$ CM frame
is 
\begin{subequations}\label{eq:boost2}%
\begin{eqnarray}\label{eq:boost2a}
\hat{k}^+ & \to & \frac{E_2-P_{\rm CM}}{M_2}\,\hat{k}^+,\\
\hat{k}^- & \to & \frac{E_2+P_{\rm CM}}{M_2}\,\hat{k}^-,\\
\hat{k}_{\perp} & \to & \hat{k}_{\perp}.\label{eq:boost2c}
\end{eqnarray}
\end{subequations}
Here,  
\begin{subequations}
\begin{eqnarray}
E_i&=&\sqrt{P_{\rm CM}^2+M_i^2}\sim Q\;,
\\
P_{\rm CM}&=&\frac{\lambda^{1/2}(s,M_1^2,\widetilde{M}_2^2)}{2\sqrt{s}}
\sim Q\;,
\\
\label{def:Mi}
M_i&=&2\sqrt{m_c^2+\hat{\bm{q}}_i^2}.
\end{eqnarray}
\end{subequations}
$\widetilde M_2=M_2$ in the case of $e^+e^-$ annihilation into two
quarkonia, and $\widetilde M_2=m_K$ in the case of $B$-meson decays. It
then follows that, in the $e^+e^-$ CM frame or the $B$-meson
rest frame,
\begin{equation}\label{eq:qCM}
\begin{array}{lcl}
P_1^+\sim Q, & \quad 
   & P_2^+=2\frac{m_c^2-q_2^2}{P_2^-}\sim \frac{m_c^2}{Q},\\
P_1^-=2\frac{m_c^2-q_1^2}{P_1^+}\sim \frac{m_c^2}{Q}, 
   & \quad & P_2^-\sim Q,\\
{\bm P}_{1\perp}=0, & \quad & {\bm P}_{2\perp}=0,\\
q_1^+\sim vQ, & \quad & q_2^+\sim \frac{vm_c^2}{Q},\\
q_1^-\sim \frac{vm_c^2}{Q}, & \quad & q_2^-\sim vQ,\\
\epsilon_1^+\sim \frac{Q}{m_c}, &\quad 
   & \epsilon_2^+\sim \frac{m_c}{Q}, \\
\epsilon_1^-\sim \frac{m_c}{Q}, & \quad 
   &\epsilon_2^-\sim \frac{Q}{m_c}, \\
\bm{q}_{i\perp}\sim m_cv, &\quad& \bm{\epsilon}_{i\perp}\sim 1.
\end{array}
\end{equation}

\section{Proof of factorization \label{sec:fact}}

\subsection{Strategy}

If we dress the lowest-order decay and production amplitudes in our
models with additional gluons, then certain regions of integration of
the gluon momenta yield contributions that are leading in powers of the
large momentum scale, $Q$. We will describe these regions in
Sec.~\ref{sec:leading} below. We wish to isolate the contributions from
the loop integrations that can be calculated in perturbation theory from
those that cannot. That is, we wish to isolate contributions in which
propagators have large virtuality, of order $Q$, from contributions with
lower virtualities. We call the large-virtuality part of the amplitude
the ``hard'' part. In order to establish factorization, we will show
that the low-virtuality contributions either cancel or can be absorbed
into nonperturbative functions. The nonperturbative functions are the
NRQCD matrix elements for the charmonia and, in the case of $B$-meson
decays, the $B$-meson-to-light-meson form factor, the light-cone
distribution amplitude for the $B$ meson, and the light-cone
distribution amplitude for the light meson. We will first demonstrate a
factorization involving quarkonium distribution amplitudes. Then, we
will argue that the distribution amplitudes can be straightforwardly
decomposed into a sum over NRQCD matrix elements multiplied by
short-distance coefficients.\footnote{For a discussion at the one-loop
level of the decomposition of quarkonium light-cone distribution
amplitudes into a sum over NRQCD matrix elements see
Refs.~\cite{Ma:2006hc,Bell:2008er}.} After the factorization of
low-virtuality contributions, the remaining hard part will depend only
on the momenta and spins of the quarks and antiquarks that enter into
the leading-order process and will be independent of the low-virtuality
properties of the external mesons.

The low-virtuality contributions arise from regions of loop integration
that are logarithmically enhanced. In these logarithmically enhanced
regions, loop integrations have logarithmic power counts and can lead to
actual infrared (IR) divergences or would-be IR divergences that
are cut off by scales smaller than $Q$, such as quark masses. In the
case of a would-be divergence that arises from the emission of a gluon
that is nearly collinear to one of the external charm quarks, the minimum
virtuality of the quark propagator is of order
$m_c^2|\bm{k}|/(2|\bm{p}|)$, where $k$ is the gluon momentum and $p$ is
the charm-quark momentum. Hence, the virtuality can be much less than
$Q^2$, and even of order $\Lambda_{\rm QCD}^2$. Therefore, we must
factor such contributions from the hard part in order to arrive at a
perturbatively calculable contribution.

One could, in principle, deal with the low-virtuality contributions by
devising a suitable subtraction scheme for the contributions that would
appear order by order in perturbation theory. That would be a
formidable task, as one would need to ensure that all such contributions
are accounted for in an arbitrarily complicated Feynman diagram, with no
double counting of contributions.

For our purposes, we can take a simpler approach. We consider the
singularities that appear in the limit $m_c\to 0$ with $q_i$ fixed.
First, we establish that the contributions from infinitesimal
neighborhoods of these singularities can be factored into
nonperturbative functions. Then, we restore $m_c$ to its physical
value and extend the regions contained in the nonperturbative
functions from the infinitesimal neighborhoods of the singularities to
regions of finite size. Then the hard part, which is defined to be
the remainder of the amplitude, contains no logarithmically enhanced
contributions.

The factorization proofs entail the use of soft and collinear
approximations, which are exact at the singular points. These
approximations are described in Secs.~\ref{sec:collinear-app} and
\ref{sec:soft-app}. The actual factorization is achieved through the use
of decoupling relations, which are based on the graphical Ward
identities of QCD. These decoupling relations are described in
Sec.~\ref{sec:decoupling}.

We note that, because our models make use of on-shell external quarks
and antiquarks, it is possible to emit collinear and nearly collinear
gluons of arbitrarily low energy from the external lines. This
situation is discussed in detail in Ref.~\cite{Bodwin:2009cb}. It is
unphysical since, in a meson, confinement cuts off gluon energies at
values of order $\Lambda_{\rm QCD}$. Nevertheless, it is important to 
establish factorization in the on-shell case in order to guarantee the 
consistency of perturbative calculations of the hard part, which are 
usually carried out in the context of on-shell amplitudes. Because the 
logarithmically enhanced contributions in the presence of a cutoff of 
order $\Lambda_{\rm QCD}$ are a subset of the logarithmically enhanced 
contributions in the case of on-shell external lines, the factorization
argument that we will present also applies in the simpler case of a
model with a cutoff. As we will see, the methods that we use to prove
factorization apply to models in which the external particles are off
their mass shells, provided that the models maintain gauge invariance.
For example, one could model the $B$ meson as an elementary,
color-singlet pseudoscalar that produces the constituent quark and
antiquark off their mass shells through a pointlike
pseudoscalar-interaction vertex that is proportional to $\gamma_5$.

\subsection{Leading momentum regions \label{sec:leading}}

In describing the regions of loop momenta that yield contributions that
are leading in powers of the large scale $Q$, we make use of the
nomenclature of Ref.~\cite{Bodwin:2009cb}. We  first describe the various
regions of momentum space, and then, in Sec.~\ref{sssec:leadmomreg}, we
specify the conditions that must be fulfilled in order for these regions
to give leading contributions to an amplitude. In a Feynman diagram,
we  call a gluon or quark, or, generically, a line that carries momentum
of type $X$ an ``$X$ gluon,'' ``$X$ quark,'' or ``$X$ line.''

\subsubsection{Hard, soft, and collinear regions}
The hard ($H$), soft ($S$), 
collinear-to-plus ($C^+$), and collinear-to-minus ($C^-$) momenta have 
components with the following orders of magnitude:

\begin{subequations}
\begin{eqnarray}
H:\,\,\,&&Q(1,1,\bm{1}_\perp),
\\
S:\,\,\,\,&&Q\epsilon_S (1,1,\bm{1}_\perp),
\\
C^+:&&Q\epsilon^+[1,(\eta^+)^2,\bm{\eta}^+_\perp],
\\
C^-:&&Q\epsilon^-[(\eta^-)^2,1,\bm{\eta}^-_\perp].
\end{eqnarray}
\end{subequations}

The energy scales of the various types of momenta are determined by
the parameters $\epsilon_S$, $\epsilon^+$, and $\epsilon^-$. The soft
region of momentum space is defined by the condition
\begin{equation}
\epsilon_S\ll 1.
\label{fact-scale-s}%
\end{equation}
The collinear regions of momentum space are defined by the conditions
\begin{eqnarray}
\epsilon^\pm &\lesssim & 1,\nonumber\\
\eta^\pm &\ll &1.
\label{fact-scale-c}%
\end{eqnarray}

In the case of $B$-meson decays, there is also a leading region that
is associated with momenta that are nearly collinear to the light-quark 
momentum $p_l$. We call this region $C^l$. It is characterized by 
momenta that scale as
\begin{equation}
C^l:\,\,\, Q\epsilon^l[{e}_l+(\eta^l)^2 \bar{e}_l+\eta^l e_l^T],
\end{equation}
where ${e}_l$,
is a unit vector that is parallel to the lightlike vector $p_l$, 
$\bar{e}_l$ is the parity inverse of ${e}_l$,
and $e_l^T$ is a unit vector 
that is transverse to $e_l$ and $\bar{e}_l$. The $C^l$ region is defined by 
\begin{eqnarray}
\epsilon^l&\lesssim& \Lambda_{\rm QCD}/Q,\nonumber\\
\eta^l&\ll& 1.
\end{eqnarray}
We assume that $p_l$ does not lie exactly in the plus or minus direction.
In order to simplify the discussion to follow, we often do not mention
$C^l$ momenta explicitly. In these instances, it may be assumed that
the lines carrying $C^l$ momenta may be treated analogously to the lines
carrying $C^\pm$ momenta.

We note that soft and collinear contributions lie in restricted
regions of phase space. Were it not for enhancements that arise from
propagators with low virtuality, soft contributions would be suppressed
by a phase-space factor $\epsilon_{S}^4$ and collinear contributions would
be suppressed by a phase-space factor $(\epsilon^\pm)^4(\eta^\pm)^4$.
The low-virtuality propagators associated with these contributions lead
to loop integrals that have a logarithmic power count and to
contributions that behave as $\epsilon_{S}^0$ and
$(\epsilon^\pm)^0(\eta^\pm)^0$. We refer to such contributions as soft
and collinear logarithmic enhancements.

The definitions given above for the $H$, $S$, and $C^i$ momentum
regions do not specify unambiguously the boundaries between them. For
instance, if the $\eta^i$ parameters in the collinear regions take
on values that are not too different from one, then the $C^i$
momenta are not  distinguished from the $S$ momenta; {\it i.e.}, it
would not be clear if a $C^i$  momentum with $\eta^{i}$ close to
1 belongs to the $C^i$ region or to the $S$ region. Hence, the
possibility of double counting arises. Analogous issues appear at the
other boundaries between the $H$, $S$, and $C^i$ regions. However, as
we have mentioned previously, this is not a problem for our proof of
factorization, which will be presented later, because the 
proof focuses on the singularities and would-be singularities, rather
than on the momentum regions. The heuristic description of regions
presented here is intended only to set the stage for the subsequent
discussions of the singular regions.

In contrast with the corresponding momentum regions, the soft and
collinear singularities {\it are}  distinct. The soft singularities
appear in the limit $\epsilon_S\to 0$ or $\epsilon^i\to 0$ and the
collinear singularities appear in the limits $\eta^i \to 0$. A
double (soft and collinear) singularity can arise if $\epsilon^i\to 
0$ and $\eta^i \to 0$ at the same time. 

\subsubsection{Endpoint region\label{sec:endpoint}}

In the case of $B$-meson decays, there is a leading contribution from
the so-called ``endpoint''  region \cite{Beneke:2000ry}. This
contribution is associated with a gluon that connects the $B$-meson and
light-meson antiquarks to the remainder of the amplitude. The
contribution in the endpoint region arises from a would-be infrared
divergence that corresponds to the singular point at $\bar y=0$. The
divergence is cut off by $q_{k}$, the residual momentum of
light-meson antiquark, and by $p_l$, the momentum of the $B$-meson 
antiquark, both of which are of order $\Lambda_{\rm QCD}$. That is,
the divergence is cut off at $\bar y\sim \Lambda_{\rm QCD}/m_b$. In our
model, the explicit diagrammatic factors yield a linearly divergent power
count, but the would-be divergence is moderated by a factor $\bar y$
from the light-meson wave function and is actually logarithmic. The
gluon that is associated with the endpoint region carries $S$ momentum
of order $\Lambda_{\rm QCD}$.\footnote{If the spectator antiquark line
that connects the $B$ meson to the light meson carries a $C^-$ momentum
whose invariant square is of order $\Lambda_{\rm QCD}^3/Q$, then that
momentum is said to be in the soft-collinear or messenger region
\cite{Becher:2003qh}. Such a momentum arises from a small part of the
phase space in which a gluon on the $B$-meson side of the soft-collinear
spectator line carries away most of $p_l$  and a gluon on the
light-meson side of the soft-collinear spectator line carries away most
of $p_{k_{\bar{q}}}$. It has been argued that the soft-collinear region
is leading only when one makes use of certain infrared regulators
\cite{Manohar:2006nz,Beneke:2003pa,Bauer:2003td}. In any case, a
contribution from the soft-collinear region does not require any special 
treatment in our factorization argument: The gluon on the            
light-meson side of the soft-collinear spectator line can be treated as 
$C^+$, and the gluon on the $B$-meson side of the soft-collinear
spectator line can be treated as $S$, as it would be in the endpoint 
region.}

If the gluon that is associated with the endpoint region attaches to
an active-quark line from the $B$ meson or the light meson or to a
heavy-quark or heavy-antiquark line from the charmonium, then its
momentum causes the propagators of those lines to be off shell by an
amount of order $m_b\Lambda_{\rm QCD}$. We call such lines ``semihard''
lines. Contributions from these lines can be calculated in perturbation
theory. We treat the semihard region as part of the hard region, and we
include lines carrying semihard momenta in the hard subdiagram that we
will describe below.

\subsubsection{Glauber region}

The ``Glauber'' region is also leading in power counting
\cite{Bodwin:1994jh,Bodwin:1984hc}. In this region, $|k^+|\ll
|k_\perp|$, $|k^-|\ll |k_\perp|$, and $k_\perp^2\ll Q^2$. In processes
with two incoming hadrons, such as Drell-Yan lepton-pair production,
pinch singularities can develop in the Glauber region for the $k^+$ and
$k^-$ contours of integration on a diagram-by-diagram basis
\cite{Bodwin:1984hc,Collins:1985ue,Collins:1989gx}. The pinches arise
when a gluon connects a spectator parton in one initial-state hadron
with a spectator parton in the other initial-state hadron. (Here, in
contrast with the terminology that is used to discuss exclusive
$B$-meson decays, ``spectator parton'' means a parton that does not
participate in the hard-scattering process.) The pinches appear
because the momentum of a gluon that attaches to a spectator-parton line
must route through the hadron wave function and the active-parton line
from that hadron to the hard process. If the gluon's momentum in the
active-parton line is in the same direction as the momentum of the
active parton, then, in the spectator-parton line, it is in the
direction opposite to the momentum of the spectator parton. Consequently,
there is a pinch in the light-cone variable that is conjugate to the
direction of the momentum of the hadron.  In contrast, in
exclusive processes, all of the partons in a hadron are connected in the
lowest-order process, either through the hard subprocess or, possibly
through a soft gluon in the case of $B$-meson decays. (See the
discussion of the endpoint region above.) Thus, if an additional gluon
carrying soft momentum attaches to a parton, one can always route that
momentum through a leading-order connection to the hard part, avoiding
routings through other partons in the hadron that could produce a
pinch. Because of this, the $k^+$ and $k^-$ contours of integration
are not pinched in the Glauber region in exclusive processes, and it
is possible to deform them out of the Glauber region on a
diagram-by-diagram basis. Therefore, we ignore the Glauber region in the
remainder of our discussion.

\subsubsection{Threshold region}

In the case of a quarkonium, there are ``threshold enhancements'' that are
associated with the exchange of a gluon between the quark and the
antiquark. (See Ref.~\cite{Bodwin:1994jh} for examples.) In the
quarkonium rest frame, the enhancement occurs when the exchanged gluon
has momentum components $\hat{k}^0\sim m_c v^2$ and 
$|\hat{\bm{k}}|\sim m_c v$. The
enhancement produces a power infrared divergence that is cut off by the
relative momentum of the quark and antiquark $\hat{\bm{q}}\sim m_c v$.
The divergence is proportional to $m_c/|\hat{\bm{q}}|\sim 1/v$. Now let us
consider the momentum of the exchanged gluon in the $e^+e^-$ CM frame in
the case of $e^+e^-$ annihilation and in the $B$-meson rest frame in the
case of $B$-meson decays. In these frames, as can be seen from the boosts
in Eqs.~(\ref{eq:boost1}) and (\ref{eq:boost2}), an exchanged
gluon in the quarkonium with momentum $P_1$ has momentum $k\sim
(Qv,m_c^2 v/Q,m_c \bm{v}_\perp)$, and an exchanged gluon in the
quarkonium with momentum $P_2$ has momentum $k\sim (m_c^2 v/Q,Qv,m_c
\bm{v}_\perp)$. Therefore, the exchanged gluons associated with
threshold enhancement have $C^+$ or $C^-$ momentum, and, in our
analysis, we do not distinguish them from other gluons with $C^+$ or
$C^-$ momentum. Because the threshold enhancements involve the gluons
and heavy quarks in a single quarkonium, in each Feynman diagram they
are {\it a priori} compatible with the factorized forms. Therefore, it
will not be necessary to manipulate the threshold contributions or to
identify them by considering the limit $v\to 0$. 

\subsubsection{Leading momentum configurations in Feynman diagrams}\label{sssec:leadmomreg}
 
Next we identify the configurations of momentum types that can yield
leading contributions in the Feynman diagrams. By ``leading'', we mean
contributions that are not suppressed as powers of ratios of momentum
components. We  follow the analysis presented in
Ref.~\cite{Bodwin:2009cb}. Here, and throughout this paper, we work in 
the Feynman gauge.

We start with a basic diagram that is just the amplitude of lowest order
that involves the external quark and antiquark from each meson. Then
we add gluons, one at a time, determining for each gluon the momentum
types that produce leading contributions. The added gluons can contain
quark, gluon, and ghost vacuum-polarization loops.

Because there are many redundant ways to obtain a given momentum
configuration in a diagram, it is useful to define a convention for the
way in which we add gluons. In order to do that, we first define
\emph{combination momenta} $\tilde{C}^{\pm}$ and $CC$. A $\tilde{C}^\pm$
momentum arises from the sum of a $C^\pm$ momentum and an $S$ momentum
with $\epsilon_S \sim \epsilon^\pm \eta^\pm$ or from the sum of a
$C^{\pm}$ momentum and a $C^{\mp}$ momenta with
$\epsilon^\pm(\eta^\pm)^2\ll\epsilon^{\mp}\ll\epsilon^{\pm}$. A $CC$
momentum arises from the sum of a $C^\pm$ momentum and a $C^\mp$
momentum with $\epsilon^+\sim \epsilon^-$. These combination momenta
have the following orders of magnitude:

\begin{subequations}
\begin{eqnarray}
\tilde{C}^+:&& Q\epsilon^+(1,\tilde{\eta}^+,\bm{\eta}^+_{\perp}),
\\
\tilde{C}^-:&& Q\epsilon^-(\tilde{\eta}^-,1,\bm{\eta}^-_{\perp}),
\\
CC:&& Q\epsilon_{CC}(1,1,\bm{\eta}_{CC\perp}),
\end{eqnarray}
\end{subequations}

where
\begin{equation}
1\gg\tilde{\eta}^\pm \gg (\eta^\pm)^2.
\end{equation}
Analogous combination momenta may be defined for combinations of $C^l$
momenta with other momenta. A $\tilde{C}^l$ momentum has dominant
component in the $e_l$ direction. A $CC^l$ momentum, which arises
from the sum of a $C^l$ momentum and a $C^\pm$ momentum with
$\epsilon^l\sim \epsilon^\pm$, has at least two components of order
$\epsilon^l$.

Now we define our convention for adding gluons to the basic diagram. We
say that a gluon with momentum $l$ can attach to a line with momentum
$p$ only if the energy scale of the momentum $p+l$ is of the same
order as the $\epsilon$ parameter of the momentum $p$ and one of
the following conditions is fulfilled:
\begin{enumerate}
\item The momentum $p+l$ is of the same type as the momentum $p$;
\item The momentum $p$ is $C^i$ and $p+l$ is $\tilde{C}^i$;
\item The momentum $p$ is $C^i$ or $\tilde{C}^i$ and $p+l$ is $CC$.
\end{enumerate}

\begin{widetext}
\begin{center}
\begin{table}
\begin{tabular}{c}
\begin{tabular}{c|ccc}
$k$\,$\backslash$\,$p$ &
   \hspace{7ex}$S$\hspace{7ex}
&\hspace{7ex} $C^\pm$\hspace{7ex}
&\hspace{7ex} $\tilde{C}^\pm$\hspace{7ex}
\\
\hline
\hspace{2ex}$S$\mbox{\hspace{2ex}}&
$\epsilon_{S_k}\sim\epsilon_{S_p}$&
$\epsilon_{p}^{\pm}(\eta^{\pm}_p)^2\lesssim \epsilon_{S_k}
\ll \epsilon_{p}^{\pm}$
&

$\epsilon_p^{\pm} \tilde{\eta}_p^{\pm} \lesssim
\epsilon_{S_k}
\ll \epsilon_{p}^{\pm}$
\end{tabular}
\\[5ex]
\begin{tabular}{c|cccc}
$k$\,$\backslash$\,$p$ &
 \hspace{7ex}$S$\hspace{7ex}
&\hspace{7ex}$C^\mp$ \hspace{7ex}
&\hspace{7ex}$\tilde{C}^\mp$\hspace{7ex}
&\hspace{7ex}$CC$\mbox{\hspace{7ex}}\\
\hline
\hspace{2ex}$C^\pm$\mbox{\hspace{2ex}}&
$\epsilon^\pm_{k}\sim\epsilon_{S_p}$&
$\epsilon_{p}^\mp(\eta^\mp_{p})^2\lesssim\epsilon^{\pm}_k\lesssim\epsilon^{\mp}_p$
&$\epsilon^\mp_{p}\tilde\eta^{\mp}_{p}\lesssim\epsilon^{\pm}_k\lesssim\epsilon^{\mp}_p$&
$\epsilon^\pm_{k}\sim\epsilon_{{CC}_p}$
\\
$CC$&
$\epsilon_{CC_k}\sim\epsilon_{S_p}$&
$\epsilon_{p}^\mp(\eta^\mp_{p})^2\lesssim\epsilon_{CC_k}\ll\epsilon^{\mp}_p$&
$\epsilon^\mp_{p}\tilde\eta^{\mp}_{p}\lesssim\epsilon_{CC_k}\ll\epsilon^{\mp}_p$&
$\epsilon_{CC_k}\sim\epsilon_{CC_p}$
\end{tabular}
\end{tabular}
\caption{Conditions that a gluon with momentum $k$ must fulfill in order
to attach to a line with momentum $p$. These conditions guarantee that
the resulting attachment is allowed according to the convention
described in the text and that it results in a leading contribution.
Here, ``leading'' means that the contribution is not suppressed as
powers of ratios of momentum components. In each table, the left-hand
column gives the momentum type of the gluon with momentum $k$, and the
top row gives the momentum type of the line with momentum $p$. The
symbol ``$\sim$'' means that quantities are of the same order. For
purposes of power counting, an $H$ line behaves as a soft line with
$\epsilon_S\sim 1$. The rules for attachment when $k$ is $\tilde{C}^\pm$
 are the same as the rules for attachment when $k$ is $C^\pm$. If $k$ is
$S$, and the lines to which it attaches have momentum $p_i$ and $p_j$,
then $p_i$ and $p_j$ cannot both be $C^+$ or $C^-$. If $k$ is $C^\pm$,
then at least one of $p_i$ and $p_j$ is $C^\pm$. Analogous
conditions exist for the attachments of gluons with $C^l$ momenta.
\label{tab:power-counting}}
\end{table}
\end{center}

\end{widetext}

The analysis in Ref.~\cite{Bodwin:2009cb} shows that, if we consider 
only the terms $2p\cdot l$ in propagator denominators, then the gluons that we
add to the basic diagram must be $S$, $C^+$, $C^-$, or $C^l$ in order
to obtain a leading contribution.  The combination momenta defined
above, arise when we add $S$, $C^{\pm}$, and $C^l$  momenta. If we 
consider, as well, the terms $p^2$ and $l^2$ in propagator denominators, 
then contributions are subleading unless 
\begin{eqnarray}
k\cdot p&\gtrsim& k^2,\nonumber\\
k\cdot p&\gtrsim& p^2.
\label{ksq-psq-conditions}
\end{eqnarray}
The constraints in Eq.~(\ref{ksq-psq-conditions}) lead to additional
restrictions on the momentum combinations that yield leading
contributions. These restrictions, combined with our conventions for
adding gluons to a diagram, result in the rules for the attachments that
yield leading contributions that are given in
Table~\ref{tab:power-counting}. The rules in
Table~\ref{tab:power-counting} also apply when the gluon attaches to
one of the fermion lines that begins as an external quark or antiquark.
In that case, one sets $\eta^\pm=0$ for the external quark or antiquark.
We have not displayed the rules for the attachments of gluons with
$C^\pm$ or $\tilde C^\pm$ momenta to lines with $C^\pm$ or $\tilde
C^\pm$ momenta because the rules for such attachments are complicated
and cannot be characterized simply in terms of the magnitudes of the
momentum components. For our purposes, it suffices to note that
necessary conditions for such attachments are given in
Eq.~(\ref{ksq-psq-conditions}).

Some of the allowed attachments in Table~\ref{tab:power-counting} change
the type of the momentum in the top row, for example, when we add an $S$
gluon to a $C^\pm$ gluon with $\epsilon_S\sim \eta^\pm \epsilon^\pm$.
That change can propagate through the Feynman diagram. In those cases one
must check that the rules in Table~\ref{tab:power-counting}  still allow
the attachments of all the vertices that are affected by the change.

The constraints in Eq.~(\ref{ksq-psq-conditions}) imply that an
attachment of a gluon to a given line is allowed only if the virtuality
that it produces on that line is of order or greater than the virtuality
that is produced by the gluons that attach to that line to the outside  
of the attachment in question. If a gluon with momentum $k$ of type
$C^\pm$, $\tilde{C}^\pm$, $S$, $C^\mp$, or $CC$ attaches to a       
$C^\pm$ line from an on-shell external quark or antiquark, then it adds
virtuality $Q^2\epsilon_{k}^\pm (\eta_{k}^\pm)^2$, $Q^2\epsilon_{k}^\pm
\tilde\eta_{k}^\pm$, $Q^2\epsilon_{S_k}$, $Q^2\epsilon_{k}^\mp$, or
$Q^2\epsilon_{CC_k}$, respectively.

\subsection{Topologies of the leading regions}

Now let us specify the diagrammatic topology that corresponds to the
leading regions. 

\begin{figure}
\includegraphics[angle=0,height=7cm]{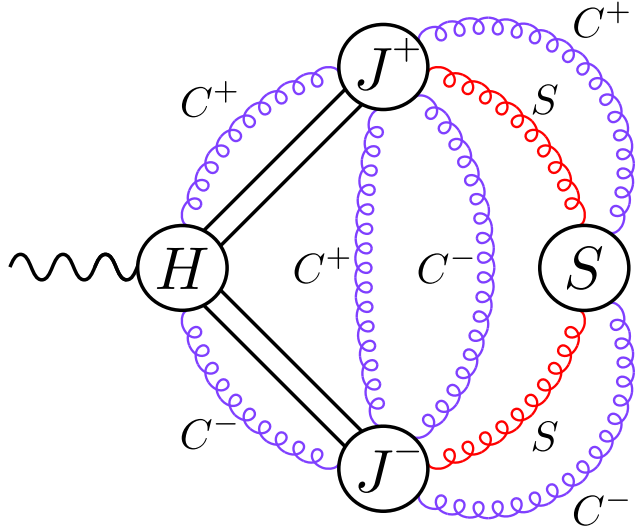}
\vspace*{8pt}
\caption{\label{fig:regionsdc}%
Leading regions for
double-charmonium production in $e^+e^-$ annihilation. The wavy line
represents the virtual photon.} 
\end{figure}

The topology of the leading regions for $e^+e^-$ annihilation into two
quarkonia is shown in Fig.~\ref{fig:regionsdc}. In this topology, there is
a hard subdiagram that includes the lowest-order process, a soft
subdiagram, and a jet subdiagram for each of the two collinear 
regions, which correspond to the two quarkonia. In the hard subdiagram, all
propagator denominators are of order $s$. The soft subdiagram includes
gluons with soft momenta and loops involving quarks and ghosts with soft
momenta. The soft subdiagram attaches to the jet subdiagrams through
any number of soft-gluon lines. Each jet subdiagram contains the quark
and antiquark lines for a given quarkonium, as well as gluons and loops
involving quarks and ghosts with momenta collinear to the meson or
quarkonium. The $J^\pm$ subdiagram attaches to the hard subdiagram
through the quark and antiquark lines and through any number of $C^\pm$
gluons. As was pointed out in Ref.~\cite{Bodwin:2009cb}, because gluons
with $C^\pm$ momenta of arbitrarily low energy can contribute at leading
power in $Q$, the $J^\pm$ subdiagram also attaches to the soft and $J^{\mp}$
subdiagrams through any number of $C^\pm$ gluons.

\begin{figure}
\includegraphics[angle=0,height=8cm]{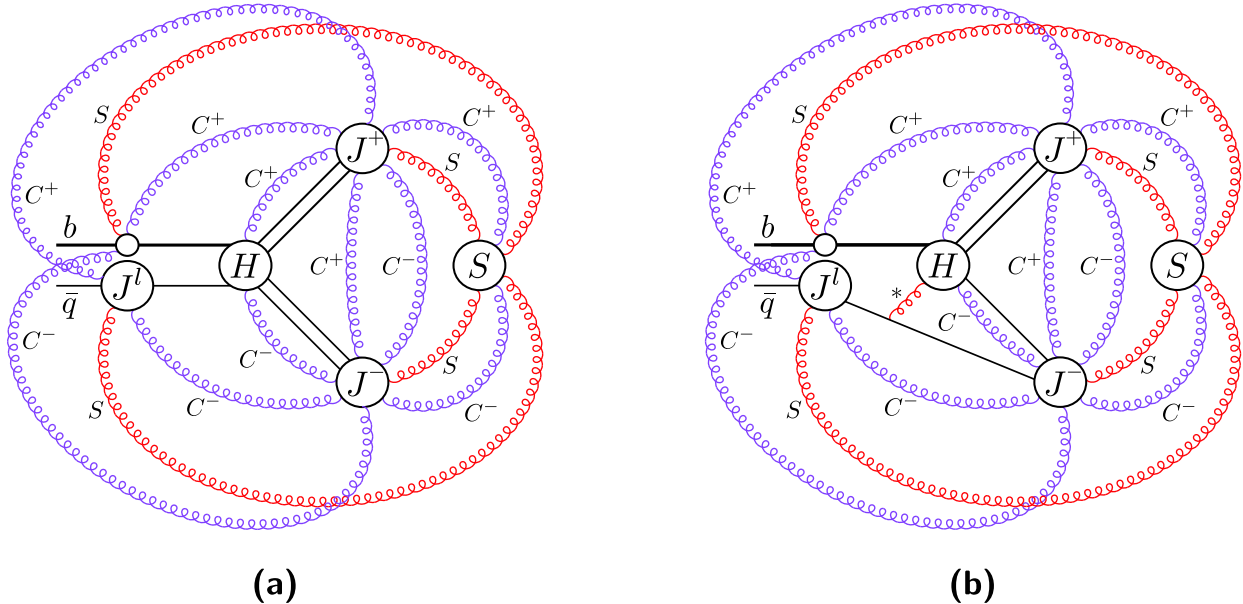}
\vspace*{8pt}
\caption{\label{fig:regions}%
Leading regions for the $B$-meson-decay case. The collinear region $J^+$
corresponds to the charmonium and collinear region $J^-$ corresponds
to the light meson.
}
\end{figure}

There are two distinct topologies in the case of $B$-meson decays: one
in which the $B$-meson and light-meson spectators participate in the hard
interaction and another in which they do not. These two topologies are
shown in Figs.~\ref{fig:regions}(a) and \ref{fig:regions}(b),
respectively. The topology of Fig.~\ref{fig:regions}(a) is appropriate
when the light-meson-antiquark momentum is outside the endpoint
region, and the topology of Fig.~\ref{fig:regions}(b) is appropriate
when the light-meson-antiquark momentum is in the endpoint region.

In each topology in Fig.~\ref{fig:regions}, there is a hard subdiagram
that includes the lowest-order parton-level process, there is a soft
subdiagram, and there is a jet subdiagram for each of the two collinear
regions, which correspond to the light meson and the quarkonium. In the
hard subdiagram, all propagator denominators are of order $m_b^2$ or
$m_b\Lambda_{\rm QCD}$. The soft subdiagram includes gluons with soft
momenta and loops involving quarks and ghosts with soft momenta. 
The soft subdiagram attaches to the jet subdiagrams and to the $B$-meson
quark and antiquark quark lines through any number of soft gluon lines.
The $J^+$ subdiagram contains the quarkonium quark and antiquark lines;
the $J^-$ subdiagram contains the light-meson quark and antiquark
lines; the $J^l$ subdiagram contains the $B$-meson spectator-quark
line. In addition, the jet subdiagrams contain gluons and loops
involving quarks, gluons, and ghosts with momenta in the $C^\pm$,
or $C^l$ regions. Each jet subdiagram contains the active- and
spectator-quark lines for a given meson or quarkonium as well as
gluons and loops
involving quarks, gluons, and ghosts with momenta collinear to the
meson or quarkonium. A $J^\pm$ or $J^l$ subdiagram attaches to the hard
subdiagram through the active- and spectator-quark lines in
the topology of Fig.~\ref{fig:regions}(a), through the active-quark
lines in the topology of Fig.~\ref{fig:regions}(b) and through any
number of gluons. (We have not shown explicitly the attachments of the
$J^l$ jet subdiagram that involve gluons with $C^l$ momenta.) As we
have already mentioned, because gluons with $C^i$ momenta of
arbitrarily low energy can contribute at leading power in $Q$, the
$J^\pm$ subdiagram also attaches to the soft, $J^{\mp}$, and $J^l$
subdiagrams through any number of $C^\pm$ gluons, and the $J^l$
subdiagram also attaches to the soft and $J^{\pm}$ subdiagrams through
any number of $C^l$ gluons \cite{Bodwin:2009cb}.

In the case of the topology of Fig.~\ref{fig:regions}(b), we show
explicitly a gluon that is marked with an asterisk. This is the gluon
that was mentioned in our discussion of the endpoint region in
Sec.~\ref{sec:endpoint}. We choose the momentum routing so that it
always carries the momentum of the $B$-meson antiquark and the
(endpoint) momentum of the light-meson antiquark, both of which
are of order $\Lambda_{\rm QCD}$. Therefore, we consider this gluon to
be part of the soft subdiagram. However, we single it out because it
must be present in our model in order for the light antiquarks
(spectators) to be connected to the remainder of the diagram and because
its momentum is fixed by the $B$-meson and light-meson antiquark
momenta. Soft gluons and low-energy $C^\pm$ gluons can connect to the
marked gluon, although we have not shown these connections explicitly.
We show the marked gluon connecting to the hard subdiagram because its
allowed connections to the jet subdiagrams or the $b$-quark line result
in propagators with semihard virtualities, of order $m_b\Lambda_{\rm
QCD}$, which are part of the hard subdiagram. As we have said, the
topology of Fig.~\ref{fig:regions}(b) applies to the endpoint region, in
which the marked gluon has virtuality of order $\Lambda_{\rm QCD}^2$.
Away from the endpoint region, the marked gluon itself has virtuality of
order $m_b\Lambda_{\rm QCD}$ and can be incorporated into the hard
subdiagram, resulting in the topology of Fig.~\ref{fig:regions}(a).

\subsection{Topologies of the Singular Regions}

In the massless sector of QCD, there are singularities that are
associated with soft and collinear divergences. (See, for example,
Refs.~\cite{Sterman:1978bi,Sterman:1978bj,Collins:1985ue,Collins:1989gx}.)
In the present case, the masses of charm quarks and antiquarks cut off
some of the collinear divergences. Some potential soft divergences are
also cut off because they are associated with a gluon that attaches to a
line that cannot go precisely to its mass shell because a quark mass
cuts off a collinear divergence. Nevertheless, as we have mentioned, we
wish to consider not only actual divergences, but also divergences
that appear only in the limit $m_c/Q\to 0$ with $q_i$ fixed, because
such divergences are associated with logarithmic enhancements. These
divergences are associated with singularities in the domain of
integrations. We call the infinitesimal neighborhoods of such
singularities ``singular regions.'' In the remainder of the discussion 
of the factorization of the contributions of the singular regions, we 
assume that we have taken the limit $m_c/Q\to 0$ with $q_i$ fixed.

The topologies of the singular regions follow from the power-counting
rules that are given in Sec.~\ref{sec:leading}. These topologies have
been discussed in Ref.~\cite{Bodwin:2009cb}. Here, we recapitulate that
discussion, describing the relationships of the topologies of the
singular regions to the topologies of the leading regions in
Figs.~\ref{fig:regionsdc}, \ref{fig:regions}(a), and \ref{fig:regions}(b).

The $C^i$ singular region is situated in the outermost part of the
$J^i$ subdiagram. (Here, ``out'' means toward the external fermion
lines.) We call this part of the $J^i$ subdiagram the $\tilde{J}^i$
subdiagram. (We denote the part of the $J^i$ subdiagram that excludes
the $\tilde{J}^i$ subdiagram as the $J^i-\tilde{J}^i$ subdiagram.) The
$S$ singular region is situated in the outermost part
of the $S$ subdiagram. We call the $S$ singular part of the $S$
subdiagram the $\tilde{S}$ subdiagram. (We denote the part of the $S$
subdiagram that excludes the $\tilde{S}$ subdiagram as the $S-\tilde{S}$
subdiagram.) $S$ singular gluons connect the $\tilde{S}$ subdiagram only
to the $\tilde{J}^i$ subdiagrams and to the external $b$-quark line. The
gluon that is marked with an asterisk in the topology of
Fig.~\ref{fig:regions}(b) is {\it not} part of the $\tilde{S}$ subdiagram
because its momentum components are fixed to be of order $\Lambda_{\rm
QCD}$. That is, it is $S$ but not $S$ singular. The $\tilde{J}^i$
subdiagrams connect to the $J^i$, $S$, and $H$ subdiagrams via $C^i$
gluons. We denote by $\tilde{H}$ the union of all of the subdiagrams in
our topology except for $\tilde{S}$, $\tilde{J}^+$, $\tilde{J}^-$, and
$\tilde{J}^l$. We note that the connections of the $\tilde{J}^\pm$
subdiagrams to the $\tilde{S}$ subdiagram via $C^i$ gluons were not
considered in the discussions in
Refs.~\cite{Collins:1985ue,Collins:1989gx}. Otherwise, the general
structure of the topologies of the singular regions that we consider are
the same as in Refs.~\cite{Collins:1985ue,Collins:1989gx}, provided that
we identify the hard subdiagram in those references with $\tilde{H}$.
 
\subsection{Collinear approximation \label{sec:collinear-app}}

We now describe the collinear approximations, which are useful in
factoring the $C^i$ singular contributions. We follow the notation
of Ref.~\cite{Bodwin:2009cb}.

Suppose that there is a gluon with momentum in the $C^i$ singular
region that attaches to a line that is not in $\tilde{J}^i$. Then, we
can apply a collinear approximation to that gluon
\cite{Bodwin:1984hc,Collins:1985ue,Collins:1989gx} without loss of
accuracy. The $C^i$ approximation consists
of replacing $g_{\mu\nu}$ in the gluon-propagator numerator as follows:

\begin{equation}\label{eq:collapp}
g_{\mu\nu}
\,\,\longrightarrow\,\,
\left\{
\begin{array}{ll}
\displaystyle
\frac{k_\mu \tilde{n}_{1\nu}}{k\cdot \tilde{n}_1-i\varepsilon}& (C^+),
\\[2ex]
\displaystyle
\frac{k_\mu \tilde{n}_{2\nu}}{k\cdot \tilde{n}_2+i\varepsilon}& (C^-),
\\[2ex]
\displaystyle
\frac{k_\mu \tilde{n}_{l\nu}}{k\cdot \tilde{n}_l-i\varepsilon}& (C^l).
\end{array}
\right.
\end{equation}
Here, the index $\mu$ corresponds to the attachment of the gluon to
the line with momentum not in the $C^i$ singular region, and the index
$\nu$ corresponds to the attachment of the gluon to the $J^i$
subdiagram. We always use the convention that $k$ flows out of a $C^+$
or $C^l$ line and into a $C^-$ line. There is a large amount
of freedom in choosing the auxiliary vectors $\tilde{n}_1$,
$\tilde{n}_2$ and $\tilde{n}_l$ in Eq.~(\ref{eq:collapp}). We need
only have $\tilde{n}_1\cdot p_{1q}> 0$ (or $\tilde{n}_1\cdot
p_{1\bar{q}}> 0$), $\tilde{n}_2\cdot p_{2q} > 0$ (or $\tilde{n}_2\cdot
p_{2\bar{q}} > 0$), and $\tilde{n}_l\cdot p_{l} > 0$ in order to
reproduce the amplitude in the collinear singular region. Our choice is
to take $\tilde{n}_1$, $\tilde{n}_2$, and $\tilde{n}_l$  to be lightlike
vectors in the minus, plus, and minus directions, respectively:
\begin{subequations}\label{tilde-n}%
\begin{eqnarray}
\tilde{n}_1&=&\bar{n}_1\equiv (1/\sqrt{2})(0,1,\bm{0}_\perp),\\
\tilde{n}_2&=&\bar{n}_2\equiv (1/\sqrt{2})(1,0,\bm{0}_\perp),\\
\tilde{n}_l&=&\bar{n}_1.
\end{eqnarray}
\end{subequations}

In order for the $C^i$ approximation to be exact in the $C^i$
limit, $j\cdot k$ must be equal to $j^\mp k^\pm$, where $j$ is the
current to which the $\mu$ index of the gluon with momentum $k$ attaches
and $n_i$ is a unit lightlike vector in the $C^i$ direction. This
requirement is met provided that the gluon does not attach with its
$\mu$ index to a line that is also carrying momentum in the $C^i$
singular region. That is, the $C^i$ approximation holds in the
collinear limit if $j$ is a current in any of the subdiagrams except for
the $\tilde J^i$ subdiagram. We note that, in the $C^i$
approximation, the gluon's polarization is longitudinal, {\it i.e.},
proportional to the gluon's momentum. This fact is essential to the
application of graphical Ward identities to derive decoupling relations.
We note also that the collinear approximation is exact, not only for the
collinear singularity, but also for the associated collinear logarithmic
enhancement.

\subsection{Soft approximation \label{sec:soft-app}}

We now describe the soft approximation, which is useful in factoring
the $S$ singular contributions. Again, we follow the
notation of Ref.~\cite{Bodwin:2009cb}.

Suppose that there is a gluon with momentum $k$ in the $S$ singular
region that attaches to a line carrying momentum $p$ that lies outside
the $S$ singular region. Then we can apply the soft approximation to
that gluon without loss of accuracy. The soft approximation
\cite{Grammer:1973db,Collins:1981uk} consists of replacing $g_{\mu\nu}$
in the gluon-propagator numerator as follows:
\begin{equation}
g_{\mu\nu}
\,\,\longrightarrow\,\, \frac{k_\mu p_\nu}{k\cdot p},
\end{equation}
where the index $\mu$ corresponds to the attachment of the gluon to
the line with momentum $p$. 

Unlike the collinear approximation, the soft approximation depends on
the momentum of the line to which the gluon attaches. However, it is
convenient to apply the same soft approximation to all of the lines in
the $\tilde{J}^\pm$ subdiagram. The lines in the $\tilde{J}^\pm$
subdiagram are collinear either to the momentum of the quark or the
momentum of the antiquark in the jet. In the case of the light-quark
jet, the quark and antiquark momenta are parallel, up to corrections of
relative order $\Lambda_{\rm QCD}/Q$. In the case of the quarkonium
jet(s), the quark and antiquark momenta $p_{iq}$ and $p_{i\bar q}$ are
parallel up to corrections of relative order $m_cv/Q$. In both cases, we
neglect the difference between the quark and antiquark momenta and
define a ``modified soft approximation'' for each jet that corresponds
to the soft approximation for the average of the quark and antiquark
momenta. The leading errors that arise in applying the modified soft
approximation to the light-meson, charmonium-1, and charmonium-2 jets
are of relative order $\bm{q}_{k\perp}/p_K^- \sim \Lambda_{\rm QCD}/Q$,
$\bm{q}_{1\perp}/P_1^+\sim m_cv/Q$, and $\bm{q}_{2\perp}/P_2^-\sim
m_cv/Q$, respectively.

It is convenient, for purposes of discussing the decoupling relations in
Sec.~\ref{sec:decoupling}, to choose lightlike vectors for the soft
approximation that correspond to the average of the quark and antiquark
momenta in the limits $\Lambda_{\rm QCD}/Q\to 0$ and $m_c/Q\to 0$. In
making this choice, we introduce an error of relative order
$\Lambda_{\rm QCD}^2/Q^2$ in the case of the light-meson jet and
of relative order $m_c^2/Q^2$ in the case of the quarkonium jet(s). These
errors are negligible in comparison with the errors that we make in
neglecting the difference between the quark and antiquark momenta. Then,
for both the light-meson and quarkonium jets we have the same soft
approximations.

For the attachment of the gluon with momentum $k$ to
any line with momentum in the $C^+$ ($C^-$) singular region, the
(modified) soft approximation consists of the following replacements in
the gluon-propagator numerator:
\begin{subequations}
\label{soft-app}
\begin{eqnarray}
g_{\mu\nu} &\longrightarrow&
\frac{k_\mu n_{1\nu}}{k\cdot n_1+i\varepsilon} (S^+),\\
\label{soft-app-plus}
g_{\mu\nu} &\longrightarrow&
\frac{k_\mu n_{2\nu}}{k\cdot n_2-i\varepsilon} (S^-),
\label{soft-app-minus}
\end{eqnarray}
\end{subequations}
where $n_1$ is a lightlike vector that is proportional to $P_1$ and
$n_2$ is a lightlike vector that is proportional to $P_2$ or $p_K$. We
normalize $n_1$ and $n_2$ so that they are the parity inverses of the
vectors $\bar{n}_1$ and $\bar{n}_2$ in Eq.~(\ref{tilde-n}),
respectively:
\begin{subequations}
\begin{eqnarray}
n_1&\equiv& (1/\sqrt{2})(1,0,\bm{0}_\perp),\\
n_2&\equiv& (1/\sqrt{2})(0,1,\bm{0}_\perp).
\end{eqnarray}
\end{subequations}

The index $\mu$ contracts
into the line carrying the momentum of type $C^+$ ($C^-$). As we have
mentioned, the modified soft approximation in Eq.~(\ref{soft-app})
accounts for the contributions in the $S$ singular region up to
corrections of relative order $\Lambda_{\rm QCD}/Q$ in the case of the
light-meson jet and up to corrections of relative order $m_cv/Q$ in the
case of quarkonium jets. We note also that the soft approximation is
valid at this accuracy not only for the soft singularity, but also for
the associated soft logarithmic enhancement.

We do not apply soft approximations to the $B$ meson because the
eikonal vectors that are associated with soft approximations for $J^l$
and the $b$-quark line are not approximately proportional to each other.
That is, a common soft approximation cannot be applied to the $B$ meson.
In consequence, the cancellations of the soft contributions that
apply in the case of the quarkonia and the light meson (described in
Sec.~\ref{sec:soft-cancellations}) fail in the case of the $B$ meson.

\subsection{Decoupling relations \label{sec:decoupling}}

\begin{figure} 
\includegraphics[angle=0,height=3.5cm]{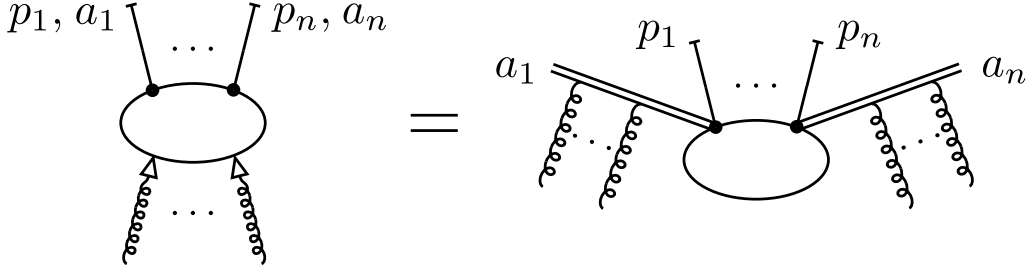}
\caption{Graphical representation of the decoupling relations for
collinear gluons and the decoupling relations for soft gluons. The
applicability of these decoupling relations is described in the text.
The relations show the decoupling of longitudinally polarized gluons,
which are represented by curly lines. The longitudinally polarized gluon
lines are to be attached in all possible ways to the Green's function
that is represented by an oval. The factors $k^\mu \bar n_i^\nu/(k\cdot
\bar n_i)$ [$k^\mu n_i^\nu/(k\cdot n_i)$] that appear in the collinear
(soft) approximations are represented by the arrows on the gluon lines.
The external lines with hash marks are truncated. In addition, the
subdiagram can include any number of untruncated on-shell external legs,
provided that the polarizations of the on-shell gluons are orthogonal to
their momentum. The $p_i$ are momenta, and the $a_i$ are color
indices. The double lines are $C^+$, $C^-$, $S^+$, or $S^-$ eikonal
lines, as is described in the text.
\label{fig:wi}}
\end{figure}

Once we have implemented a collinear approximation or a soft
approximation, the associated gluons are longitudinally polarized.
This allows us to make use of decoupling relations to factor gluons
with momenta in the soft or collinear singular regions from certain
parts of the amplitude. The general graphical form of the decoupling
relations for longitudinally polarized gluons is shown in
Fig.~\ref{fig:wi}.
A decoupling relation of this form applies when any number of
longitudinally polarized gluons attach to a subdiagram in all possible
ways, provided that the gluon momenta are all proportional to each
other.\footnote{In the case of an Abelian theory, such as quantum
electrodynamics, a decoupling relation of this form holds even if the
gluon momenta are not proportional to each other.} 

If the external gluons all have momenta in one of the $C^i$ singular
regions, then the gluon momenta are all proportional to each other, and
a decoupling relation of the form in Fig.~\ref{fig:wi} holds, once the
$C^i$ approximation has been implemented to render the gluon
polarizations longitudinal. The subdiagram can have any number of
truncated legs and any number of untruncated on-shell external legs (not
shown in the figure), provided that the polarizations of the untruncated
on-shell gluons are orthogonal to their momentum. The eikonal (double)
lines in this decoupling relation have the Feynman rules in the $C^+$,
$C^-$, or $C^l$ cases that a vertex is $\mp igT_a \bar n_{1\mu}$, $\pm
igT_a \bar n_{2\mu}$, or $igT_a \bar n_{l\mu}$ and a propagator is
$i/(k\cdot \bar n_1-i\varepsilon)$, $i/(k\cdot \bar n_2+i\varepsilon)$,
$i/(k\cdot \bar n_l-i\varepsilon)$, respectively, where the upper
(lower) sign in the vertex is for eikonal lines that attach to quark
(antiquark) lines. Here, $T_a$ is an $SU(3)$ color matrix in the
fundamental representation. (Our convention is that a QCD gluon-quark
vertex is $igT_a\gamma_\mu$.) We call these eikonal lines ``$C^i$
eikonal lines.'' The Feynman rules for the eikonal lines in these
decoupling relations are summarized in the first, second, and third
lines of Table~\ref{tab:cs-eikonal}.
 
If the external gluons all have momenta in the $S$ singular region,
then the decoupling requirement that the gluon momenta be
proportional to each other is not necessarily satisfied. However, if
the subdiagram into which the $S$ singular gluons enter is a
$\tilde{J}^+$ ($\tilde{J}^-$) subdiagram, then a soft momentum $k$
entering that subdiagram contracts only into currents proportional to
$n_1$ ($n_2$), up to corrections of relative order
$\bm{q}_{k2}/p_K^-\sim \Lambda_{\rm QCD}/Q$ for a light-meson jet and
relative order $\bm{q}_{1\perp}/P_1^+\sim \bm{q}_{2\perp}/P_2^- \sim
m_cv/Q$ for a charmonium jet. Consequently, at these levels of accuracy,
we can make the replacement
\begin{subequations}
\begin{equation}
k\to \tilde{k}_1=\bar{n}_1 \frac{n_1\cdot k}{n_1\cdot \bar{n}_1}
\end{equation} 
in the $\tilde{J}^+$ subdiagram and associated soft approximation and 
the replacement

\begin{equation}
k\to \tilde{k}_2=\bar{n}_2 \frac{n_2\cdot k}{n_2\cdot \bar{n}_2}
\end{equation}
\end{subequations}
in the $\tilde{J}^-$ subdiagram and associated soft approximation
\cite{Collins:1985ue,Collins:1989gx}.
Since
\begin{subequations}
\label{n-tilde-k}%
\begin{eqnarray}
n_1\cdot\tilde{k}_1&=&n_1\cdot k,\\
n_2\cdot\tilde{k}_2&=&n_2\cdot k,
\end{eqnarray}
\end{subequations}
these replacements do not change the amplitude, up to corrections of 
relative order $m_c/Q$ and $\Lambda_{\rm QCD}/Q$.
In subsequent discussions, we consider these replacements to be part
of the modified soft approximation. After these replacements have been
made, the gluon momenta entering the $\tilde{J}^+$ ($\tilde{J}^-$)
subdiagram are all proportional to each other, and a decoupling relation
of the form in Fig.~\ref{fig:wi} holds. In these decoupling relations,
the eikonal lines have the Feynman rules that a vertex is $\pm igT_a
n_{1\mu}$ ($\mp igT_a  n_{2\mu}$) and a propagator is $i/(k\cdot
n_1+i\varepsilon)$ [$i/(k\cdot  n_2-i\varepsilon)$] when the subdiagram
is $C^+$ ($C^-$). These rules follow from Eq.~(\ref{n-tilde-k}). We call
these eikonal lines $S^+$ and $S^-$ eikonal lines, respectively. The
Feynman rules for the eikonal lines in the soft decoupling relations are
summarized in the fourth and fifth lines of
Table~\ref{tab:cs-eikonal}, respectively.
\begin{table}
\begin{center}
\begin{tabular}{c|cc}
Type&
Vertex&Propagator\\
\hline
&&\\[-1.5ex]
$C^+$&
\hspace{2ex}
$\mp i g T_a \bar{n}_{1\mu}$
\hspace{2ex}
&
\hspace{2ex}
$\displaystyle \frac{i}
                                   {k\cdot \bar{n}_1-i\varepsilon}$ 
\hspace{2ex}
\\[2.5ex]
$C^-$&
\hspace{2ex}
$\pm i g T_a \bar{n}_{2\mu}$
\hspace{2ex}
&
\hspace{2ex}
$\displaystyle \frac{i}
                                   {k\cdot \bar{n}_2+i\varepsilon}$ 
\hspace{2ex}
\\[2.5ex]
$C^l$&
\hspace{2ex}
$i g T_a \bar{n}_{l\mu}$
\hspace{2ex}
&
\hspace{2ex}
$\displaystyle \frac{i}
                                   {k\cdot \bar{n}_l-i\varepsilon}$ 
\hspace{2ex}
\\[2.5ex]
$S^+$&
\hspace{2ex}
$\pm i g T_a n_{1\mu}$
\hspace{2ex}
&
\hspace{2ex}
$\displaystyle \frac{i}
                                   {k\cdot {n}_1+i\varepsilon}$ 
\hspace{2ex}
\\[2.5ex]
$S^-$&
\hspace{2ex}
$\mp i g T_a n_{2\mu}$
\hspace{2ex}
&
\hspace{2ex}
$\displaystyle \frac{i}
                                   {k\cdot {n}_2-i\varepsilon}$ 
\hspace{2ex}
\end{tabular}
\caption{
Feynman rules for the collinear 
($C^\pm$ and $C^l$) and soft ($S^\pm$)  eikonal lines. 
The upper (lower) sign is for 
the eikonal line that attaches to a quark (antiquark) line.
\label{tab:cs-eikonal}}
\end{center}
\end{table}

\subsection{Factorization of the singular regions\label{sec:singular-fact}}

Now we summarize the factorization of the singular regions. We refer the
reader to Ref.~\cite{Bodwin:2009cb} for detailed arguments. 

In analyzing the singular regions, we wish to identify the momentum
configurations that yield singular contributions. We can do so by making
use of the power-counting rules that we have outlined in
Sec.~\ref{sec:leading} and invoking the following specific
interpretations of those rules: the symbol $\sim$ and the phrase ``of
the same order'' mean that quantities differ by a finite factor, while
the phrases ``much less than'' and ``much greater than'' mean that
quantities differ by an infinite factor. It follows that, for gluons in
the singular regions, our convention that an allowed attachment of a
gluon cannot change the essential nature of the momentum of the line to
which it attaches has the following meaning: The attaching gluon cannot
have an energy that is greater by an infinite factor than the energy of
the line to which it attaches.

The rules in Sec.~\ref{sec:leading} lead to complicated relationships
between the allowed momenta of gluons in a given diagrammatic topology.
However, there is a general principle, which we have already mentioned,
that allows us to organize the discussion: The attachments of gluons to
a given line must be ordered so that a given attachment produces a
virtuality along the line that is of order or greater than the
virtualities that are produced by the attachments that lie to the
outside of it. In particular, the virtuality that a $C^i$, or $S$
singular gluon produces on a $C^j$ line with $j\neq i$ or an $S$ line is
of order the energy of gluon times the energy of the line to which it
attaches.

Our goal is to factor $C^\pm$ contributions from all subdiagrams
except $\tilde{J}^\pm$, to factor all $C^l$ singular contributions from
all subdiagrams except $\tilde{J}^l$ and the external $b$-quark line
and to factor all $S$ singular contributions from the $\tilde{J}^\pm$
subdiagrams. We will show that the factored soft contributions that are
associated with the external-quark and external-antiquark lines in
$\tilde{J}^\pm$ ultimately cancel.

Note that we do not factor $S$ singular contributions from
$\tilde{J}^l$ or from the external $b$-quark line. Nor do we factor
$C^l$ singular contributions from  the external 
$b$-quark line in the $B$-meson subdiagram. In principle, we could carry 
out such
factorizations. However, because the soft approximations are different
for the $b$ quark and the light antiquark in the $B$ meson, we do not
expect the factored soft contributions that are associated with the
$b$-quark and light-antiquark lines to cancel. Furthermore, it will
prove convenient, for purposes of expressing our results in terms of a
$B$-meson light-cone distribution, not to factor the $C^l$ singular
contributions from the $b$-quark line in the
$B$-meson subdiagram.

In the singular limits $\epsilon_S\to 0$, $\eta^i\to 0$, $\epsilon^i\to
0$, an infinite hierarchy of energy scales emerges. The energy scales of
the various levels in the hierarchy are separated by infinite factors.
We characterize each level in the hierarchy by the energy scale of
the $S$ singular gluons in that level. We call this scale the nominal
energy scale of that level. Collinear singular gluons in a level may
have energies that are of the nominal energy scale or energies that are
infinitely larger than the nominal scale, but still infinitesimal in
comparison with the nominal energy scale of the next higher level. We
call the latter gluons ``large-scale collinear singular gluons.'' We
carry out the factorization iteratively, starting with the level with
the largest nominal energy scale. As we shall see, this ordering of
the factorization procedure is convenient because it allows us to apply
the decoupling relations rather straightforwardly to decouple gluons
whose connections lie toward the inside of the Feynman diagrams before
we decouple gluons whose connections lie to the outside of the Feynman
diagrams.

\begin{figure}
\includegraphics[angle=0,height=7cm]{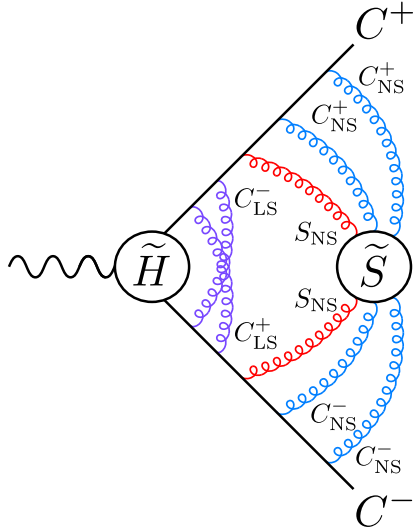}
\vspace*{8pt}
\caption{Diagram to illustrate the factorization of large-scale
collinear gluons and nominal-scale soft and collinear gluons for the
case of double-charmonium production in $e^+e^-$ annihilation.
$C^i_{\textrm{LS}}$ denotes a large-scale $C^i$ singular gluon, 
$C^i_{\textrm{NS}}$ denotes a nominal-scale $C^i$ singular gluon, and 
$S_{\textrm{NS}}$ denotes a nominal-scale $S$ singular gluon. 
\label{fig:fact1}%
}
\end{figure}

We will illustrate the factorization of the large-scale collinear
gluons and the nominal-scale soft and collinear gluons for the case of
double-charmonium production in $e^+e^-$ annihilation by referring to
the diagram that is shown in Fig.~\ref{fig:fact1}. In this diagram, we
have suppressed gluons with energies that are much less than the nominal
scale. These gluons have connections that lie to the outside of the
connections of the gluons that are shown explicitly. In the diagram in
Fig.~\ref{fig:fact1}, each gluon represents any finite number of
gluons, including zero gluons. For clarity, we have suppressed the 
antiquark lines in each meson and we have shown explicitly only the
connections of the gluons to the quark line in each meson and only a
particular ordering of those connections. However, we take the diagram
in Fig.~\ref{fig:fact1} to represent a sum of many diagrams, which
include all of the connections that we specify in the arguments below of
the singular gluons to the quark and antiquark in each meson, to other
singular gluons, and to the $\tilde{H}$ subdiagram.

\subsubsection{Factorization of the large-scale $C^i$ singular 
gluons\label{sec:large-c}}

First, we factor the large-scale $C^i$ singular gluons. In the first 
step of the iteration, these include gluons with finite energies, as 
well as infinitesimal energies. In subsequent steps, only gluons with 
infinitesimal energies are involved. There is a hierarchy in the energy 
scales of the large-scale $C^i$ singular gluons. We factor these gluons 
iteratively, beginning with the largest energy scale.  

We apply the $C^i$ approximations and the $C^i$ decoupling relations. In
applying the $C^\pm$ decoupling relations, we include the attachments
that are allowed by our conventions to all subdiagrams outside of
$\tilde{J}^\pm$, and, in applying the $C^l$ decoupling relations, we
include the attachments that are allowed by our conventions to all
subdiagrams outside of $\tilde{J}^l$ and the external $b$-quark line. We
also include, formally some attachments that may yield vanishing
contributions in the singular limits. These are attachments to
$\tilde{H}$ and attachments that lie to the inside of the allowed
attachments to $\tilde{J^j}$ for $j\neq i$.  We include in this class
attachments to the interior of $C^j$ eikonal lines. (``Interior'' means
to the inside of attachments of $C^j$ gluons.) 

The outermost allowed attachment of $C^i$ gluon to a $C^j$ singular
line in $\tilde{J}^j$ ($i\neq j$) generally lies to the
inside of attachments of additional gluons that have infinitesimally
smaller energy scales. While the propagator immediately to the outside
of the outermost allowed attachment of $C^i$ gluon is not precisely on
the mass shell, it is on the mass shell, up to relatively infinitesimal
corrections. Furthermore, if it is a gluon propagator, then its
polarization is orthogonal to its momentum, up to relatively
infinitesimal corrections. Therefore, when we apply the  $C^+$
decoupling relation, no eikonal-line contribution appears at this point.

The result of the application of the $C^i$ decoupling relations to
the large-scale $C^i$ singular gluons with the largest energies is
that the connections of these gluons to subdiagrams other than
$\tilde{J}^i$ and the $b$-quark line are replaced with connections to
$C^i$ eikonal lines. The $C^\pm$ eikonal lines attach to the $C^\pm$
external-fermion lines just to the outside of $\tilde{H}$. (Here, and in
subsequent discussions, ``external-fermion lines'' denote the fermion
lines that originate in the external quarks and antiquarks that are
associated with the mesons in our model.) The $C^l$ eikonal lines attach
to the external $b$-quark line and the external light-quark line from
the $B$ meson just to the outside of $\tilde{H}$.

\begin{figure}
\includegraphics[angle=0,height=7cm]{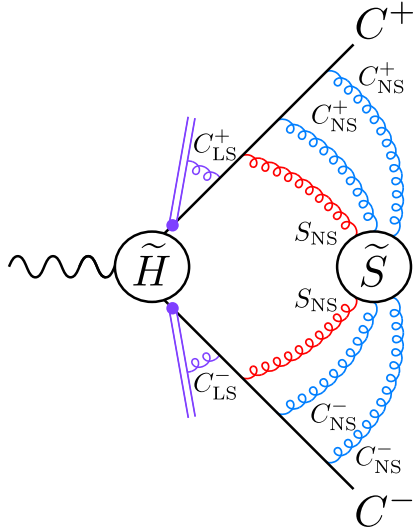}
\vspace*{8pt}
\caption{Diagram representing the sum of diagrams 
that occurs after one applies the decoupling of the large-scale 
collinear gluons that is described in 
Sec.~\ref{sec:large-c}.
\label{fig:fact2}%
}
\end{figure}

We can iterate this procedure for large-scale $C^i$ singular gluons with
successively lower energy scales. After each iteration, there is
a new $C^i$ eikonal line that attaches to each $C^i$
external-fermion line just to the inside of the $C^i$ eikonal line from
the previous iteration. It is easy to see that, for each
external-fermion line, the new eikonal line can be combined with the
eikonal line from the previous iteration to form a single eikonal
line, on which the $C^i$ singular gluons with lower energy scale
attach to the outside of the $C^i$ singular gluons with higher energy
scales. Other orderings of the attachments yield vanishing
contributions. We continue iteratively in this fashion until we have
factored all of the large-scale $C^i$ gluons. After this decoupling
step, the sum of diagrams represented by Fig.~\ref{fig:fact1} becomes a
sum of diagrams represented by Fig.~\ref{fig:fact2}.

\subsubsection{Initial factorization of the nominal-scale $C^i$ gluons
\label{sec:nom-c}}

\begin{figure}
\includegraphics[angle=0,height=7cm]{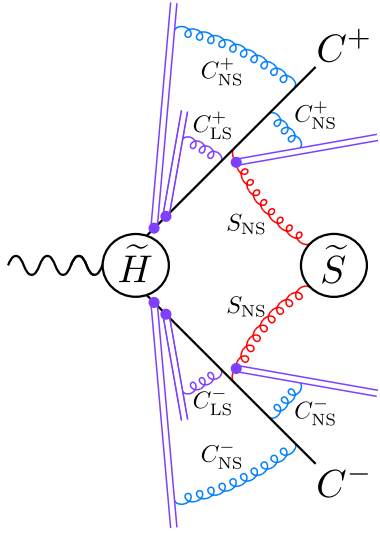}
\vspace*{8pt}
\caption{Diagram representing the sum of diagrams
that occurs after one applies the initial decoupling of the
nominal-scale collinear gluons that is described in
Sec.~\ref{sec:nom-c}.
\label{fig:fact3}%
}
\end{figure}

Next, we factor the nominal-scale $C^i$ singular gluons. In addition to
the attachments enumerated in the case of the large-scale $C^i$ gluons,
we include attachments to the nominal-scale $S$ gluons. Then, the
application of the $C^\pm$ decoupling relations leads to $C^\pm$ eikonal
lines that attach to the following locations: to the $C^\pm$
external-fermion lines just to the inside of the large-scale
$C^\pm$ eikonal lines from the previous step; to the nominal-scale $S$
singular gluon lines just to the inside of the connections of those
lines to the $C^\pm$ external-fermion lines. After this decoupling step, 
the sum of diagrams represented by Fig.~\ref{fig:fact2} becomes a 
sum of diagrams represented by Fig.~\ref{fig:fact3}.
Application of the $C^l$
decoupling relation leads to $C^l$ eikonal lines that attach to the
following locations: to the external-fermion lines from the $B$ meson
just to the inside of the large-scale $C^l$ eikonal lines from the
previous step; to the nominal-scale $S$ singular gluon lines just to the
inside of the connections of those lines to the external-fermion lines
from the $B$ meson. The $C^l$ eikonal line that attaches to a given
external-fermion line from the $B$ meson can be combined with the
large-scale $C^l$ eikonal line from the previous step to form a
single $C^l$ eikonal line.

\subsubsection{Factorization of the nominal-scale $S$ 
gluons\label{sec:nom-s}}

We now wish to apply the soft decoupling relations to factor the
nominal-scale soft gluons. In order to do this, we implement the $S^\pm$
approximations for the allowed attachments of the soft gluons to
$\tilde{J}^\pm$. (Recall that we do not apply the soft approximations or
the soft decoupling relations to the attachments of the soft gluons to
the external $b$-quark line or to $\tilde{J}^l$.) On the connections to
the $\tilde{J}^\pm$ subdiagrams, we modify the soft approximation in the
following way: We combine the momentum of the nominal-scale soft gluon
with the total momentum of the attached nominal-scale $C^\pm$ eikonal
line from the previous step. Then, when we implement the $S^\pm$
decoupling relations, the nominal-scale $C^\pm$ eikonal lines are carried
along with the nominal-scale soft-gluon attachments. We apply the
$S^\pm$ decoupling relations to the allowed attachments of the soft
gluons to $\tilde{J}^\pm$. We also include vanishing connections of the
nominal-scale soft gluons to the interior of the large-scale
$C^\pm$ eikonal lines \cite{Collins:1985ue}. The propagator that lies to
the outside of the outermost allowed connection of a nominal-scale soft
gluon to a line in $\tilde{J}^\pm$ is on shell, up to relative
corrections of infinitesimal size. Furthermore, if it is a gluon
propagator, its polarization is transverse to its momentum, up to
relative corrections of infinitesimal size.  Therefore, when we apply
the $S^\pm$ decoupling relations, no $S^\pm$ eikonal lines appear at
those points.

\begin{figure}
\includegraphics[angle=0,height=7cm]{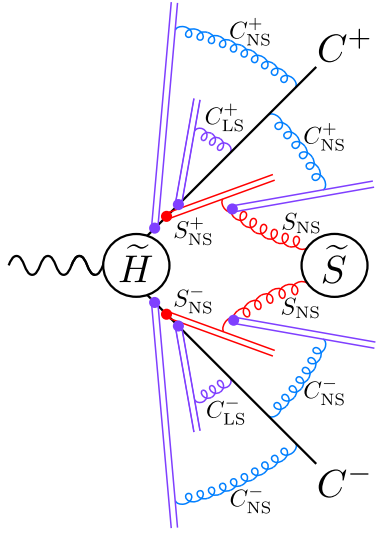}
\vspace*{8pt}
\vspace*{8pt}
\caption{Diagram representing the sum of diagrams
that occurs after one applies the decoupling of the nominal-scale soft
gluons that is described in Sec.~\ref{sec:nom-s}.
\label{fig:fact4}%
}
\end{figure}

The result of applying the $S^\pm$ decoupling relations is that soft
gluons attach to $S^\pm$ eikonal lines, to the external $b$-quark line
and to $\tilde{J}^l$. The $S^\pm$ eikonal lines attach to the $C^\pm$
external-fermion lines just to the outside of the nominal-scale $C^\pm$
eikonal lines and just to the inside of the large-scale $C^\pm$
eikonal lines. Associated with each connection of a nominal-scale soft
gluon to an $S^\pm$ eikonal line is a $C^\pm$ eikonal line. Associated
with each connection of a nominal-scale soft gluon to the external
$b$-quark line or to $\tilde{J}^l$ is a $C^l$ eikonal line. Our
sample diagram is now given by Fig.~\ref{fig:fact4}.

\subsubsection{Further factorization of the nominal-scale $C^\pm$ 
gluons\label{sec:nom-c-f}}

\begin{figure}
\includegraphics[angle=0,height=7cm]{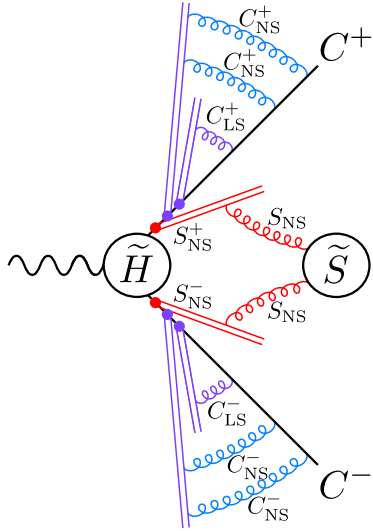}
\vspace*{8pt}
\caption{Diagram representing the sum of diagrams 
that occurs after one applies the further decoupling of the 
nominal-scale collinear gluons that is described in
Sec.~\ref{sec:nom-c-f}.
\label{fig:fact5}%
}
\end{figure}

We next factor the nominal-scale $C^\pm$ gluons from the $S^\pm$ eikonal
lines. In order do this, we include formally the vanishing contributions
that arise when one connects the nominal-scale $C^\pm$ gluons to all
points on the $S^\pm$ eikonal lines that lie to the inside of the
outermost connection of the nominal-scale soft gluons. We also make use
of the following facts: a nominal-scale $C^\pm$ eikonal line that
attaches to one of the $C^\pm$ external-fermion lines is identical to the
eikonal line that one would obtain by applying the $C^\pm$ decoupling
relation to the attachments of the nominal-scale $C^\pm$ gluons to an
on-shell fermion line (that does not have exactly $C^{\pm}$ momentum); a
nominal-scale $C^\pm$ eikonal line that attaches to a nominal-scale gluon
is identical to the eikonal line that one would obtain by applying the
$C^\pm$ decoupling relation to the attachments of nominal-scale $C^\pm$
gluons to an on-shell nominal-scale soft-gluon line. Then, applying
the $C^\pm$ decoupling relation, we find that the nominal-scale $C^\pm$
gluons attach to $C^\pm$ eikonal lines that attach to the
external-fermion lines just to the inside of the large-scale
$C^\pm$ eikonal lines. This situation is represented by the diagram 
that is shown in Fig.~\ref{fig:fact5}.

\begin{figure}
\includegraphics[angle=0,height=7cm]{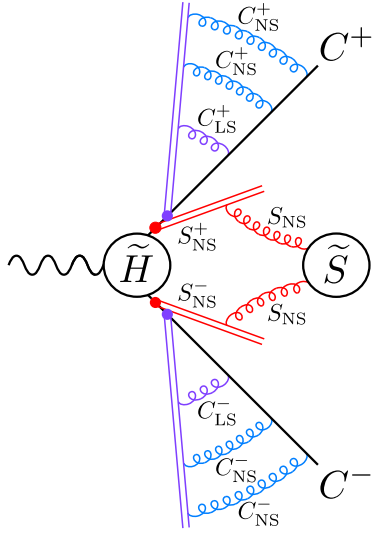}
\vspace*{8pt}
\caption{Diagram representing the sum of diagrams 
that occurs after one completely decouples the large-scale collinear 
gluons and the nominal-scale soft and collinear gluons.
\label{fig:fact6}%
}
\end{figure}

The nominal-scale $C^\pm$ eikonal lines can then
be combined with the large-scale $C^\pm$ eikonal lines. After
performing those steps we arrive at the final factorized form for our
sample diagram, which is given in Fig.~\ref{fig:fact6}.

\subsubsection{Completion of the factorization}

Now we can iterate the procedure that we have given in
Secs.~\ref{sec:large-c}--\ref{sec:nom-c-f}, taking the nominal scale to
be the next smaller soft-gluon scale. In these subsequent iterations, we
include the connections of soft and collinear gluons that have already
been described. In addition, we include formally, in the steps of
Secs.~\ref{sec:large-c} and \ref{sec:nom-c}, the vanishing contributions
from the connections of the large-scale and nominal-scale $C^i$ gluons
to the soft gluons of higher energies and to the $S^\pm$ eikonal lines
that are associated with those soft gluons.

Proceeding iteratively through all of the soft-gluon scales, we produce
new nominal-scale $S^\pm$ eikonal lines at each step that
connect to the external $C^\pm$ fermion lines just to the outside of the
existing $S^\pm$ eikonal lines. Each gluon that attaches to a
nominal-scale $S^\pm$ eikonal line has attached to it a $C^\pm$ eikonal
line.  In addition, there are nominal-scale $C^\pm$ eikonal lines from
the steps of Sec.~\ref{sec:nom-c} that attach to the $C^\pm$
external-fermion lines just to the inside of the nominal-scale $S^\pm$
eikonal lines. After the further factorization of the nominal-scale
$C^\pm$ gluons that is described in Sec.~\ref{sec:nom-c-f}, both of the
$S^\pm$ eikonal lines that attach to a given external-fermion line can be
combined into a single $S^\pm$ eikonal line.

At each step in the iteration, new $C^l$ eikonal lines appear that
attach to the external-fermion lines from the $B$ meson just to the
inside of the $C^l$ eikonal lines from the previous step. For each
external-fermion line, the new $C^l$ eikonal line can be combined with
the $C^l$ eikonal line from the previous step to form a single eikonal
line. Similarly, at each step in the iteration, new $C^l$ eikonal lines
appear that attach to the nominal-scale $S$ singular gluon lines that
attach to the external fermion lines from the $B$ meson. These new $C^l$
eikonal lines attach just to the inside of the $C^l$ eikonal
lines from the previous iteration. Again, for each external-fermion
line, the new $C^l$ eikonal line can be combined with the previous $C^l$
eikonal line to form a single $C^l$ eikonal line.

\begin{figure}
\includegraphics[angle=0,height=7cm]{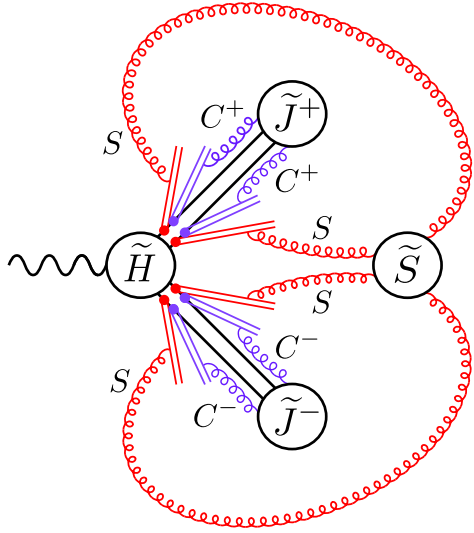}
\vspace*{8pt}
\caption{Illustration of the factorization  for the
case of $e^+e^-$ annihilation. After the use of the decoupling
relations, gluons with momenta in the $S$ singular region attach to
$S^{\pm}$ eikonal lines and gluons with momenta in the $C^{\pm}$
singular regions attach to $C^{\pm}$ eikonal
lines.\label{fig:softfa}%
}
\end{figure}

Following this procedure, we arrive at the factorized form for the
singular contributions. The $\tilde{S}$ subdiagram now connects only to
$S^\pm$ eikonal lines, to the external $b$-quark line, and to
$\tilde{J}^l$. The $S^\pm$ eikonal lines attach to the $C^\pm$
external-fermion lines just outside of $\tilde{H}$. All of the $C^\pm$
singular contributions are contained in the $J^\pm$ subdiagram and the
associated $C^\pm$ eikonal lines, which attach to the $C^\pm$
external-fermion lines just outside of the $S^\pm$ eikonal lines. All of
the $C^l$ contributions are contained in the $\tilde{J}^l$ subdiagram
and associated $C^l$ eikonal lines. These $C^l$ eikonal lines attach to
the external-fermion lines from the $B$ meson just to the outside of the
$\tilde{H}$ subdiagram and to $S$ singular gluon lines just to the inside of the
connections of those lines to the $b$-quark line and to $\tilde{J}^l$. 
This factorization is illustrated, for the case of $e^+e^-$
annihilation, in Fig.~\ref{fig:softfa}.

\subsection{Forms of the $\tilde{S}$ and $\tilde{J^\pm}$ functions and
cancellations of eikonal lines\label{cancel-eikonal}}

\subsubsection{Cancellations of the soft eikonal 
lines\label{sec:soft-cancellations}}

At this point, in the case of  $e^+e^-$ annihilation into two quarkonia,
the $\tilde{S}$ subdiagram and associated
soft eikonal lines, which we call $\bar S$, take the form of the
vacuum-expectation value of a time-ordered product of four eikonal lines:
\begin{equation}
\bar{S}(x_{1 q},x_{1\bar q},x_{2 q},x_{2\bar q})
=\langle 0\vert T\{[x_{1\bar q},\infty^+][\infty^+,x_{1 q}]
\otimes
[x_{2\bar q},\infty^-][\infty^-,x_{2 q}]\} \vert 0 \rangle_S,
\label{S-tilde-me}
\end{equation}
where $x_{i q}$ and $x_{i\bar q}$ are the points at which the eikonal 
lines attach to the quark and antiquark external lines from meson~1 and 
meson~2, respectively. $[y,x]$ is the eikonal line that is defined in
Eq.~(\ref{eikonal-line}), $\infty^+=(\infty,0,\bm{0}_\perp)$, and
$\infty^-=(0,\infty,\bm{0}_\perp)$. 
The symbol $\otimes$ indicates a direct product of the color factors 
that are associated with the soft-gluon attachments to meson~1 and the 
soft-gluon attachments to meson~2.
The $S$ subscript on the matrix element indicates that  only the
contributions from the $S$ singular region are kept.

In the case of $B$-meson decays, the $\tilde{S}$ subdiagram is still
connected to the $B$ meson and takes the form
\begin{equation}
\bar{S}_B(x_{1 q},x_{1\bar q},x_{2 q},x_{2\bar q})=
\left<0\vert T\{[x_{1\bar q},\infty^+][\infty^+,x_{1 q}]
\otimes
[x_{2\bar q},\infty^-][\infty^-,x_{2 q}]\}
\bar{\Psi}_l\Gamma_{Bm}\Psi_b\vert B\right>_S.
\end{equation}
Here, we have suppressed the $C^l$ eikonal lines that are associated 
with the $B$ meson. No soft gluons attach to those lines.

Because the $\tilde{H}$ subdiagram is insensitive to a
momentum in the $S$ singular region that flows through it, one can ignore the difference between $x_{1q}$ and $x_{1\bar{q}}$ and the difference between $x_{2q}$ and $x_{2\bar{q}}$.  Therefore, in
consequence of the fact that the external mesons are color singlets, the
$S^+$ quark and antiquark eikonal lines cancel, and the $S^-$ quark and
antiquark eikonal lines cancel. In the case of $e^+e^-$ annihilation into
two quarkonia, this cancellation implies that the $\tilde{S}$ subdiagram
is completely disconnected, and, therefore, can be ignored. In the
case of $B$-meson decays, the remaining $\tilde{S}$ subdiagram now
connects only to the external $b$-quark line and to $\tilde{J}^l$.

\subsubsection{Rearrangement of the $B$-meson singular contributions}

As we have noted, there are $C^l$ eikonal lines associated with the $B$
meson. These $C^l$ eikonal lines attach to the external-fermion lines
from the $B$ meson just to the outside of the $\tilde{H}$ subdiagram and to $S$
singular gluon lines just to the inside of the connections of those
lines to the external $b$-quark line and to $\tilde{J}^l$. We can now
remove the latter class of eikonal lines as follows. We note that,
because the $\tilde{S}$ subdiagram now connects only to the external
$b$-quark line and to $\tilde{J}^l$, the $C^l$ eikonal lines that attach
to the $S$ singular gluons are precisely the $C^l$ eikonal lines that
would appear if one were to factor connections of $C^l$ singular gluons
from $\tilde{S}$. (One can carry out the factorization
iteratively, level-by-level, factoring the nominal-scale $C^l$ gluons
from the nominal-scale $S$ singular gluons.) Therefore, we restore
the connections of the $C^l$ singular gluons to $\tilde{S}$ and drop the
$C^l$ eikonal lines that attach to $S$ singular gluons.

\subsubsection{Forms of the meson distributions}

We make a Fierz rearrangement to decouple the
color structures of the $\tilde{J}^{\pm}$ subdiagrams, the 
$b$-quark and $\tilde{J}^l$ subdiagram, and their associated collinear 
eikonal lines. Then, these subdiagrams and their eikonal lines are 
given by the following matrix elements:
\begin{equation}
\bar{J}_{\alpha\beta}^+(\bar q_1)=
\int_{-\infty}^{+\infty}d^4 (2x)
\exp[ - 2i\bar q_1\cdot x]
\langle H_1(P_1)\vert \bar\Psi_{\alpha}(x) 
T\{[x,\infty^-][\infty^-,-x]\}\Psi_{\beta} (-x)\vert 0\rangle_{C^+}
\label{J-plus-sing}
\end{equation}
for the $C^+$ quarkonium,
\begin{equation}
\bar{J}_{\alpha\beta}^-(\bar q_2)
=
\int_{-\infty}^{+\infty}d^4 (2x)
\exp[ - 2i\bar q_2\cdot x]
\langle H_2(P_2)\vert \bar\Psi_{\alpha}(x) T\{[x,\infty^+]
[\infty^+,-x]\}\Psi_{\beta}(-x)\vert 0\rangle_{C^-}
\label{J-minus-sing}
\end{equation}
for the $C^-$ quarkonium,
\begin{equation}
\bar{J}_{K\alpha\beta}(\bar r_k)
=\int_{-\infty}^{+\infty}d^4 (2x)
\exp[ - 2i\bar r_k\cdot x]
\langle  K(p_{K})\vert \bar\Psi_{\alpha}(x) T\{[x,\infty^+]
[\infty^+,-x]\}\Psi_{\beta}(-x)\vert 0\rangle_{C^-}
\label{J-K-sing}
\end{equation}
for the light meson, and
\begin{equation}
\bar{J}_{B\alpha\beta}(\bar p_l)
=\int_{-\infty}^{+\infty}d^4 x
\exp[ i\bar p_l\cdot x]
\langle 0\vert \bar\Psi_{l\beta}(x) T\{[x,\infty^-]
[\infty^-,0]\}\Psi_{b\alpha}(0)\vert B(p_B)\rangle_{S,C^l}.
\label{J-B-sing}
\end{equation}
for the $B$ meson. 
In Eqs.~(\ref{J-plus-sing})--(\ref{J-B-sing}), 
$\alpha$ and $\beta$ are Dirac indices.
It is understood that the fields $\Psi$ and $\bar{\Psi}$ in each 
matrix element are in a color-singlet state.
In these distributions, the arguments $\bar q_1$,
$\bar q_2$, and $\bar r_k$ are each half
the difference between the quark momentum and the antiquark 
momentum at the points at which they enter $\tilde{H}$,
and the argument $\bar q_l$ is the antiquark momentum at the point at
which it enters $\tilde{H}$. We have suppressed the dependences on the
total meson momenta $P_1$, $P_2$, $p_K$, and $p_B$ in the arguments on
the left sides of Eqs.~(\ref{J-plus-sing})--(\ref{J-B-sing}). The
subscripts $S$, $C^+$, $C^-$ and $C^l$ on the matrix elements indicate
that we are retaining only the $S$, $C^+$, $C^-$, and $C^l$ singular
contributions.

\subsubsection{Light-cone distributions and cancellations of the collinear eikonal lines}

Away from the endpoint region, we can simplify the factorized
expression further. 

In $\tilde{H}$, away from the endpoint region, we can approximate
the momenta of the quark and antiquark in the light meson by their minus
components. The leading relative errors in this approximation are of
order $q_k/p_K\sim \Lambda_{\rm QCD}/Q$. Then, integrating $\bar{J}_K$
over $\bar{r}_k^+$ and $\bar{\bm{r}}_{k\perp}$, we obtain
\begin{eqnarray}
\bar{J}_{K\alpha\beta}(y)&\equiv&
\frac{p_{K}^-}{2\pi}
\int_{-\infty}^{+\infty}
\frac{d\bar{r}_k^+\, d^2\bar{\bm{r}}_{k\perp}}
     {(2\pi)^3}\, 
\bar{J}_{K\alpha\beta}(\bar{r}_k)
\nonumber\\
& = & \frac{p_{K}^-}{\pi}\int_{-\infty}^{+\infty} d x^+
\exp[ - i(2y-1)p_{K}^- x^+] 
\nonumber\\ &&\qquad\times\,
\langle K(p_{K})\vert \bar{\Psi}_\alpha(x^+) 
T\{[x^+,\infty^+ ]
 [\infty^+ ,-x^+]\}\Psi_\beta(-x^+)\vert 
 0\rangle_{C^-}.
\label{J-K-lc}
\end{eqnarray}

The quantity $\tilde{H}$ has been analyzed in the context of
soft-collinear effective theory (SCET) for the case of $B$-meson decays
into a lepton pair plus a photon \cite{Bosch:2003fc} and for the
contribution to $B$-meson decays into two light mesons that arises away
from the endpoint region \cite{Beneke:2003pa}. The conclusion of these
analyses is that $\tilde{H}$ is given, to leading order in $\Lambda_{\rm
QCD}/Q$, by a matrix element of a SCET operator that depends only on the
plus component of the momentum of the light antiquark in the $B$
meson.\footnote{These analyses are based on Lorentz (reparametrization)
invariance and power counting in $\sqrt{\Lambda_{\rm QCD}/Q}$. The
next-to-leading-order spectator-scattering contributions to $B$-meson
decays to light mesons have been computed in
Refs.~\cite{Beneke:2005vv,Pilipp:2007mg, Kivel:2006xc,
Beneke:2006mk,Jain:2007dy} and confirm the general analysis for this
process.} Furthermore, the SCET operator has a Dirac-matrix structure
such that only the $B$-meson light-cone distribution $\Phi_{B1}$
contributes. We assume that a similar SCET analysis holds in the case of
$B$ meson decays to a quarkonium plus a light meson away from the
endpoint region. Then, integrating $\bar{J}_{B}$ over $p_l^-$ and
$\bm{p}_{l\perp}$, we obtain
\begin{eqnarray}
\bar{J}_{B\alpha\beta}(\xi)
&\equiv&
\frac{p_B^+}{2\pi}
\int_{-\infty}^{+\infty}
\frac{ dp_l^- \,d^2\bar{\bm{p}}_{l\perp}}{(2\pi)^3} \,
\bar{J}_{B\alpha\beta}(p_l)
\nonumber\\
& = & \frac{p_B^+}{2\pi}\int_{-\infty}^{+\infty} d x^-
\exp[ i\xi p_B^+ x^-]
\nonumber\\ &&\qquad\times\,
\langle 0\vert \bar{\Psi}_{l\beta}(x^-) 
T\{[x^-,\infty^- ]
 [\infty^- ,0]\}\Psi_{b\alpha}(0)\vert 
 B(p_B)\rangle_{C^+},
\label{J-B-lc}
\end{eqnarray}
where $\xi=p_l^+/Q$.

We do not approximate the momenta of the heavy-quark and heavy antiquark
in the quarkonia by their dominant momentum components because, in so
doing, we would introduce errors of relative order $m_cv/Q$ for each
quarkonium. As we will explain in
Sec.~\ref{sec:factorization-corrections}, such an error would be larger
than the errors that arise from the approximations that we have used to
derive the factorization result.

Now, we can see that there is a partial cancellation of the $C^-$ quark
and antiquark eikonal lines in Eq.~(\ref{J-K-lc}) and a partial
cancellation of the $C^l$ quark and antiquark eikonal lines in
Eq.~(\ref{J-B-lc}). The cancellations would be complete, were it not for
the fact that the $\tilde{H}$ subdiagram is sensitive the routing of
collinear momenta through it. This sensitivity corresponds to the
separation in space-time of the points $x^+$ and $-x^+$ in
Eq.~(\ref{J-K-lc}) and the points $x^-$ and $0$ in
Eq.~(\ref{J-B-lc}). The quark and antiquark eikonal lines in
Eqs.~(\ref{J-K-lc}) and (\ref{J-B-lc}) cancel where they overlap,
leaving an eikonal line that runs directly between the quark and the
antiquark:
\begin{subequations}\label{eikonal-cancellation}%
\begin{eqnarray}
\label{eikonal-cancellation1}
\bar{J}_{K\alpha\beta}(y)
& = & 
 \frac{p_{K}^-}{\pi}\int_{-\infty}^{+\infty} d x^+
\exp[ - i(2y-1)p_{K}^-x^+] 
\langle K(p_{K})\vert \bar{\Psi}_\alpha(x^+) 
P[x^+,-x^+]\Psi_\beta(-x^+)\vert 
 0\rangle_{C^-}
\nonumber\\
&\equiv&\sum_j 
\Phi_{Kj}(y)\left[\Gamma_{Kj}\right]_{\alpha\beta},
\\
\label{eikonal-cancellation2}
\bar{J}_{B\alpha\beta}(\xi)
& = & \frac{p_B^+}{2\pi}\int_{-\infty}^{+\infty} d x^-
\exp[ i\xi p_B^+ x^-] 
\langle 0\vert \bar{\Psi}_{l\beta}(x^-) 
P[x^-,0]\Psi_{b\alpha}(0)\vert 
 B(p_B)\rangle_{C^+}
\nonumber\\
&\equiv&\sum_m \Phi_{Bm}(\xi) 
\left[\Gamma_{Bm}\right]_{\alpha\beta},
\end{eqnarray}
\end{subequations}
where we have written the time-ordered product of the exponentiated line
integral as a path-ordered product.\footnote{Reference \cite{Bodwin:2008nf}
contains an incorrect statement that the eikonal lines in
Eq.~(\ref{eikonal-cancellation}) cancel completely.} The expressions in
Eqs.~(\ref{eikonal-cancellation}) have the form of the conventional
light-meson and $B$-meson light-cone distributions, but, at this stage,
they contain only the singular contributions to those light-cone
distributions. Since the integrations over $y$ and $\xi$ have a finite range
of support in $\tilde{H}$, the typical separation of the points $x^{+}$
and $-x^{+}$ in Eq.~(\ref{eikonal-cancellation1}) and the points
$x^{-}$ and $0$ in Eq.~(\ref{eikonal-cancellation2}) is of order
$1/Q$.

\subsection{Factorized form}\label{sec:factorized-form}

\subsubsection{Factorization of the logarithmic enhancements}

At this point, we have established that the contributions from the soft
singular region decouple completely from the $\tilde{J}^\pm$
subdiagrams (leaving no residual eikonal lines). We have also
established that the contributions from the collinear singular regions
factor from the $\tilde{H}$ subdiagram and are contained
entirely in the $\bar J^\pm$, $\bar J_K$ and $\bar J_B$ subdiagrams.
As we have mentioned, in the case of $e^+e^-$ annihilation into
two quarkonia, the $\tilde{S}$ subdiagram is now completely disconnected,
and can be ignored. In the case of $B$-meson decays, the $\tilde{S}$
subdiagram is still connected to the external $b$-quark line and to
$\tilde{J}_l$ ({\it i.e.}, to $\bar{J}_B$).

Now let us restore $m_c$ to its nonzero physical value. Then, some of
the soft and collinear singularities become would-be soft and collinear
singularities. However, the would-be singularities are still contained
in the $\bar J^\pm$, $\bar J_K$ and $\bar J_B$ subdiagrams. Therefore,
there are no actual or would-be collinear singularities in the
$\tilde{H}$ subdiagram. Furthermore, there are no actual or would-be
soft singularities in the $\tilde{H}$ subdiagram. In the case of
$B$-meson decays, there are, however, soft contributions from the
endpoint region in the $S-\tilde{S}$ subdiagram, and, hence, in the
$\tilde{H}$ subdiagram. As we have emphasized, these endpoint
contributions are associated with the topology of
Fig.~\ref{fig:regions}(b).

Next let us redefine $\bar J^\pm$, $\bar J_K$ and $\bar J_B$ by
extending the ranges of integration from the infinitesimal $C^\pm$,
$C^l$ singular regions and, in the case of $\bar J_B$, the $S$ singular
region, to finite regions that are defined by an ultraviolet cutoff
$\mu_F\sim Q$ on the logarithmic integrals. $\tilde{H}$ is then
redefined to be the remainder of the amplitude. One can think of $\mu_F$
as an infrared cutoff on the soft and collinear enhancements in
$\tilde{H}$. This redefinition has the effect of absorbing the collinear
logarithmic enhancements that are associated with the collinear
singularities into $\bar J^\pm$, $\bar J_K$ and $\bar J_B$. It also has
the effect of absorbing soft enhancements that are associated with soft
singularities into $\bar J_B$.

One might worry that, in making such an extension, we could introduce
new singularities and logarithmic enhancements in $\bar J^\pm$, $\bar
J_K$ and $\bar J_B$ that are associated with their collinear eikonal
lines. The lightlike eikonal lines that are parametrized by the vectors
$\bar n_1$, $\bar n_2$, and $\bar n_l$ could, in principle, be sources
of gluons that are collinear to the minus, plus, and $\bar e_l$
directions, respectively, as well as sources of soft gluons. In fact,
this does not happen in the case of the light-meson light-cone
distribution [Eq.~(\ref{eikonal-cancellation1})] or the $B$-meson
light-cone distribution [Eq.~(\ref{eikonal-cancellation2})]. As we
have noted, there is a partial cancellation between the quark and
antiquark eikonal lines in these light-cone distributions. The remaining
eikonal-line segment is typically of length $1/Q$. Therefore, only modes
with virtuality of order $Q$ can propagate along it, and no collinear or
soft singularities or logarithmic enhancements are associated with it.

In the case of the $\bar J^\pm$ distributions in Eqs.~(\ref{J-plus-sing})
and (\ref{J-minus-sing}) and the $\bar{J}_K$ and $\bar{J}_B$
distributions in Eqs.~(\ref{J-K-sing}) and (\ref{J-B-sing}), which are
appropriate when the light-meson momentum is in the endpoint region, we
make use of a trick to prevent collinear singularities and
enhancements from developing along the eikonal lines: In each case, we
replace the lightlike eikonal lines with spacelike eikonal
lines. That is, we replace the eikonal-line vectors $\bar n_1$, $\bar
n_2$, and $\bar n_l$ with a vector
$n_{z}=(1/\sqrt{2})(1,-1,\bm{0}_\perp)$, which points in the $z$
direction. Because of the freedom in choosing the collinear eikonal
vectors that we described in Sec.~\ref{sec:collinear-app}, this
replacement has no effect on the $C^+$, $C^-$, and $C^l$ singular
contributions in $\bar J^+$, $\bar J^-$, $\bar{J}_K$, and 
$\bar{J}_B$, respectively. Furthermore, the soft singularities (and
enhancements) that arise from soft-gluon attachments to the quark and
antiquark eikonal lines in Eqs.~(\ref{J-plus-sing}),
(\ref{J-minus-sing}), (\ref{J-K-sing}), and (\ref{J-B-sing}) cancel.
This cancellation derives from the following facts: The $S$-singular
attachments lie to the exterior of any non-$S$-singular attachments to
the eikonal lines; any non-$S$-singular attachments are within $1/Q$ of
the eikonal-line endpoints; the endpoints $-x$ and $x$ in
Eqs.~(\ref{J-plus-sing}), (\ref{J-minus-sing}), and (\ref{J-K-sing}) and
$0$ and $x$ in Eq.~(\ref{J-B-sing}) are within $1/Q$ of each other.
Hence, one can argue, as in Sec.~\ref{sec:soft-cancellations},
that the segments of the quark and antiquark eikonal lines that contain
$S$-singular-gluon attachments cancel.

We have argued that there are neither soft nor collinear logarithmic
enhancements in the $\tilde{H}$ subdiagram. Therefore, in the cases of
$e^+e^-$ annihilation and $B$-meson decay in the topology of
Fig.~\ref{fig:regions}(a), the $\tilde{H}$ subdiagram involves only
momenta of order $Q$. The lower-virtuality momenta are contained in
the distributions $\bar J^\pm$ in Eqs.~(\ref{J-plus-sing}) and
(\ref{J-minus-sing}), $\Phi_K$ in Eq.~(\ref{eikonal-cancellation1}),
and $\Phi_B$ in  Eq.~(\ref{eikonal-cancellation2}).

\subsubsection{Further factorization of the endpoint 
contributions\label{sec:endpoint-fact}}

In the case of $B$-meson decays in the topology of
Fig.~\ref{fig:regions}(b), the $\tilde{H}$ subdiagram is also free of
soft and collinear logarithmic enhancements, but it still contains
gluons with momenta of order $\Lambda_{\rm QCD}$ that arise from the
endpoint region. These gluons consist of the gluon that is marked with
an asterisk in Fig.~\ref{fig:regions}(b) and gluons that are radiated
from it. They are the part of $S-\tilde{S}$ that remains after
$\tilde{S}$ has been extended to include soft enhancements. They can
connect to active-quark or active-antiquark lines (those that
participate in the weak interaction). However, they cannot connect to
any part of the $\bar{J}^+$ or $\bar{J}^-$ subdiagrams, which reside to
the outside of the connections of the soft gluons to the active-quark or
active-antiquark lines. Because these soft gluons connect the $B$ meson
and light meson to the quarkonia, they potentially violate the
factorized form in the second term of Eq.~(\ref{B-meson-fact}). 

However, we can make a further decoupling of the connections of the
endpoint soft gluons from the active-quark and active-antiquark lines in
the quarkonium. We apply a modified soft approximation to these gluons.
Because the soft gluons have a finite soft momentum of order
$\Lambda_{\rm QCD}$, rather than a soft singular momentum, there are
errors associated with the application of the soft approximation to the
quark or antiquark line that are order $\Lambda_{\rm QCD}/Q$. These
errors are negligible in comparison with the errors that are associated
with the modified soft approximation for the average of the quark and
antiquark momenta. Next we apply the $S^+$ decoupling relation. Then,
the soft gluons attach to eikonal lines that attach to the heavy-quark
and heavy-antiquark lines just outside the $\tilde{H}$ subdiagram.
Because the remaining part of $\tilde{H}$ is insensitive to routing of
the soft momenta through it, the quark and antiquark eikonal lines
cancel, up to corrections of order $\Lambda_{\rm QCD}/Q$. Then, the
endpoint contributions are contained entirely in a subdiagram $BK$,
which consists of $S-\tilde{S}$ (after $\tilde{S}$ has been extended to
include soft enhancements), the parts of the $B$-meson and light-meson
quark and antiquark lines to which $S-\tilde{S}$ attaches, $\bar{J}_K$
in Eq.~(\ref{J-K-sing}), and $\bar{J}_B$ in Eq.~(\ref{J-B-sing}).. The
$\tilde{H}$ subdiagram now contains only momenta of order $Q$.
Consequently, we can contract $\tilde{H}$ to a point with respect to
the soft interactions in $BK$. Then, decoupling the Dirac and color
indices of $BK$ from $\tilde{H}$ by making Fierz
rearrangements, we obtain the $B$-meson-to-light-meson form factors
in Eq.~(\ref{form-factor}) from $BK$ and short-distance
coefficients $\tilde{H}_e$ from $\tilde{H}$.

\subsubsection{NRQCD decomposition of the quarkonium distribution amplitudes}

At this stage, we have achieved the factorized forms of
Eqs.~(\ref{ep-em-fact}) and (\ref{B-meson-fact}), except that the
quarkonium factors are expressed in terms of quarkonium 
distribution amplitudes, instead of NRQCD matrix elements. We now argue
that the quarkonium  distribution amplitudes can be expanded
as a sum of products of NRQCD matrix elements times short-distance
coefficients.

The $\bar{J}^\pm$ distribution amplitude describes the local creation
of a quark-antiquark pair, followed by its evolution, through QCD
interactions, into a quarkonium. The gluons in the $\bar J^\pm$
distribution amplitude, which have $C^\pm$ momentum in the $e^+e^-$ CM
frame or the $B$-meson rest frame, have hard, soft, and threshold
(potential) momenta in the quarkonium rest frame. If the $\bar{J}^\pm$
distribution amplitude involved only a heavy quark, a heavy antiquark,
and any number of gluons and light-quark-antiquark pairs, then it is
clear that it could be written as a standard NRQCD decomposition of a
full QCD amplitude. That is, it could be written as a sum over
products of short-distance coefficients times matrix elements of local
NRQCD operators. Lines with virtualities of order $m_c$ or greater lie
to the inside of the lower virtuality lines, and could be integrated
out to yield the local NRQCD operators times short-distance
coefficients. Lines with virtualities less than of order $m_c$ are well
described by NRQCD and would be accounted for by the NRQCD matrix
elements of these local operators between the vacuum state and the
quarkonium state.

A complication to this picture arises because the distribution
amplitudes also contain eikonal lines, which are not a part of QCD.
However, the attachments of the eikonal lines to the external
heavy-quark and heavy-antiquark lines at the points $x_{iq}$ and
$x_{i\bar q}$ are separated in space-time by a distance of order $1/Q$.
Hence, only high-virtuality modes can propagate on these lines.
Therefore, they too can be integrated out to yield local operators times
short-distance coefficients. These operators would involve the gauge
field, as well as the quark and antiquark fields.

Therefore, we can write
\begin{subequations}
\label{eq:Jbar-fact}
\begin{eqnarray}
\bar{J}^+(\bar q_1)&=&\sum_{i} a_{1i}(\bar q_1)
\langle H_1\vert {\cal O}_i \vert 
0\rangle,\\
\bar{J}^-(\bar q_2)&=&\sum_{i} a_{2i}(\bar q_2)
\langle H_2\vert {\cal O}_i \vert 
0\rangle,
\end{eqnarray}
\end{subequations}
where $a_{1i}$ and $a_{2i}$ are short-distance
coefficients. $a_{1i}$ and $a_{2i}$ each have two Dirac indices, 
corresponding to the quark line and the antiquark line in $\bar{J}^+$ 
and $\bar{J}^-$, respectively. We suppress those indices. 
We then make the following identifications for the cases of $e^+e^-$
annihilation, $B$ decay in the topology of Fig.~\ref{fig:regions}(a), and
$B$ decay in the topology of Fig.~\ref{fig:regions}(b), respectively:
\begin{subequations}
\label{eq:Jbar-SD-coeff}
\begin{eqnarray} 
A_{ij}&=&\int 
\frac{d^4\bar q_1}{(2\pi)^4}\, 
\frac{d^4\bar q_2}{(2\pi)^4}\, 
\tilde{H}_{}(\bar q_1,\bar q_2)\,
a_{1i}(\bar q_1)\, a_{2j}(\bar q_2),\\
A'_{ije}(y,\xi)&=&\int 
\frac{d^4\bar q_1}{(2\pi)^4}\, 
\tilde{H}_{je}(\bar q_1,y,\xi)\, 
a_{1i}(\bar q_1),\\ 
A_{ie}&=&\int 
\frac{d^4\bar q_1}{(2\pi)^4}\, 
\tilde{H}_{e}\,
a_{1i}(\bar q_1).
\end{eqnarray}
\end{subequations}
Here, we have also suppressed the Dirac indices on $\tilde{H}$, which are
contracted into the (suppressed) Dirac indices on $a_{1i}$ and
$a_{2j}$. The identifications in Eq.~(\ref{eq:Jbar-SD-coeff}) lead
directly to the factorization formulas in Eqs.~(\ref{ep-em-fact}) and
(\ref{B-meson-fact}).

\subsection{Corrections to factorization
\label{sec:factorization-corrections}}

Now let us discuss the corrections to the factorized form. The most
important corrections to the factorized form arise because of the
approximate nature of the cancellations of the couplings of soft-singular
gluons to the color-singlet quarkonia. These cancellations hold only up
to the errors in the modified soft approximation. We wish to compare the
sizes of these errors relative to the factorized contributions. In
some cases, the contributions from the modified soft approximation,
which ultimately cancel, simply scale with $m_c$, $Q$, and $v$ in the same way
as the factorized contribution. However, there can be exceptions to this
scaling because of the specific quantum numbers of the final states in a 
given process. We give some examples of such exceptions below.

Note that, in the case of $e^+e^-$ annihilation into two quarkonia,
violations of factorization arise only in contributions involving the
corrections to the modified soft approximation for {\it both} quarkonia.
The reason for this is that, if the $\tilde{S}$ subdiagram decouples from
quarkonium~$i$, but not from quarkonium~$j$, then 
the $\tilde{S}$ subdiagram can be absorbed
into the definition of the $\tilde{J}$ subdiagram for meson~$j$. (See 
Ref.~\cite{Bodwin:2009cb} for a more detailed discussion of this point.)

Now let us discuss the dependence of the relative size of the
corrections to factorization on the orbital angular momenta of the
produced quarkonia. The leading errors in the modified soft
approximation are proportional to $\bm{q}_{i\perp}/P_i\sim m_cv/Q$.
Because of their proportionality to $\bm{q}_{i\perp}$, the leading
errors in the modified soft approximation contribute one unit of orbital
angular momentum. Consequently, in order to yield a $Q\bar Q$ pair in
the quarkonium angular-momentum state, they must be accompanied by an
additional factor $\bm{q}_{i\perp}/m_c\sim v$ from the
short-distance production process in the case of an $S$-wave
quarkonium and an additional factor $(\bm{q}_{i\perp}/m_c)^{L-1}\sim
v^{L-1}$ from the short-distance production process in the case of
an $L$-wave quarkonium with $L>0$. The factorized contributions contain
a factor $v^L$ for each $L$-wave quarkonium. Therefore, the
factorization-violating contributions are suppressed, relative to the
factorized contributions, by a factor $f_i\sim m_c v^2/Q$ for each
quarkonium~$i$ in an $S$-wave state and by a factor $f_i\sim m_c/Q$ for
each quarkonium~$i$ in a higher orbital-angular-momentum
state.\footnote{In the case of the factorized form for $B$-meson decays
in the first term of Eq.~(\ref{B-meson-fact}), there are also errors in
the cancellation of the couplings of the soft-singular gluons to the
light meson. These errors are of order $q_k/Q\sim \Lambda_{\rm
QCD}/Q$ relative to the factorized contributions. In addition, there
are errors of relative order $\Lambda_{\rm QCD}/Q$ that arise when one
expresses the amplitude in terms of the light-cone distributions for the
light meson [Eq.~(\ref{J-K-lc})] and $B$ meson [Eq.~(\ref{J-B-lc})].
These errors arise because one neglects in $\tilde{H}$ the plus and
transverse components of the momenta of the quark and the antiquark in
the light meson and the minus and transverse components of the momentum
of the antiquark in the $B$ meson. We neglect these errors in comparison
with the errors in the cancellation of the couplings of soft-singular
gluons to the quarkonia.} The suppressions of the
factorization-violating contributions that we find here are consistent
with those that were found in Ref.~\cite{Bodwin:2008nf}. However, in
Ref.~\cite{Bodwin:2008nf}, powers of $v$ in the factorization-violating
contributions were ignored.

The relative sizes of the corrections to factorization can depend on
additional quantum numbers, beyond the orbital angular momenta of the
quarkonia. Let us mention a few examples. In the case of production of
$S$-wave quarkonia, the factorized production process can be suppressed
by powers of $m_c/Q$ if it involves a helicity flip. (See, for example
Ref.~\cite{Braaten:2002fi}.) However, we expect such a helicity
suppression to apply to the factorization-violating contributions, as
well, and so it should not affect the relative size of the
factorization-violating contributions. In order $\alpha_s^0$,
$B$-meson decays do not produce a $\chi_{c0}$ or $\chi_{c2}$ charmonium
(a $J=0$ or $J=2$ $P$-wave state). Those processes are allowed only in
order $\alpha_s$. On the other hand, the factorization-violating
corrections to $B$-meson decays {\it do} produce $\chi_{c0}$ and
$\chi_{c2}$ charmonia in order $\alpha_s^0$. Therefore, the
factorized process for $\chi_{c0}$ or $\chi_{c2}$ production is
suppressed by a power of $\alpha_s$, relative to the
factorization-violating process, and may not be dominant. Since the
factorization-violating contributions arise from diagrams in which at
least one gluon has been added to the leading-order process, there can
also be a dependence of the relative size of the factorization-violating
contributions on the color structure of the hard subprocess.

In perturbation theory, the factorization-violating contributions
may be enhanced by logarithms of $Q^2/m_c^2$. Furthermore, they are
infrared divergent. In reality, these infrared divergences are cut off by
nonperturbative effects associated with confinement. Our analysis does
not determine the size of these factorization-violating
contributions: It only shows that they vanish as one or two powers
of $f$ as $f$ approaches zero. One might use the small parameter $f$ as
an estimate of the size of the factorization-violating
contributions. However, the size of the factorization-violating
contributions is an issue that, at present, must be settled through
experiment or, perhaps, lattice simulations.

We note that, at lowest order in $v$, one sets $q_i=0$, and the
cancellation of the couplings of soft-singular gluons to each quarkonium
is exact.\footnote{This result falsifies the conjecture in
Ref.~\cite{Bodwin:2008nf} that the soft cancellation might be inexact in
higher orders in $\alpha_s$, even at lowest order in $v$.} At lowest
order in $v$, only $S$-wave quarkonium production is possible. An
explicit calculation of the one-loop corrections to $S$-wave quarkonium
production in $B$-meson decays at lowest order in $v$ \cite{Chay:2000xn}
confirms our expectation that these corrections are free of infrared
divergences. In the case of double quarkonium production in $e^+e^-$
annihilation, the exact cancellation of the couplings of soft-singular
gluons holds for each quarkonium that is treated at lowest order in $v$.
If only one quarkonium is treated at lowest order in $v$, then  
$\tilde{S}$ can be absorbed into a re-definition of the 
distribution function of the remaining quarkonium, and the cancellation of
factorization-violating infrared divergences is expected to be exact.
This expectation is confirmed by an explicit calculation of the one-loop
corrections to $\sigma[e^+e^- \to J/\psi+\chi_{cJ}]$, where the $J/\psi$
is treated at lowest order in $v$ \cite{Zhang:2008gp}. Even in the case
of $S$-wave quarkonium production, we expect infrared divergence to
appear at higher orders in $v$, accompanied by a suppression factor $f$,
as discussed above.

Finally, we mention that we could have written the collinear functions
$\bar{J}^\pm$ that are associated with each quarkonium in terms of
light-cone distributions instead of NRQCD matrix elements. The
derivation of this result would entail the use of collinear
approximations for the momenta of the heavy-quark and heavy-antiquark in
meson~1~(2) in which one neglects the minus (plus) and transverse
components in comparison with the plus (minus) components. These
approximations introduce an error of relative order $f_i$ for each
quarkonium, and, in the case of double-quarkonium production, these
errors must be {\it added}, rather than {\it multiplied}, in order to
obtain the error for the complete amplitude. In the resulting
factorized expression, no power-suppressed soft divergences would
appear in $\tilde{H}$ because, when the quark and antiquark momenta in
each quarkonium are taken to be collinear to each, the cancellation of
soft divergences between the quark and antiquark in each quarkonium is
exact. In contrast, in the factorized expressions involving NRQCD
matrix elements that we have presented, power-suppressed soft
divergences {\it do} appear in $\tilde{H}$ and must be discarded in
order to obtain the factorized expression. However, as we have said,
these divergences are suppressed as $f_if_j$ (rather than $f_i+f_j$) in
double-quarkonium production. Hence, in the case of double-quarkonium
production, a factorized expression involving light-cone distributions
for the quarkonia would be less accurate than the expression involving
NRQCD matrix elements that we have presented.

\section{One-loop examples}\label{sec:1lex}

In this section, we illustrate some of the features of the
factorization result by presenting some one-loop examples for
double-charmonium production in $e^+e^-$ annihilation and for production
of a charmonium and a light meson in $B$-meson decay. In our examples,
we wish only to identify the soft divergences, and so we consider the
loop gluon to be soft, and we make use of the soft approximation in our
calculations.

\subsection{Soft approximation\label{sec:softapp}}

In the general factorization argument, we have taken the soft
approximation to be the replacement
\begin{equation}
g_{\mu\nu} \,\,\longrightarrow\,\,\frac{k_{\mu}p_{\nu}}{k\cdot p},
\end{equation}
in the gluon-propagator numerator. Recall that $k$ is the soft-gluon
momentum, that $p$ is the momentum of the line in the collinear
subdiagram, and that $\mu$ and $\nu$ correspond to the attachments of the
gluon to the collinear and soft subdiagrams, respectively. Consider now,
for instance, an initial-state quark with momentum $p$ that absorbs a
gluon with soft momentum $k$. Then, we can make use of the graphical
Ward identity (Feynman identity) to rewrite the soft approximation:
\begin{eqnarray}
&&\left(\frac{k_{\mu}p_{\nu}}{k\cdot p+i\varepsilon}\right)
\frac{\left(p\slsh+k\slsh\right)+m}{
\left(p+k\right)^2-m^2+i\varepsilon}\gamma_{\mu}u(p)\nonumber\\
&&\qquad =\frac{p_{\nu}}{k\cdot p+i\varepsilon}
\frac{\left(p\slsh+k\slsh\right)+m}
{\left(p+k\right)^2-m^2+i\varepsilon}
\left[p\slsh+k\slsh-m-\left(p\slsh-m\right)\right]u(p)
=\left(\frac{p_{\nu}}{k\cdot p+i\varepsilon}\right)u(p).\nonumber\\
\label{eq:softkp}
\end{eqnarray}
Since we are interested only in identifying the soft divergences, we 
eliminate any ultraviolet divergences in loop integrals, without 
affecting the soft divergences, by reintroducing 
the $k^2$ terms in the quark and antiquark denominators.  
That is, we make the substitution
\begin{equation}
2k\cdot p +i\varepsilon \to k^2+2k\cdot p+i\varepsilon
\end{equation}
in the denominator of the last line of Eq.~(\ref{eq:softkp}).
This is the form of the soft approximation that we will use.
Therefore, if a quark or antiquark line with physical momentum $p_i$
absorbs a gluon with soft momentum $k$, color $a$, and vector index
$\mu$, then the full amplitude is approximated as
\begin{equation}
\label{soft-approx}
\mathcal{A}[Q(p_i)+g(k)]
\approx
g_sT^a\frac{2b_i p_i^\mu}{k^2+2a_ik\cdot p_i+i\varepsilon}
\mathcal{A}[Q(p_i)],
\end{equation}
where
\begin{subequations}
\begin{eqnarray}
a_i&=&
\begin{cases}
+1&\textrm{initial-state particle} 
\\
-1&\textrm{final-state particle}
\end{cases},
\\
b_i&=&
\begin{cases}
+1&\textrm{quark} 
\\
-1&\textrm{antiquark}
\end{cases}.\label{eq:bi}
\end{eqnarray}
\end{subequations}

Some of the calculations that we present below  involve soft-gluon loop
corrections in which a gluon can be emitted or absorbed by two
particles, each of which can be a quark or an antiquark. [See
Figs.~\ref{fig:DoubleH} and \ref{fig:Bdecay}(a)--(d)]. If we choose the
sense of the loop momentum $k$ such that it is absorbed by the line with
momentum $p_i$, then application of Eq.~(\ref{soft-approx}) yields the
soft loop factor
\begin{equation}
I(p_i,p_j)=-ig_s^2\int \frac{d^dk}{(2\pi)^d}
\frac{4b_ib_jp_i\cdot p_j}{[k^2+2a_ik\cdot p_i+i\varepsilon]
[k^2-2a_jk\cdot p_j+i\varepsilon][k^2+i\varepsilon]},
\end{equation}
where we have regulated the soft divergence by using dimensional 
regularization, with $d=4-2\epsilon$.
The infrared-divergent part of the multiplicative correction factor from
this soft loop factor is given by 
\begin{equation}\label{eq:softfact}
I(p_i,p_j)=
\frac{\alpha_s}{4\pi\epsilon_{\textrm{IR}}}
\frac{a_ia_jb_ib_j}{\bar{\beta}_{ij}}
\Big[
\ln\left(\frac{1-\bar{\beta}_{ij}}{1+\bar{\beta}_{ij}}\right)
    +2\pi i \theta_{ij}
\Big],
\end{equation}
where $p_i$ is the physical momentum of the particle $i$,
\begin{subequations}
\begin{eqnarray}
\bar{\beta}_{ij}&\equiv&\bar{\beta}\left(p_i,p_j\right)=
\sqrt{1-\frac{p_i^2 p_j^2}{(p_i\cdot p_j)^2}},\\
\theta_{ij}&=&\frac{1}{2}(1+a_ia_j).
\end{eqnarray}
\end{subequations}
(See also Ref.~\cite{Nayak:2007zb}).

\subsection{Exclusive double quarkonium production\label{sec:DoubleH}}
\begin{figure}
\includegraphics[angle=0,height=7cm]{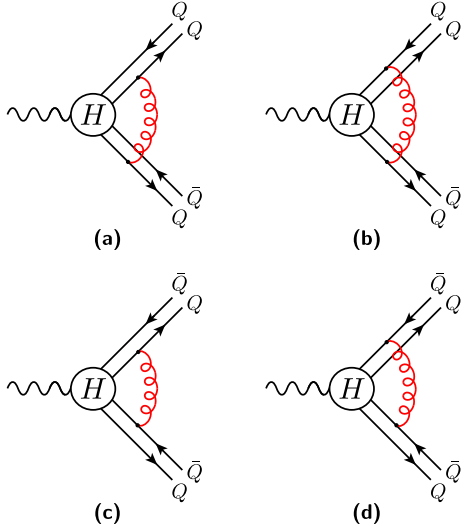}
\vspace*{8pt}
\caption{\label{fig:DoubleH}%
One-gluon corrections to the double-quarkonium production amplitude. The
blob labeled $H$ represents the lowest-order
hard-scattering process in which two heavy quark-antiquark pairs are
created.
}
\end{figure}
In this section we consider double quarkonium production in $e^+e^-$
annihilation, $\gamma^*\to H_1(P_1)+H_2(P_2)$. $H_1$ and $H_2$ are the
produced quarkonium states. For definiteness, we will assume that the
produced quarkonia are charmonium states.

Recall that there are QCD and QED contributions to the Born-level
hard-scattering process for double-quarkonium production in $e^+e^-$
annihilation. (See Ref.~\cite{Braaten:2002fi} for details.) However,
since this is an exclusive process, only color-singlet $Q\bar{Q}$ pairs
can contribute. Therefore, the soft-gluon loop corrections to the QED
diagrams are zero at one loop. Nonvanishing corrections to the QED
diagrams appear only at two-loop order.

We now write the amplitude as
\begin{equation}
\mathcal{A}_1^{\textrm{soft}}=R_{H_1+H_2}
\mathcal{A}_0\textrm{C},
\end{equation}
where $\textrm{C}$ is the appropriate color factor and $R_{H_1+H_2}$ is
the soft loop factor, which is given by
\begin{equation}
R_{H_1+H_2}=
I(p_{1q},p_{2q})+I(p_{1\bar q},p_{2q})+I(p_{1q},p_{2\bar q})
+I(p_{1\bar q},p_{2\bar q}).
\end{equation}
We obtain $I(p_{1i},p_{2j})$ from Eq.~(\ref{eq:softfact}). Retaining 
only those terms with one or fewer factors of $q_1$ or $q_2$ and 
discarding terms containing two or more powers of $z=4m_c^2/s$, we 
obtain
\begin{eqnarray}
I(p_{1i},p_{2j})&=&
\frac{\alpha_s}{4\pi\epsilon_{\rm IR}}\nonumber\\
&&\hbox{}\hskip -5em\times\left[2b_ib_j\pi 
i+b_ib_j\left(2\ln z+4z\right)-4b_{i}\frac{P_1\cdot q_2}{P_1\cdot P_2}
-4b_{j}\frac{P_2\cdot q_1}{P_1\cdot P_2}-8\frac{q_1\cdot q_2}{P_1\cdot P_2}
+8\frac{P_1\cdot q_2 P_2\cdot q_1}{(P_1\cdot P_2)^2}\right],\nonumber\\
\label{double-ir-factor}
\end{eqnarray}
where $b_i$ is defined in Eq.~(\ref{eq:bi}).

Now let us evaluate the various scalar products that appear in
Eq.~(\ref{double-ir-factor}). From Eqs.~(\ref{eq:def-Pandq}),
(\ref{eq:boost1}), and (\ref{eq:boost2}), we have
\begin{subequations}
\begin{eqnarray}
P_1\cdot P_2&=&\frac{1}{2}(s-M_1^2-M_2^2),\\
P_1\cdot q_2&=&
-\frac{Q}{M_2}P_{\rm CM}\hat{q}_{2}^z,
\\
P_2\cdot q_1&=&
+\frac{Q}{M_1}P_{\rm CM}\hat{q}_{1}^z,
\\
q_1\cdot q_2&=&
-
\left[
(P_{\rm CM}^2+E_1E_2)\frac{\hat{q}_{1}^z\hat{q}_{2}^z}{M_1M_2}
+\bm{q}_{1\perp}\cdot \bm{q}_{2\perp}\right]
\nonumber\\
&=&
-P_1\cdot P_2 \frac{\hat{q}_{1}^z\hat{q}_{2}^z}{M_1M_2}
-\bm{q}_{1\perp}\cdot \bm{q}_{2\perp},
\end{eqnarray}
\end{subequations}
where
\begin{equation}
P_{\rm CM}^2=\frac{1}{4s}[(s-M_1^2-M_2^2)^2-4M_1^2M_2^2].
\end{equation}
Thus, we see that all of the invariants are of order $Q^2=s$ and that
the various terms in Eq.~(\ref{double-ir-factor}) are of order
$(m_c/Q)^0$. If we add the contributions of Figs.~\ref{fig:DoubleH}(a) and
\ref{fig:DoubleH}(b) or Figs.~\ref{fig:DoubleH}(c) and
\ref{fig:DoubleH}(d), then the first, second, and third terms in
Eq.~(\ref{double-ir-factor}) cancel. Similarly, if we add the
contributions of Figs.~\ref{fig:DoubleH}(a) and \ref{fig:DoubleH}(c) or 
Figs.~\ref{fig:DoubleH}(b) and \ref{fig:DoubleH}(d), then
the first, second, and fourth terms in Eq.~(\ref{double-ir-factor})
cancel. Therefore, we obtain
\begin{equation}\label{eq:Rbevsc}
R_{H_1+H_2}=
\frac{8\alpha_s}{\pi\epsilon_{\textrm{IR}}}
\left[
-\frac{q_1\cdot q_2}{P_1\cdot P_2}
+\frac{q_1\cdot P_2 q_2\cdot P_1}{(P_1\cdot P_2)^2}
\right].
\end{equation}
Now,
\begin{subequations}
\begin{eqnarray}
-\frac{q_1\cdot q_2}{P_1\cdot P_2}&=&
\frac{\hat{q}_{1}^z\hat{q}_{2}^z}{M_1M_2}
+\frac{2\bm{q}_{1\perp}\cdot \bm{q}_{2\perp}}
      {s-M_1^2-M_2^2},\hspace{1cm}
\\
\frac{q_1\cdot P_2 q_2 \cdot P_1}{(P_1\cdot P_2)^2}&=&
-\frac{\hat{q}_{1}^z\hat{q}_{2}^z}{M_1M_2}
\frac{4sP_{\rm CM}^2}{[s-M_1^2-M_2^2]^2}.
\end{eqnarray}
\end{subequations}
Therefore, we have
\begin{eqnarray}
\label{eq:RH1H2scale}
R_{H_1+H_2}&=& \frac{16\alpha_s}{\pi\epsilon_{\textrm{IR}}}
\Big[\frac{\bm{q}_{1\perp}\cdot \bm{q}_{2\perp}}
 {s-M_1^2-M_2^2}+\hat{q}_1^z\hat{q}_2^z\frac{2M_1M_2}{[s-M_1^2-M_2^2]^2}
\Big]\nonumber\\
 & \approx& \frac{16\alpha_s}{\pi\epsilon_{\textrm{IR}}}\frac{\bm{q}_{1\perp}
\cdot \bm{q}_{2\perp}}{M^2}z \sim \frac{(m_cv)^2}{s},
\end{eqnarray}
where we have used the fact that the components of $\bm{q}_{i\perp}$ are
of order $m_c v$. Equation~(\ref{eq:RH1H2scale}) shows explicitly the
suppression of the soft divergence that is expected from the
factorization proof.

We can see that the soft divergent terms are proportional to one
power of $q_1$ and one power of $q_2$. Therefore, double-$S$-wave
production ($\gamma^*\to J/\psi +\eta_c$) and $S$-wave/$P$-wave
production ($\gamma^*\to J/\psi+\chi_c,$ $\eta_c +h_c$) at one-loop and at
leading order in $v$ are free of soft divergences. This is confirmed
by the explicit one-loop calculations of
Refs.~\cite{Zhang:2005cha,Zhang:2008gp}, for the $J/\psi + \eta_c$ and
$J/\psi + \chi_c$ cases. 

Note that, if we consider only the contributions from the diagrams of
Figs.~\ref{fig:DoubleH}(a) and \ref{fig:DoubleH}(b) or
Figs.~\ref{fig:DoubleH}(c) and \ref{fig:DoubleH}(d), then the fourth term in
Eq.~(\ref{double-ir-factor}) survives. Similarly, if we consider only
the contributions from the diagrams of Figs.~\ref{fig:DoubleH}(a) and
\ref{fig:DoubleH}(c) or Figs.~\ref{fig:DoubleH}(b) and \ref{fig:DoubleH}(d),
then the third term in Eq.~(\ref{double-ir-factor}) survives. Each of
these terms are of order $(m_c/Q)^0$. At first sight, this seems
puzzling, since we expect soft divergences to cancel up to terms of
order $m_c/Q$ when we add the contributions from the connections of a
soft gluon to the quark and antiquark in a quarkonium. The failure of
that cancellation in the present case can be understood because the soft
approximation that we have taken contains enhancements that arise when
the gluon momentum is nearly collinear to either of the quarkonia
momenta. Actual collinear divergences (logarithms of $m_c$) are absent
in the third and fourth terms in Eq.~(\ref{double-ir-factor}) because,
in the one-loop case, they appear with equal strength in the
contributions in which the gluon connects to the quark or the antiquark
in a quarkonium. However, a residual finite piece of the collinear
enhancement survives because the soft approximations for the quark and
antiquark lines are not equal when the gluon momentum is in the
collinear region. This failure of the soft cancellation when the soft
function contains collinear enhancements was noted in
Ref.~\cite{Bodwin:2009cb}. In the proof of factorization that we have
given, such collinear enhancements are removed from the soft function
$S$ and reside in the $J^\pm$ functions and associated eikonal lines. In
our one-loop example, we have neglected the dependence of the hard
function on the momentum of the gluon. Therefore, the $C^+$ eikonal
lines that arise from the $C^+$ enhancements in the diagrams of
Figs.~\ref{fig:DoubleH}(a) and \ref{fig:DoubleH}(c) or
Figs.~\ref{fig:DoubleH}(b) and \ref{fig:DoubleH}(d) cancel, and the $C^-$
eikonal lines that arise from the $C^-$ enhancements in the diagrams of
Figs.~\ref{fig:DoubleH}(a) and \ref{fig:DoubleH}(b) or
Figs.~\ref{fig:DoubleH}(c) and \ref{fig:DoubleH}(d) cancel. The sum of all
four diagrams in Fig.~\ref{fig:DoubleH} is therefore free of collinear
enhancements that could spoil the soft cancellation and is in accord with
the results of the factorization proof.

\subsection{$\bm{B}$-meson decays \label{sec:Bdecays}}

\begin{figure}
\includegraphics[angle=0,height=7cm]{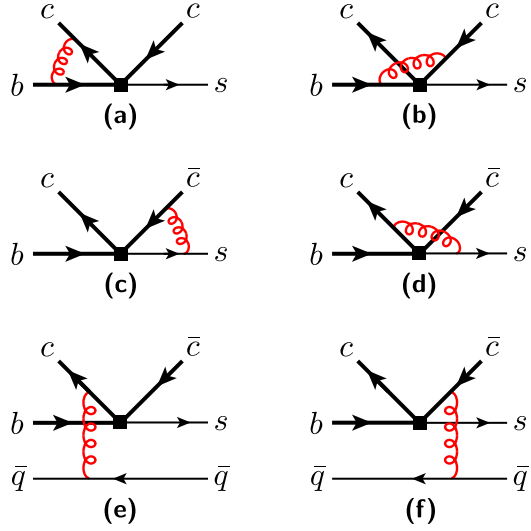}
\vspace*{8pt}
\caption{\label{fig:Bdecay}%
One-gluon corrections to the lowest-order $B$ decay amplitude. The black
square represents the electroweak interaction.
}
\end{figure}
In this section we consider decay of a $B$ meson into a light meson
plus a charmonium state. For definiteness, we take the light meson to
be a $K$ meson. Therefore we have the process $B(p_B)\to H(P_1)+K(p_K)$.
$H$ is the produced charmonium state, which we take to be a ${}^3P_J$
state. The
one-soft-gluon corrections to the lowest-order decay amplitude are
represented in Fig.~\ref{fig:Bdecay}. We  refer to diagrams (a)--(d)
as vertex corrections and to diagrams (e) and (f) as spectator
contributions.

\subsubsection{Vertex corrections\label{sec:vertex}}

In discussing soft contributions to the vertex corrections, we assume
that the light quark has a small mass $m_s$, and we work in the limit
$m_s\to 0$. In order to make contact with the results in
Ref.~\cite{Song:2003yc}, we neglect $q_k=-yp_K+p_{k_{q}}$ in comparison
with $p_K$, and we neglect $\bm{q}_B\sim\Lambda_{\rm QCD}$ in
comparison with $m_b$. We keep terms containing zero or one power of
$q_1$.

For the vertex corrections in which the soft gluon
attaches to the light-quark line [Figs.\ref{fig:Bdecay}(c) and
\ref{fig:Bdecay}(d)], we obtain
\begin{equation}
\bar{\beta}\Big(p_{k_{q}},p_{1q(\bar q)}\Big) =1-8\frac{m_c^2m_{s}^2}
{y^2\left(m_b^2-4m_c^2\right)^2}\left(1\mp 8\frac{p_{b}\cdot
q_1}{m_b^2-4m_c^2}\right).
\end{equation}
The upper sign corresponds to the quark with momentum $p_{1q}$,
and the lower sign corresponds to the antiquark with momentum
$p_{1\bar q}$. The soft loop factor is then
given by
\begin{equation}\label{softs}
I\left(p_{k_{q}},p_{1q(\bar q)}\right)  =  \pm\frac{\alpha_s}
{4\pi\epsilon_{\textrm{IR}}}\left[\ln\left(8\frac{m_c^2m_{s}^2}
{y^2\left(m_b^2-4m_c^2\right)^2}\right)
 \mp 8\frac{p_{b}\cdot q_1}{\left(m_b^2-4m_c^2\right)}+ 2\pi 
i\right].
\end{equation}
This expression contains collinear divergences, which manifest
themselves when we take $m_{s}=0$. The divergences arise because
the soft expressions contain contributions from $C^-$ momentum. However,
as we have explained in Sec.~\ref{sec:DoubleH}, at one-loop order, the
collinear divergences cancel when one sums over the connections to the
quark and the antiquark in the charmonium. Computing that sum, we
obtain
\begin{equation}\label{eq:2slines}
I\left(p_{k_{q}},p_{1q}\right)+I\left(p_{k_{q}},p_{1\bar q}\right)  =  
-\frac{16\alpha_s}{4\pi\epsilon_{\textrm{IR}}}
\frac{p_{b}\cdot q_1}{\left(m_b^2-4m_c^2\right)}
 = -\frac{16\alpha_s}{4\pi\epsilon_{\textrm{IR}}}
\frac{p_{b}\cdot q_1}{m_b^2\left(1-z\right)},
\end{equation}
where, again, $z=4m_c^2/m_b^2$. 

For the vertex corrections in which
the soft gluon attaches to the $b$-quark line
[Figs.~\ref{fig:Bdecay}(a) and \ref{fig:Bdecay}(b)], we obtain
\begin{equation}
\bar{\beta}\Big(p_{b},p_{1q(\bar q)}\Big)=\beta_1(1\pm\delta_1),
\end{equation}
with
\begin{equation}
\beta_1=\frac{m_b^2-4m_c^2}{m_b^2+4m_c^2},\; 
\delta_1=\frac{64m_c^2m_b^2p_{b}\cdot q_1}{(m_b^2+4m_c^2)(m_b^2-4m_c^2)^2}.
\end{equation}
The corresponding soft loop factors are given by
\begin{eqnarray}\label{softb}
I\Big(p_{b},p_{1q(\bar q)}\Big) & = &\frac{\alpha_s}
{4\pi\epsilon_{\textrm{IR}}}\!\left\{\pm\frac{1}{\beta_1}
\ln\left(\frac{1+\beta_1}{1-\beta_1}\right)\right.\\
&&\left.+\frac{\delta_1}{\beta_1}\!\!\left[\!\ln\!
\left(\frac{1-\beta_1}{1+\beta_1}\right)\!+\frac{2\beta_1}{1-\beta_1^2}\right]
\right\}\!.\nonumber
\end{eqnarray}
Summing over both diagrams in 
Figs.~\ref{fig:Bdecay}(a) and \ref{fig:Bdecay}(b), we obtain
\begin{eqnarray}\label{eq:2blines}
I\left(p_{b},p_{1q}\right)+I\left(p_{b},p_{1\bar q}\right) & = & 
\frac{2\alpha_s}{4\pi\epsilon_{\textrm{IR}}}
\frac{\delta_1}{\beta_1}\left[\!\ln\!
\left(\frac{1-\beta_1}{1+\beta_1}\right)\!
+\frac{2\beta_1}{1-\beta_1^2}\right]\nonumber\\
 & = & \frac{16\alpha_s}{4\pi\epsilon_{\textrm{IR}}}
\frac{p_{b}\cdot q_1}{m_b^2}\frac{2z}{(1-z)^3}
\left(\ln z+\frac{1-z^2}{2z}\right).
\end{eqnarray}

As we have explained in Sec.~\ref{sec:DoubleH}, it is necessary
to add the contributions of all four diagrams in order obtain the
suppression of the infrared-divergent terms because the soft expressions
contain $C^+$ contributions that spoil the soft cancellation. Those
$C^+$ contributions, which correspond to $C^+$ eikonal lines in the
general factorization proof, cancel at one-loop order when one sums over
the connections to the $b$ quark and the light quark. When we add the
contributions of all four diagrams, {\it i.e.}, Eqs.~(\ref{eq:2slines})
and (\ref{eq:2blines}), the leading term does indeed cancel, and we
obtain
\begin{equation}
I\left(p_{k_{q}},p_{1q}\right)+I\left(p_{k_{q}},p_{1\bar q}\right)
+I\left(p_{b},p_{1q}\right)+I\left(p_{b},p_{1\bar q}\right) 
=\frac{16\alpha_s}{4\pi\epsilon_{\textrm{IR}}}\frac{p_{b}\cdot q_1}{m_b^2}2z
\frac{1-z+\ln z}{(1-z)^3}.
\label{vertex-IR}
\end{equation}
The remaining infrared-divergent terms are suppressed at least as
$z\ln z$. We generally expect a suppression of the
factorization-violating contributions by only a factor $\sqrt{z}$.
However, as we have mentioned in
Sec.~\ref{sec:factorization-corrections}, the suppression factor can
become $z$ for production of $P$-wave quarkonia if one neglects the
transverse momenta of the constituents of the $B$ meson and the light
meson, as we are doing in the present example. As was noted in
Ref.~\cite{Song:2003yc}, the expression in Eq.~(\ref{vertex-IR}) gives a
nonvanishing contribution to the production of a ${}^3P_J$ charmonium
only if $J=0$ or $J=2$. The Born-level cross section to produce a
${}^3P_J$ charmonium with $J=0$ or $J=2$ vanishes, and so the violations
of factorization are suppressed only as $z\ln z/\alpha_s$ with respect
to the leading factorizing terms.

Finally, we mention that, when we include the Born factors in the
amplitude, along with the soft factor, and decompose $q_1$ and the
quarkonium spin polarization $\epsilon^\star$ into the $J=0$ and $J=2$
angular-momentum tensors, then we obtain agreement with the results in
Eqs.~(14) and (16) of Ref.~\cite{Song:2003yc}.

\subsubsection{Spectator contributions\label{sec:spectator}}

In the spectator contributions of Figs.~\ref{fig:Bdecay}(e) and
\ref{fig:Bdecay}(f), we initially assume that $p_l$, $q_k$ and
$\bar{y}p_K$ are all of order $\Lambda_{\rm QCD}$. Then, because the
gluon in Figs.~\ref{fig:Bdecay}(e) and \ref{fig:Bdecay}(f) carries
momentum $p_{k_{\bar{q}}}-p_l\sim \Lambda_{\rm QCD}$, the heavy-quark or
heavy-antiquark propagator is off shell by order $m_b\Lambda_{\rm QCD}$.
That is, its momentum is in the semihard region. Therefore, in this
example, we are discussing the further factorization of the gluons with 
endpoint soft momenta that was described in
Sec.~\ref{sec:endpoint-fact}. Eventually, we wish to make contact with
the results in Ref.~\cite{Song:2003yc}. In that work,
$q_k=\bar{y}p_K-p_{k_{\bar{q}}}$ was neglected in comparison with 
$\bar{y}p_K$.
Neglecting $q_k$ generates endpoint divergences in $\bar{y}$ that are
cut off in our model when $\bar{y}p_K\sim q_k\sim \Lambda_{\rm QCD}$.
Our discussion in this example also applies to the cancellation of those
endpoint divergences.

Since the gluon in Figs.~\ref{fig:Bdecay}(e) and \ref{fig:Bdecay}(f)
carries momentum of order $\Lambda_{\rm QCD}$ or less, we can apply
the soft approximation to the connections of the gluon to the
heavy-quark and heavy-antiquark lines. Keeping terms up to order $q_1$,
we find that the soft factor for the heavy quark and antiquark lines for
sum of the diagrams in Figs.~\ref{fig:Bdecay}(e) and
\ref{fig:Bdecay}(f) is
\begin{equation}\label{eq:spect}
S_\mu=\frac{1}{(p_{k_{\bar{q}}}-p_l)^2}
\frac{4}{P_1\cdot (p_{k_{\bar{q}}}-p_l)}q_\mu' ,
\end{equation}
where
\begin{equation}
q'= 
q_{1}-P_{1}
\frac{q_1\cdot (p_{k_{\bar{q}}}-p_l)}
     {P_1\cdot (p_{k_{\bar{q}}}-p_l)} ,
\end{equation}
and we have included the factor $1/(p_{k_{\bar{q}}}-p_l)^2$ from the
gluon propagator in $S_\mu$. We note that $S_\mu$ is proportional to
$q\sim m_cv$. We will concern ourselves only with the contributions to
the production a $P$-wave quarkonium at leading order in $v$. Therefore,
we can retain only terms of leading order in $v$ in the remaining
factors that are associated with the heavy quarkonium.

In computing the remaining factors in the spectator amplitudes, we make
use of the quark-antiquark spin-projection operators that are given in the
Appendix. In the projector for heavy quark and
antiquark, we take the spin-triplet case, which corresponds to the
calculation for the $\chi_{cJ}$ in Ref.~\cite{Song:2003yc}. The
projector for a charm-quark pair production can be obtained by setting
$m_q=m_{\bar{q}}=m_c$ and
$E_q+E_{\bar{q}}=2\sqrt{m_c^2+\hat{\bm{q}}_i^2}$  in
Eq.~(\ref{pi3prod}). Retaining only the terms of leading order in $v$ 
and using relativistic normalization, we have
\begin{equation}
\bar{\Pi}_3^{\rm onium}
\approx
-\frac{1}{2\sqrt{2}}
\,\epsilon\slsh^\star 
(/\!\!\!\!P_{1}+2m_c).
\end{equation}               

We obtain the $B$-meson projector by setting $m_q=m_b$, $E_q=m_b$,
$m_{\bar{q}}=m_l$, and $E_{\bar{q}}=E_l$ in Eq.~(\ref{pi1})  and
retaining the terms of leading order in $\Lambda_{\rm QCD}$. Using 
relativistic normalization, we have
\begin{equation}
\Pi_1^B \approx
C_B
(p\slsh_b+m_b)\gamma_5\left(1-\frac{p\slsh_l}{m_l}\right),
\end{equation}
where the light-antiquark mass $m_l$ is of order $\Lambda_{\rm QCD}$
and the factor $C_B$ is defined by
\begin{equation}
C_B= \frac{1}{2} \sqrt{\frac{m_l}{m_b(1+E_l/m_l)}}.
\end{equation}
Similarly, we obtain the $K$-meson projector by retaining the terms
in Eq.~(\ref{pi1prod}) of leading order in $\Lambda_{\rm QCD}$. Using
relativistic normalization, we have

\begin{equation}                                        
\Pi_1^K \approx
C_K\left\{
\left[1+\left(\frac{m_{\bar{q}}}{m_q}-1\right)y\right]
p\slsh_K\gamma_5
-\frac{1}{m_q}p\slsh_K q\slsh_k\gamma_5\right\},
\end{equation}
where we have retained small masses $m_q$ and $m_{\bar{q}}$ of order
$\Lambda_{\rm QCD}$ for the quark and antiquark, respectively.
The coefficient $C_K$ is defined by
\begin{equation}
C_K=\frac{m_q(E_q+m_q+E_{\bar{q}}+m_{\bar{q}})}{2\sqrt{2(E_q+m_q)
(E_{\bar{q}}+m_{\bar{q}})}(E_q+E_{\bar{q}})}.
\end{equation}

Now, the trace over the heavy quark and antiquark lines is
\begin{equation}
U_\rho={\rm Tr}[\Pi_3^{\rm onium}\gamma_\rho(1-\gamma_5)]
\approx -2\sqrt{2}
m_c\epsilon^\star_\rho.
\end{equation} 
The factor $\gamma_\rho(1-\gamma_5)$ comes from the $V-A$ weak vertex.
The trace over the $B$-meson and light-meson quark and antiquark lines 
is
\begin{eqnarray}\label{eq:trprop}
L_\rho^\mu&=&
\mathrm{Tr}\left[\Pi_1^K\gamma_{\rho}
\left(1-\gamma_5\right)\Pi_1^B\gamma^{\mu}\right]\nonumber\\
&\approx&
C_BC_K
\mathrm{Tr}
\left[\left(p\slsh_K\gamma_5
-\frac{1}{m} p\slsh_{K}q\slsh_{k}\gamma_5\right)
\gamma_{\rho}\left(1-\gamma_5\right)\left(p\slsh_b+m_b\right)
\gamma_5\left(1-\frac{p\slsh_l}{m}\right)\gamma^{\mu}\right],
\end{eqnarray}
where the factor $\gamma_{\rho}(1-\gamma_5)$ comes from the $V-A$ weak 
vertex, and, for simplicity, we have set 
$m_{q}=m_{\bar{q}}=m_l=m$. The complete 
amplitude corresponding to the diagrams in Figs.~\ref{fig:Bdecay}(e) and 
\ref{fig:Bdecay}(f) is 
\begin{eqnarray}
A^{\rm spectator}&=&\frac{-ig^2C_FC_{\rm EW}}{N_c^{3/2}}
S_\mu U^\rho L_\rho^\mu\nonumber\\
&\approx&\frac{-ig^2 C_FC_BC_KC_{\rm EW}}{N_c^{3/2}}\frac{8\sqrt{2}m_c}
{(p_l-p_{k_{\bar{q}}})^2 P_1\cdot (p_{k_{\bar{q}}}-p_l)}\nonumber\\
&&\times \mathrm{Tr}
\left[\left(-p\slsh_K\gamma_5+\frac{p\slsh_{K}q\slsh_k\gamma_5}{m}\right)
\epsilon\slsh^*\left(1-\gamma_5\right)\left(p\slsh_b+m_b\right)
\gamma_5\left(1-\frac{p\slsh_l}{m}\right)q\slsh'\right].
\label{A-spec}
\end{eqnarray}
Here, $C_{\rm
EW}=(G_F/\sqrt{2})[V_{cb}V_{cs}^* C_1-V_{tb}V_{ts}^*
(C_4+C_6)]$, where
$G_F$ is the Fermi constant, the $V_{q_1q_2}$ are the CKM matrix
elements, and the $C_i$ are the Wilson coefficients of the effective
electroweak Hamiltonian. (See, for example, Ref.~\cite{Song:2003yc} for
details.)

In Eq.~(\ref{A-spec}), the terms proportional to the spatial components
of $p_l$ vanish upon integration over the angles that are associated
with those spatial components. Then, using the fact that
$\gamma_0\approx p\slsh_b/m_b$, up to terms of relative order
$\Lambda_{\rm QCD}/m_b$, we can write
\begin{eqnarray}    
\label{eq:trprop2}%
A^{\rm spectator}&\approx&
\frac{-ig^2 C_FC_B(1+E_l/m)C_KC_{\rm EW}}{N_c^{3/2}}
\frac{8\sqrt{2}m_c}{(p_{k_{\bar{q}}}-p_l)^2P_1\cdot(p_{k_{\bar{q}}}-p_l)}
\nonumber\\
&&\times 
\mathrm{Tr}\left[
\left(-p\slsh_K\gamma_5+\frac{p\slsh_{K}q\slsh_k\gamma_5}{m}\right) 
\epsilon\slsh^*\left(1-\gamma_5\right)
(p\slsh_b+m_b) \gamma_5 q\slsh'\right].
\label{A-spec2}
\end{eqnarray}
The gamma-matrix factors in this expression that are associated with the
$B$ meson correspond to the leading-twist $B$-meson light-cone
$\Phi_{B1}$ in Eq.~(\ref{eq:PHI-B}). Expanding terms inside the
trace, we have
\begin{eqnarray}    
\label{traces}%
A^{\rm spectator}&\approx&
\frac{-ig^2 C_FC_B(1+E_l/m)C_KC_{\rm EW}}{N_c^{3/2}}
\frac{8\sqrt{2}m_c}{(p_{k_{\bar{q}}}-p_l)^2P_1\cdot(p_{k_{\bar{q}}}-p_l)}
\nonumber\\
&&\times 
\left\{-\mathrm{Tr}\left[p\slsh_K \epsilon\slsh^*\left(1-\gamma_5\right)
p\slsh_b q\slsh'\right]
-(m_b/m)\mathrm{Tr}\left[p\slsh_{K}q\slsh_k\epsilon\slsh^*
\left(1-\gamma_5\right)
q\slsh'\right]\right\}\nonumber\\
&\equiv&
\frac{-ig^2 C_FC_B(1+E_l/m)C_KC_{\rm EW}}{N_c^{3/2}}\frac{8\sqrt{2}m_c}
{(p_{k_{\bar{q}}}-p_l)^2 P_1\cdot (p_{k_{\bar{q}}}-p_l)}
(T_1+T_2).
\end{eqnarray}

In evaluating the sizes of the contributions to $A^{\rm spectator}$, 
we make use of the orders of magnitude of the components of the various 
four vectors in the $B$-meson rest frame. Some of these are given in 
Eqs.~(\ref{pk}), (\ref{qk}), (\ref{eq:qCM}), and (\ref{q-prime}). In the 
case of $q'$, we see from Eq.~(\ref{eq:boost1}) that, in the $B$-meson 
rest frame,
\begin{subequations}\label{q-prime}
\begin{eqnarray}
(q')^+&\sim& vm_c,\\
(q')^-&\sim& vm_c^2/m_b,\\
\bm{q}'_{\perp} &\sim& vm_c.
\end{eqnarray}
\end{subequations}
This is in contrast with either $P$ or $q$, which, as can be seen from
Eq.~(\ref{eq:qCM}), have plus components that are of order $m_b$ and
$vm_b$, respectively. The suppression of the plus component of $q'$
is a consequence of the soft cancellation.

Now consider the contribution of $T_1$ in Eq.~(\ref{traces}). $T_1$
comes from the $-p\slsh_K\gamma_5$ term in the first parenthesis in
Eq.~(\ref{A-spec2}), which corresponds to the leading-twist light-meson
light-cone distribution. The contribution of $T_1$ is proportional
to the one that was considered in Ref.~\cite{Song:2003yc}. Evaluating
the trace in $T_1$, we obtain
\begin{equation}
T_1=-4\left(p_K\cdot \epsilon^\star p_b\cdot q'
-p_K\cdot p_b \epsilon^\star \cdot q'
+p_K\cdot q'\epsilon^\star \cdot p_b
-i\epsilon^{\alpha\rho\beta\mu}
p_{k\alpha}\epsilon^\star_\rho p_{b\beta}q_\mu'\right),
\label{T1}
\end{equation}
where we have used the convention 
$\textrm{Tr}[\gamma_5/\!\!\!a/\,\!\!\!b\,/\!\!\!c\,/\!\!\!d]=4i
\epsilon_{\alpha\beta\gamma\delta} a^\alpha b^\beta c^\gamma d^\delta$,
with $\epsilon_{0123}=-\epsilon^{0123}=1$. It is now easily seen that
the terms in Eq.~(\ref{T1}) are of order $vm_b^3$, $vm_cm_b^2$,
$vm_b^3$, and $vm_cm_b^2$, respectively. We can compare this
contribution with the individual contributions that appear before we
apply the soft cancellation or with the  contributions in which the
gluon in Figs.~\ref{fig:Bdecay}(e) and \ref{fig:Bdecay}(f) attaches at
its upper end to the quark lines from the $B$ meson or the $K$
meson. In the cancelling contributions, $q'$ in $T_1$ is replaced
with $P$, and $T_1$ becomes of order $m_b^4/m_c$. Thus, we see that
Eq.~(\ref{T1}) is suppressed as $vm_c/m_b$ relative to the cancelling
contributions, in agreement with what we expect for the soft
cancellation from the general factorization proof. In the contributions
in which the gluon in Figs.~\ref{fig:Bdecay}(e) and \ref{fig:Bdecay}(f)
attaches at its upper end to the quark lines from the $B$ meson or the
$K$ meson, which are factorizing contributions that contribute to
the $B$-meson-$K$-meson form factor, $q'$ in $T_1$ is replaced
with $p_b$ or $p_K$. With this replacement, $T_1$ is again of order
$m_b^4/m_c$. However, as we have mentioned, the Born-level factorizing
contributions to the production of a ${}^3P_J$ charmonium with $J=0$ or
$J=2$ vanish, and so the contributions of Eq.~(\ref{T1}) are suppressed
as $m_c v/(\alpha_s m_b)$ relative to the leading factorizing
contributions.

Next consider the contribution of $T_2$ in Eq.~(\ref{traces}), which
comes from the $p\slsh_K q\slsh_k\gamma_5/m$ term in the first
parenthesis in Eq.~(\ref{A-spec2}). The leading contributions in
$T_2$ are of order $vm_b^3$ and are suppressed by a factor
$m_cv/(\alpha_s m_b)$ relative to the factorizing terms. However,
the leading contributions in $T_2$ are proportional to the
transverse components of $q_k$. Upon integration of $T_2$ over the
angles of the transverse components of $q_k$, these leading
contributions vanish. From Eq.~(\ref{qk}), we see that the minus
component of $q_k$ vanishes and that the plus component of $q_k$ is
suppressed by a factor $\Lambda_{\rm QCD}/m_b$ relative to the
transverse components. Hence, the contributions of $T_2$ are
suppressed by a factor $\Lambda_{\rm QCD}m_c v/(\alpha_s m_b^{2})$
relative to the factorizing terms, and are negligible in comparison
with the other factorization-violating contributions.

Now let us retain only the leading contribution to $A^{\rm
spectator}$, which is proportional to $T_1$. As we have mentioned, this
contribution is the one that was considered in
Ref.~\cite{Song:2003yc}. In that calculation, light-quark masses were
taken to be zero, $q_k=\bar{y}p_K-p_{k_{\bar{q}}}$ was neglected in
comparison with $\bar{y}p_K$, and $p_l$ was taken to have only a plus
component, which is written as $p_l^+=\xi p_B^+\approx \xi p_b^+$. Under
these assumptions, $p_{k_{\bar{q}}}=\bar{y}p_K$, which has only a minus
component that is nonzero, $q'=q_1-P_1(q_1\cdot p_K)/(P_1\cdot p_K)$,
and $(q')^+=0$. Then, the resulting contribution is
\begin{eqnarray}
A^{\rm spectator}&\approx&
\frac{-ig^2 C_FC_B(1+E_l/m)C_KC_{\rm EW}}{N_c^{3/2}}
\frac{16\sqrt{2}m_c}{\xi \bar{y}^2 p_K\cdot p_b P_1\cdot p_K}\nonumber\\
&&\times\left(p_K\cdot \epsilon^\star p_b\cdot q'
-p_K\cdot p_b \epsilon^\star \cdot q'
-i\epsilon^{\alpha\rho\beta\mu}
p_{k\alpha}\epsilon^\star_\rho p_{b\beta}q_\mu'\right),
\label{T1-approx}
\end{eqnarray}
which yields an endpoint divergence, owing to the factor $\bar{y}^2$ in
the denominator. The terms in parentheses in
Eq.~(\ref{T1-approx}) are all of order $vm_cm_b^2$. That is, there is a
suppression factor, relative to the cancelling terms, of order
$m_c^2/m_b^2$.  From the arguments of
Sec.~\ref{sec:factorization-corrections}, we expect such a suppression
because we are neglecting the transverse momenta of the constituents of
the $B$ meson and the light meson. In order to make contact with the
calculation of Ref.~\cite{Song:2003yc}, we make the replacements
$C_B(1+E_l/m)\to (-i/4)f_B \Phi_B(\xi)$ and $C_{K}\to (i/4)f_{K}
\Phi_{K}(y)$, multiply by a factor $\sqrt{2m_{\chi_c}/(2m_c)^2}\approx
\sqrt{1/m_c}$ to compensate for the normalization of the quarkonium
state relative to the normalizations of the quark and antiquark states,
multiply by a $P$-wave quarkonium spatial wave function, and integrate
over the wave-function momentum. Then, decomposing $q$ and
$\epsilon^\star$ into $J=0$ and $J=2$ angular-momentum tensors, we
obtain agreement between Eq.~(\ref{T1-approx}) and the divergent terms
in Eqs.~(15) and (17) of Ref.~\cite{Song:2003yc}.\footnote{We have also
checked that, if we keep the exact expression, rather than taking the
soft approximation, then we obtain the finite terms in Eqs.~(15) and
(17) of Ref.~\cite{Song:2003yc}.}
We find in the $J=1$ case that 
\begin{equation}
f_{\rm II}=-\frac{4\sqrt{2}\epsilon^*\cdot p_b z}{m_b^2(1-z)^2}
\int_0^1 d\xi\, \frac{\Phi_B(\xi)}{\xi}\, \int_0^1 dy\, 
\frac{\Phi_{K}(y)}{\bar{y}^2},
\end{equation}
where $f_{\rm II}$ is defined in Ref.~\cite{Song:2003yc} and we have 
retained only the infrared-divergent terms.

\section{Summary and Discussion\label{sec:concl}}

In this paper, we have given detailed proofs, valid to all orders in
$\alpha_s$, of factorization theorems for two exclusive
quarkonium-production processes: the production of two quarkonia in
$e^+e^-$ annihilation and the production of a charmonium and a
light-meson in $B$-meson decays. We have supplemented our proofs with
one-loop examples of the factorization and cancellation of soft
singularities. (See Sec.~\ref{sec:1lex}.) Proofs of these factorization
theorems were sketched in Ref.~\cite{Bodwin:2008nf}. In the present
paper, we have provided more detailed arguments. The proofs in
Ref.~\cite{Bodwin:2008nf} did not consider the possibility that on-shell
lines could emit gluons with arbitrarily small momenta. Such a
possibility arises, for example, when one computes short-distance
coefficients by making use of on-shell matching conditions. In the
present paper, we have shown that factorization still holds when one
takes into account this possibility. We have also given more refined
estimates of the violations of factorization than were given in
Ref.~\cite{Bodwin:2008nf}, by considering the dependence of such
violations on the velocity $v$ of the heavy quark or antiquark in the
quarkonium rest frame. We note that, although our proofs are
demonstrated in models in which external lines are taken to be on
the mass shell, the methods of these proofs would apply to off-shell
models as well, provided that the models maintain gauge invariance.

In the proofs of factorization, our general strategy has been to
identify soft singularities, collinear singularities, and would-be
collinear singularities that appear in the limit of zero heavy-quark
mass. By demonstrating the factorization of these singularities and
would-be singularities, we are able to argue that the associated
logarithmic enhancements also factorize. Once the logarithmic
enhancements have been removed, the remainder of the production
amplitude can depend only on the hard scale and, hence, is
perturbatively calculable.

In demonstrating the factorization of singularities and would-be
singularities, we have made use of standard techniques (see, for example,
Refs.~\cite{Sterman:1978bi,Sterman:1978bj,Collins:1985ue,Collins:1989gx}),
but we have had to augment them in order to deal with the situation in
which low-energy collinear gluons attach to soft gluons. For this
purpose, we made use of the approach developed in
Ref.~\cite{Bodwin:2009cb} in the context of the production of light
mesons in $e^+e^-$ annihilation. The methods of proof that we have given
here should, generally, be applicable to proofs of factorization for
other exclusive processes in QCD.

Our factorized form for exclusive production of two quarkonia in
$e^+e^-$ annihilation is given in Eq.~(\ref{ep-em-fact}). The expression
in Eq.~(\ref{ep-em-fact}) has been used in leading-order and
next-to-leading-order calculations of exclusive double-charmonium
production. It is generally referred to as the ``NRQCD factorization''
formula.

Our factorized form for the exclusive production of a charmonium and a
light meson in $B$ decays is given in Eq.~(\ref{B-meson-fact}). An
expression of the form in Eq.~(\ref{B-meson-fact}) was suggested in
Ref.~\cite{Beneke:2000ry} on the basis of an analysis of $B$-meson decays
to light mesons. (In Ref.~\cite{Beneke:2000ry}, the factorized form was
written in terms of the quarkonium light-cone distribution, rather than
in terms of NRQCD matrix elements.) In Ref.~\cite{Beneke:2000ry}, it was
conjectured that the violations of factorization should vanish in the limit
$m_c\to 0$, but a detailed analysis of the scaling of the violations of
factorization with $m_c$, $m_b$, and $v$ was not given. 

We find, generally, that the violations of factorization are suppressed
by a factor $f_i=m_c v^2/Q$ for each charmonium $i$ in an
$S$-wave state and by a factor $f_i=m_c/Q$ for each charmonium
$i$ in a higher orbital-angular-momentum state, where
$Q=\sqrt{s}$ is the CM energy in $e^+e^-$ annihilation and
$Q=m_{B}$ in $B$-meson decays. Because the violations of
factorization are proportional to $v$, they vanish (up to corrections
that are proportional to $\Lambda_{\rm QCD}/Q$) if one works to
order $v^0$ in one charmonium. This statement has been confirmed in
calculations at order $\alpha_s$ (Refs.~\cite{Chay:2000xn,Zhang:2008gp}).

In the case of $B$-meson decays, the error-suppression factors for 
$S$-wave and $P$-wave
charmonia are $f_i=v^2m_c/m_b$ and $f_i=m_c/m_b$,
respectively. These are not particularly small, and the violations of
factorization may well be comparable to the factorized contributions.
In the case of $e^+e^-$ annihilation, 
the error-suppression factors are smaller by
a factor $m_{b}/\sqrt{s}$ than in the case of $B$-meson decays.
Furthermore, there is a suppression factor for each
quarkonium in the process. Hence, in the case of $e^+e^-$ annihilation, the 
errors are likely to be sufficiently small that the factorization formula 
would be useful. Since the coefficients of the suppression factors are
nonperturbative quantities, their sizes must be determined, at present,
through phenomenological studies.

In special cases, the relative sizes of the violations of factorization
may be enhanced because of quantum-number considerations. For example,
in $B$-meson decays, the production of ${}^3P_0$ or ${}^3P_2$ charmonia
through factorized contributions is not allowed in order
$\alpha_s^0$. The production of ${}^3P_0$ or ${}^3P_2$ charmonia
through factorization-violating contributions \emph{does} occur in
order $\alpha_s^0$. Therefore, in these cases, the violations of
factorization are enhanced by a factor $1/\alpha_s$ relative to the
factorized contributions.

Finally, we mention that we could have written the collinear functions
$\bar{J}^\pm$ that are associated with each quarkonium in terms of
light-cone distributions instead of NRQCD matrix elements. As we have
explained in Sec.~\ref{sec:factorization-corrections}, such an approach
yields a hard-scattering function $\tilde{H}$ that is manifestly free of
soft divergences. In contrast, in the factorized expressions
involving NRQCD matrix elements that we have presented, power-suppressed
soft divergences {\it do} appear in $\tilde{H}$ and must be discarded.
In the case of double-charmonium production, these soft divergences are
suppressed as $f_if_j$, while the corrections to the
light-cone-distribution factorization formula are suppressed only as
$f_i+f_j$. Therefore, in the case of double-charmonium production, the
factorized expression involving NRQCD matrix elements that we have
presented is more accurate than a factorized expression involving
light-cone distributions for the quarkonia.

\begin{acknowledgments}
We thank Martin Beneke for several helpful discussions. We thank
In-chol Kim for his assistance in preparing the figures in this paper.
The work of G.T.B.\ and X.G.T.\ was supported by the U.S. Department of
Energy, Division of High Energy Physics, under Contract
No.~DE-AC02-06CH11357. The research of X.G.T.\ was also supported by Science
and Engineering Research Canada. The work of J.L.\ was supported by the
Korea Ministry of Education, Science, and Technology through the
National Research Foundation under Contract No.~2009-0086383.

\end{acknowledgments}

\appendix

\section{Spin projectors}\label{app:mesproj}
In this appendix we derive quark-antiquark spin projectors for the 
case in which the quark and antiquark have different masses. 
We take the momentum of the quark to be $p_{q}$ and the momentum
of the antiquark to be $p_{\bar{q}}$,
with both the quark and the antiquark on 
shell: $p_q^2=m_q^2$, $p_{\bar{q}}^2=m_{\bar{q}}^2$.
Therefore, in the quark-antiquark CM frame we have
\begin{subequations}
\begin{eqnarray}\label{p12}
\hat{p}_{q}&=&(E_{q},+\hat{\bm{q}}),
\\
\hat{p}_{\bar{q}}&=&(E_{\bar{q}},-\hat{\bm{q}}),
\end{eqnarray}
\end{subequations}
where 
$E_{q}=\sqrt{m_{q}^2+\hat{\bm{q}}^2}$,
$E_{\bar{q}}=\sqrt{m_{\bar{q}}^2+\hat{\bm{q}}^2}$,
 and $\hat{\bm{q}}$ is the quark
three-momentum in the CM frame. It follows that
\begin{subequations}
\begin{eqnarray}
p_{q}&=&\frac{E_{q}}{E_{q}+E_{\bar{q}}}P+q,
\\
p_{\bar{q}}&=&\frac{E_{\bar{q}}}{E_{q}+E_{\bar{q}}}P-q,
\end{eqnarray}
\end{subequations}
where $P=p_{q}+p_{\bar{q}}$ and, in the CM frame, 
$P$ and $q$ are given by
\begin{subequations}
\begin{eqnarray}
\hat{P}&=&(M,\bm{0}),
\\
\hat{q}&=&(0,\hat{\bm{q}}),
\end{eqnarray}
\end{subequations}
where $M=E_{q}+E_{\bar{q}}$.

The quark and antiquark spinors are given by 
\begin{subequations}
\begin{eqnarray}
u(p_q,s_q)&=&N_q
\begin{pmatrix}
(E_q+m_q)\xi(s_q)\\
\bm{q}\cdot \bm{\sigma}\xi(s_q)
\end{pmatrix},
\\
v(p_{\bar{q}},s_{\bar{q}})&=&N_{\bar{q}}
\begin{pmatrix}
-\bm{q}\cdot \bm{\sigma}\eta(s_{\bar{q}})\\
(E_{\bar{q}}+m_{\bar{q}})\eta(s_{\bar{q}})
\end{pmatrix},
\end{eqnarray}
\end{subequations}
where the normalization factors are 
\begin{equation}
N_i=
\left\{
\begin{array}{l}
{[2E_i(E_i+m_i)]}^{-\frac{1}{2}}\textrm{nonrelativistic},
\\
{[E_i+m_i]}^{-\frac{1}{2}}\textrm{relativistic},
\end{array}
\right.,
\end{equation}
for $i=q$ or $\bar{q}$. Here,
$\xi(s_q)$ and $\eta(s_{\bar{q}})$
are the two-component spinors for the spin states 
$s_q$ and $s_{\bar{q}}$, respectively, 
with $\eta$ in a representation that is conjugate to 
that of $\xi$. It then follows straightforwardly that the 
spin-singlet projector is given by
\begin{eqnarray}\label{pi1}
\Pi_1(p_q,m_q,p_{\bar{q}},m_{\bar{q}})&=&
\sum_{s_q,s_{\bar{q}}}u(p_q,s_q)\bar{v}(p_{\bar{q}},s_{\bar{q}})
\langle\textstyle{\frac{1}{2}}s_q,\textstyle{\frac{1}{2}}s_{\bar{q}}|00\rangle
\nonumber\\
&=&-\frac{N_qN_{\bar{q}}}{2(E_q+E_{\bar{q}})\sqrt{2}}
(p\slsh_q+m_q)
(P\!\slsh+E_q+E_{\bar{q}})\gamma_5
(p\slsh_{\bar{q}}-m_{\bar{q}})\nonumber\\
 &=&-
 \frac{\left(E_q+m_q+E_{\bar{q}}+m_{\bar{q}}\right)
N_qN_{\bar{q}}}{2\sqrt{2}(E_q+E_{\bar{q}})}
\,(p\slsh_q +m_q)\gamma_5(p\slsh_{\bar{q}}-m_{\bar{q}}),
\end{eqnarray}
and the spin-triplet projector is given by
\begin{eqnarray}\label{pi3}         
\Pi_3(p_q,m_q,p_{\bar{q}},m_{\bar{q}},\lambda)
&=&\sum_{s_q,s_{\bar{q}}}u(p_q,s_q) \bar{v}(p_{\bar{q}},s_{\bar{q}})
\langle\textstyle{\frac{1}{2}}s_q,
\textstyle{\frac{1}{2}}s_{\bar{q}}|1\lambda\rangle
\nonumber\\                      
&=&\frac{N_qN_{\bar{q}}}{2(E_q+E_{\bar{q}})\sqrt{2}}                      
(p\slsh_q+m_q)               
(P\!\slsh+E_q+E_{\bar{q}})\,/\!\!\!\epsilon\,(\lambda)
(p\slsh_{\bar{q}}-m_{\bar{q}}),
\end{eqnarray} 
where $\epsilon(\lambda)$ 
is the polarization vector of the spin-triplet state
whose components in the quarkonium rest frame are

\begin{subequations}
\begin{eqnarray}
\bm{\epsilon}(\pm)&=&\mp\frac{1}{\sqrt{2}}(1,\pm i,0),
\\
\bm{\epsilon}(0)&=&(0,0,1).
\end{eqnarray}
\end{subequations}
The results in Eqs.~(\ref{pi1}) and (\ref{pi3}) are equivalent, in the 
equal-mass case to those in Ref.~\cite{Bodwin:2002hg}.

Note that the spin projectors $\Pi_i$ in Eqs.~(\ref{pi1}) and (\ref{pi3})
are for the decay of a $q\bar{q}$ pair. The projectors $\bar{\Pi}_i$
for the production of a $q\bar{q}$ pair can be obtained in a similar
manner as

\begin{subequations}
\label{pi13prod}
\begin{eqnarray}
\label{pi1prod}
\bar{\Pi}_1(p_q,m_q,p_{\bar{q}},m_{\bar{q}})
&=&
\sum_{s_q,\,s_{\bar{q}}}
\langle \tfrac{1}{2}\, s_{q},\,
        \tfrac{1}{2}\, s_{\bar{q}}|
00\rangle
v(p_{\bar{q}},s_{\bar{q}})\bar{u}(p_q,s_q)
\nonumber\\
&=&
\frac{N_qN_{\bar{q}}}{2(E_q+E_{\bar{q}})\sqrt{2}}
(p\slsh_{\bar{q}}-m_{\bar{q}})
\gamma_5
(P\!\slsh+E_q+E_{\bar{q}})
(p\slsh_q+m_q)
\nonumber\\
 &=&
 \frac{\left(E_q+m_q+E_{\bar{q}}+m_{\bar{q}}\right)
N_qN_{\bar{q}}}{2\sqrt{2}(E_q+E_{\bar{q}})}
\,(p\slsh_{\bar{q}}-m_{\bar{q}})\gamma_5(p\slsh_q +m_q),
\\
\label{pi3prod}
\bar{\Pi}_3
(p_q,m_q,p_{\bar{q}},m_{\bar{q}},\lambda)
&=&
\sum_{s_q,\,s_{\bar{q}}}
\langle \tfrac{1}{2}\, s_{q},\,
        \tfrac{1}{2}\, s_{\bar{q}}|
1\lambda\rangle
v(p_{\bar{q}},s_{\bar{q}})\bar{u}(p_q,s_q)
\nonumber\\
&=&\frac{N_qN_{\bar{q}}}{2(E_q+E_{\bar{q}})\sqrt{2}}                      
(p\slsh_{\bar{q}}-m_{\bar{q}})
\,/\!\!\!\epsilon^{\,*}(\lambda)
(P\!\slsh+E_q+E_{\bar{q}})
(p\slsh_q+m_q).
\end{eqnarray}
\end{subequations}

The relationship between $\Pi_i$ and $\bar{\Pi}_i$ is
\begin{equation}
\bar{\Pi}_i=
\gamma^0\Pi_i^\dagger\gamma^0.
\end{equation}


\begin{thebibliography}{}
\bibitem{Bodwin:1994jh}
 G.~T.~Bodwin, E.~Braaten, and G.~P.~Lepage,
 Phys.\ Rev.\  D {\bf 51}, 1125 (1995)
 [Erratum-ibid.\  D {\bf 55}, 5853 (1997)]
 [arXiv:hep-ph/9407339].

\bibitem{Nayak:2005rt}
 G.~C.~Nayak, J.~W.~Qiu, and G.~Sterman,
 Phys.\ Rev.\  D {\bf 72}, 114012 (2005)
 [arXiv:hep-ph/0509021].

\bibitem{Nayak:2005rw}
 G.~C.~Nayak, J.~W.~Qiu, and G.~Sterman,
 Phys.\ Lett.\  B {\bf 613}, 45 (2005)
 [arXiv:hep-ph/0501235].

\bibitem{Nayak:2007mb}
 G.~C.~Nayak, J.~W.~Qiu, and G.~Sterman,
 Phys.\ Rev.\ Lett.\  {\bf 99}, 212001 (2007)
 [arXiv:0707.2973 [hep-ph]].

\bibitem{Nayak:2007zb}
 G.~C.~Nayak, J.~W.~Qiu, and G.~Sterman,
 Phys.\ Rev.\  D {\bf 77}, 034022 (2008)
 [arXiv:0711.3476 [hep-ph]].

\bibitem{Abe:2004ww}
 K.~Abe {\it et al.}  [Belle Collaboration],
 Phys.\ Rev.\  D {\bf 70}, 071102 (2004)
 [arXiv:hep-ex/0407009].

\bibitem{Aubert:2005tj}
 B.~Aubert {\it et al.}  [BABAR Collaboration],
 Phys.\ Rev.\  D {\bf 72}, 031101 (2005)
 [arXiv:hep-ex/0506062].

\bibitem{Braaten:2002fi}
  E.~Braaten and J.~Lee,
  Phys.\ Rev.\  D {\bf 67}, 054007 (2003)
  [Erratum-ibid.\  D {\bf 72}, 099901 (2005)]
  [arXiv:hep-ph/0211085].

\bibitem{Liu:2002wq}
  K.~Y.~Liu, Z.~G.~He, and K.~T.~Chao,
  Phys.\ Lett.\  B {\bf 557}, 45 (2003)
  [arXiv:hep-ph/0211181].

\bibitem{Hagiwara:2003cw}
  K.~Hagiwara, E.~Kou, and C.~F.~Qiao,
  Phys.\ Lett.\  B {\bf 570}, 39 (2003)
  [arXiv:hep-ph/0305102].

\bibitem{Bodwin:2007ga}
 G.~T.~Bodwin, J.~Lee, and C.~Yu,
 Phys.\ Rev.\  D {\bf 77}, 094018 (2008)
 [arXiv:0710.0995 [hep-ph]].

\bibitem{Beneke:2000ry}
 M.~Beneke, G.~Buchalla, M.~Neubert, and C.~T.~Sachrajda,
 Nucl.\ Phys.\  B {\bf 591}, 313 (2000)
 [arXiv:hep-ph/0006124].

\bibitem{Chay:2000xn}
 J.~Chay and C.~Kim,
 arXiv:hep-ph/0009244.

\bibitem{Bobeth:2007sh}
 C.~Bobeth, B.~Grinstein, and M.~Savrov,
 Phys.\ Rev.\  D {\bf 77}, 074007 (2008)
 [arXiv:0712.1953 [hep-ph]].

\bibitem{Bodwin:2008nf}
 G.~T.~Bodwin, X.~Garcia i Tormo, and J.~Lee,
 Phys.\ Rev.\ Lett.\  {\bf 101}, 102002 (2008)
 [arXiv:0805.3876 [hep-ph]].

\bibitem{Bodwin:2009cb}
 G.~T.~Bodwin, X.~Garcia i Tormo, and J.~Lee,
Phys.\ Rev.\  D {\bf 81}, 114005 (2010)
 [arXiv:0903.0569 [hep-ph]].

\bibitem{Beneke:2008pi}
 M.~Beneke and L.~Vernazza,
 Nucl.\ Phys.\  B {\bf 811}, 155 (2009)
 [arXiv:0810.3575 [hep-ph]].

\bibitem{Ma:2006hc}
 J.~P.~Ma and Z.~G.~Si,
 Phys.\ Lett.\  B {\bf 647}, 419 (2007)
 [arXiv:hep-ph/0608221].

\bibitem{Bell:2008er}
 G.~Bell and T.~Feldmann,
 JHEP {\bf 0804}, 061 (2008)
 [arXiv:0802.2221 [hep-ph]].

\bibitem{Becher:2003qh}
 T.~Becher, R.~J.~Hill, and M.~Neubert,
 Phys.\ Rev.\  D {\bf 69}, 054017 (2004)
 [arXiv:hep-ph/0308122].

\bibitem{Manohar:2006nz}
 A.~V.~Manohar and I.~W.~Stewart,
 Phys.\ Rev.\  D {\bf 76}, 074002 (2007)
 [arXiv:hep-ph/0605001].

\bibitem{Beneke:2003pa}
 M.~Beneke and T.~Feldmann,
 Nucl.\ Phys.\  B {\bf 685}, 249 (2004)
 [arXiv:hep-ph/0311335].

\bibitem{Bauer:2003td}
 C.~W.~Bauer, M.~P.~Dorsten, and M.~P.~Salem,
 Phys.\ Rev.\  D {\bf 69}, 114011 (2004)
 [arXiv:hep-ph/0312302].

\bibitem{Bodwin:1984hc}
 G.~T.~Bodwin,
 Phys.\ Rev.\  D {\bf 31}, 2616 (1985)
 [Erratum-ibid.\  D {\bf 34}, 3932 (1986)].

\bibitem{Collins:1985ue}
 J.~C.~Collins, D.~E.~Soper, and G.~Sterman,
 Nucl.\ Phys.\  B {\bf 261}, 104 (1985).

\bibitem{Collins:1989gx}
 J.~C.~Collins, D.~E.~Soper, and G.~Sterman,
 Adv.\ Ser.\ Direct.\ High Energy Phys.\  {\bf 5}, 1 (1988)
 [arXiv:hep-ph/0409313].

\bibitem{Sterman:1978bi}
 G.~Sterman,
 Phys.\ Rev.\  D {\bf 17}, 2773 (1978).

\bibitem{Sterman:1978bj}
 G.~Sterman,
 Phys.\ Rev.\  D {\bf 17}, 2789 (1978).

\bibitem{Grammer:1973db}
 G.~J.~Grammer and D.~R.~Yennie,
 Phys.\ Rev.\  D {\bf 8}, 4332 (1973).

\bibitem{Collins:1981uk}
 J.~C.~Collins and D.~E.~Soper,
 Nucl.\ Phys.\  B {\bf 193}, 381 (1981)
 [Erratum-ibid.\  B {\bf 213}, 545 (1983)].

\bibitem{Bosch:2003fc}
  S.~W.~Bosch, R.~J.~Hill, B.~O.~Lange and M.~Neubert,
  Phys.\ Rev.\  D {\bf 67}, 094014 (2003)
  [arXiv:hep-ph/0301123].

\bibitem{Beneke:2005vv}
  M.~Beneke and S.~Jager,
  Nucl.\ Phys.\  B {\bf 751}, 160 (2006)
  [arXiv:hep-ph/0512351].

\bibitem{Pilipp:2007mg}
  V.~Pilipp,
  Nucl.\ Phys.\  B {\bf 794}, 154 (2008)
  [arXiv:0709.3214 [hep-ph]].


\bibitem{Kivel:2006xc}
  N.~Kivel,
  JHEP {\bf 0705}, 019 (2007)
  [arXiv:hep-ph/0608291].

\bibitem{Beneke:2006mk}
  M.~Beneke and S.~Jager,
  Nucl.\ Phys.\  B {\bf 768}, 51 (2007)
  [arXiv:hep-ph/0610322].

\bibitem{Jain:2007dy}
  A.~Jain, I.~Z.~Rothstein, and I.~W.~Stewart,
  arXiv:0706.3399 [hep-ph].

\bibitem{Zhang:2008gp}
 Y.~J.~Zhang, Y.~Q.~Ma, and K.~T.~Chao,
 Phys.\ Rev.\  D {\bf 78}, 054006 (2008)
 [arXiv:0802.3655 [hep-ph]].

\bibitem{Zhang:2005cha}
 Y.~J.~Zhang, Y.~J.~Gao, and K.~T.~Chao,
 Phys.\ Rev.\ Lett.\  {\bf 96}, 092001 (2006)
 [arXiv:hep-ph/0506076].

\bibitem{Song:2003yc}
 Z.~Z.~Song, C.~Meng, Y.~J.~Gao, and K.~T.~Chao,
 Phys.\ Rev.\  D {\bf 69}, 054009 (2004)
 [arXiv:hep-ph/0309105].

\bibitem{Bodwin:2002hg}
 G.~T.~Bodwin and A.~Petrelli,
 Phys.\ Rev.\  D {\bf 66}, 094011 (2002)
 [arXiv:hep-ph/0205210].
\end{thebibliography}
\end{document}